\documentclass[twocolumn]{aastex63}
\usepackage{threeparttable}

\setcitestyle{citesep={,}}

\newcommand{\vsini}{$v\sin(i)$}
\newcommand{\prot}{$P_{rot}$}
\newcommand{\oi}{{[\ion{O}{1}]}$~\lambda$6300}

\newcommand{\halpha}{$H_{\alpha}$}

\newcommand{\mdot}{$\dot{M}_{acc}$}
\newcommand{\mwind}{$\dot{M}_{wind}$}
\newcommand{\msun}{$M_{\odot}$}

\newcommand{\msunyr}{\rm{$M_{\odot}\, yr^{-1}$}}
\newcommand{\kms}{$\,$\rm{km}$\,s^{-1}$}
\newcommand{\angstrom}{\text{\normalfont\AA}}
\usepackage{float}

\shorttitle{Rotational Evolution of CTTS}
\shortauthors{Serna et al.}

\graphicspath{{./}{figures/}}

\begin{document}

\title{Rotational Evolution of Classical T Tauri Stars: Models and Observations \footnote{\today}}

\correspondingauthor{Javier Serna}
\email{jserna@astro.unam.mx}

\author[0000-0001-7351-6540]{Javier Serna}
\affil{Instituto de Astronom\'{i}a, Universidad Aut\'{o}noma de M\'{e}xico,
Ensenada, B.C, M\'{e}xico}

\author[0000-0001-9147-3345]{Giovanni Pinz\'on}
\affiliation{Observatorio Astron\'omico Nacional, Universidad Nacional de Colombia,
Bogot\'a, Colombia}

\author[0000-0001-9797-5661]{Jes\'us Hern\'andez}
\affil{Instituto de Astronom\'{i}a, Universidad Aut\'{o}noma de M\'{e}xico,
Ensenada, B.C, M\'{e}xico}

\author[0000-0001-6647-862X]{Ezequiel Manzo-Mart\'{\i}nez}
\affil{Instituto de Astronom\'{i}a, Universidad Aut\'{o}noma de M\'{e}xico,
Ensenada, B.C, M\'{e}xico}

\author[0000-0001-8284-4343]{Karina Mauco}
\affil{Instituto de Astronom\'{i}a, Universidad Aut\'{o}noma de M\'{e}xico,
Ensenada, B.C, M\'{e}xico}
\affil{European Southern Observatory, Karl-Schwarzschild-Strasse 2, 85748 Garching bei München, Germany}

\author[0000-0001-8600-4798]{Carlos G. Rom\'an-Z\'u\~niga}
\affil{Instituto de Astronom\'{i}a, Universidad Aut\'{o}noma de M\'{e}xico,
Ensenada, B.C, M\'{e}xico}

\author[0000-0002-3950-5386]{Nuria Calvet}
\affiliation{Department of Astronomy, University of Michigan, 1085 South University Avenue, Ann Arbor, MI 48109, USA}

\author[0000-0001-7124-4094]{Cesar Brice\~{n}o}
\affiliation{Cerro Tololo Inter-American Observatory/NSF’s NOIRLab, Casilla 603, La Serena, Chile}

\author[0000-0002-7795-0018]{Ricardo L\'opez-Valdivia}
\affil{Instituto de Astronom\'{i}a, Universidad Aut\'{o}noma de M\'{e}xico,
Ensenada, B.C, M\'{e}xico}

\author[0000-0002-5365-1267]{Marina Kounkel}
\affil{Department of Physics and Astronomy, Vanderbilt University, VU Station 1807, Nashville, TN 37235, USA}

\author[0000-0003-1479-3059]{Guy S. Stringfellow}
\affiliation{Center for Astrophysics and Space Astronomy, Department of Astrophysical and Planetary Sciences, University of Colorado, Boulder, CO, 80309, USA}

\author[0000-0002-5365-1267]{Keivan G.\ Stassun}
\affil{Department of Physics and Astronomy, Vanderbilt University, VU Station 1807, Nashville, TN 37235, USA}

\author[0000-0002-7549-7766]{Marc Pinsonnealt}
\affil{Department of Astronomy, The Ohio State University, Columbus, OH 43210, USA}

\author[0000-0002-6328-6099]{Lucia Adame}
\affil{Instituto de Astronom\'{i}a, Universidad Aut\'{o}noma de M\'{e}xico,
Ensenada, B.C, M\'{e}xico}

\author[0000-0002-8849-9816]{Lyra Cao}
\affil{Department of Astronomy, The Ohio State University, Columbus, OH 43210, USA}

\author[0000-0001-6914-7797]{Kevin Covey}
\affil{Department of Physics and Astronomy, Western Washington University, 516 High St, Bellingham, WA 98225}

\author[0000-0001-7868-7031]{Amelia Bayo}
\affiliation{Instituto de F\'{\i}sica y Astronom\'{\i}a, Universidad de Valpara\'{\i}so, Chile}
\affiliation{European Southern Observatory, Karl-Schwarzschild-Strasse 2, 85748 Garching bei München, Germany}

\author[0000-0002-1379-4204]{Alexandre Roman-Lopes}
\affiliation{Departamento de Astronom{\'i}a, Universidad de La Serena, 1700000 La Serena, Chile}

\author[0000-0003-4752-4365]{Christian Nitschelm}
\affil{Centro de Astronom{\'i}a (CITEVA), Universidad de Antofagasta, Avenida Angamos 601, Antofagasta 1270300, Chile}

\author[0000-0003-1805-0316]{Richard R. Lane}
\affil{Centro de Investigaci\'{o}n en Astronom\'{i}a, Universidad Bernardo O'Higgins, Avenida Viel 1497, Santiago, Chile}

\begin{abstract}

\noindent We developed a grid of stellar rotation models for low-mass and solar-type Classical T Tauri stars (CTTS) ($0.3M_{\odot}<M_{\ast}<1.2M_{\odot}$). These models incorporate the star-disk interaction and magnetospheric ejections to investigate the evolution of the stellar rotation rate as a function of the mass of the star $M_{\ast}$, the magnetic field ($B_{\ast}$), and stellar wind ($\dot{M}_{wind}$).
We compiled and determined stellar parameters for 208 CTTS, such as projected rotational velocity {\vsini}, mass accretion rate {\mdot}, stellar mass $M_{\ast}$, ages, and estimated rotational periods using TESS data. We also estimated a representative value of the mass-loss rate for our sample using the {\oi} spectral line.
Our results confirm that {\vsini} measurements in CTTS agree with the rotation rates provided by our spin models in the accretion-powered stellar winds (APSW) picture. In addition, we used the Approximate Bayesian Computation (ABC) technique to explore the connection between the model parameters and the observational properties of CTTS. We find that the evolution of {\vsini} with age might be regulated by variations in (1) the intensity of $B_{\ast}$ and (2) the fraction of the accretion flow ejected in magnetic winds, removing angular momentum from these systems. The youngest stars in our sample ($\sim $1 Myr) show a median branching ratio $\dot{M}_{wind}/\dot{M}_{acc}\sim$ $0.16$ and median $B_{\ast}\sim$ 2000 G, in contrast to $\sim 0.01$ and 1000 G, respectively, for stars with ages $\gtrsim 3$ Myr.\\

\end{abstract}

\keywords{\href{http://astrothesaurus.org/uat/252}{Classical T Tauri stars (252)}  -- \href{http://astrothesaurus.org/uat/1599}{Stellar evolution (1599)} -- \href{http://astrothesaurus.org/uat/1610}{Stellar magnetic fields (1610)} -- \href{http://astrothesaurus.org/uat/1624}{Stellar properties (1624)} -- \href{http://astrothesaurus.org/uat/1629}{Stellar rotation (1629)} -- \href{http://astrothesaurus.org/uat/1636}{Stellar winds (1636)}}

\accepted{in ApJ, \today}

\section{Introduction} \label{sec:intro}

Low mass ($<1.2M_{\odot}$), Classical T Tauri Stars (CTTS) are young systems, still contracting and accreting gas from their circumstellar disks through the stellar magnetic field. Despite crossing a phase of gravitational contraction, the rotation rates of the majority of CTTS lie well below $10$\% of their break-up limit \citep{Bouvier1993,Herbst_2002,Jayawardhana_2006,Nguyen_2009,Bayo_2012,Pinzon2021,Serna_2021}, indicating that these stars lose angular momentum expeditiously during the first million years (Myr).\\

In the so-called Disk-Locking (DL) scenario, angular momentum in CTTS is transferred outward from the star to the disk regions beyond the corotation radius through the stellar magnetic field closed lines anchored to the star surface \citep{Ghosh_1979,CollierCameron_1993}.  Nevertheless, the  stellar spin-down efficiency in the DL model is substantially reduced due to the turbulent diffusion of the field inside the disk, which results in more realistic magnetic field topologies with open field regions \citep{Matt_2005a}. These regions allow the appearance of stellar winds, which are an alternative source of angular momentum loss \citep{Matt_2008a}. \\

CTTS exhibit energetic stellar and disk winds traced through the blue-shifted absorption profiles of their most prominent lines and by the presence of intense forbidden line emission of low ionization species in their spectra \citep{Kuhi1964,Hartmann_1982,Alencar2000,Edwards2003}. The correlation between the forbidden line emission and stellar accretion suggests that such winds are somehow powered by disk accretion \citep{Hartigan1995}. However, how the energy is transferred from the accretion flow to a stellar wind through the corona is still unknown.\\

Magnetohydrodynamic (MHD) simulations of CTTS, including stellar and disk wind scenarios, have shown that stellar rotation can reach an equilibrium around 10\% of the break-up limit at the end of the Hayashi track for X-wind models \citep{Shu_1988}. Mass loss mechanisms such as stellar wind models \citep{Tout1992}, accretion-powered stellar winds models \citep[APSW,][]{Matt_2008a,Matt_2008b}, and disk wind models \citep{Zanni2013,Ercolano_2021} can be involved to reach this equilibrium.
While all these models rely on the basis  that a fraction  of the accretion luminosity powers the stellar wind, only APSW establishes an upper limit for the so-called branching ratio between the mass loss rate in the stellar wind and the disk accretion rate ($\chi\equiv\dot{M}_{wind} / \dot{M}_{acc}$). This upper limit of $\chi\sim0.6$  has been confirmed by \cite{Watson_2016} using infrared forbidden line emission from the Spitzer and Herschel observations. The authors obtain a distribution of $\chi$ centered close to $0.1$ with a tail towards lower values ($\chi<0.6$), whose effect on the angular momentum evolution of CTTS has not been properly explored.\\

In this work, we computed synthetic rotational velocities for low-mass stars ($0.3M_{\odot}<M_{\ast}<1.2M_{\odot}$) in the framework of APSW models with the aim of obtaining the branching ratio distribution of a well-characterized CTTS sample with available rotation rates and accretion indicators. We use a Bayesian approach to investigate the connection between the branching-ratio parameter, the stellar magnetic fields, rotational evolution, and internal structure.
This paper is organized as follows. In \S \ref{sec:sample}, we describe the observations and data sources used to study our sample of CTTS. In \S \ref{sec:results}, we estimate the rotation periods using the TESS light curves. We derive the stellar mass accretion rate based on the $H_{\alpha}$ spectral line and identify wind tracers (e.g., {\oi}) to estimate the mass-loss rate by winds ({\mwind}). We also build and present the spin models for CTTS and use the Approximate Bayesian Computation (ABC) technique to investigate the connection between the model parameters and available observable distributions of CTTS.
In \S \ref{sec:discussion}, we discuss the implications of our results. Finally, in \S\ref{sec:conclusions}, we present the conclusions of this work.\\

\section{T Tauri Stars Sample And Data Sources} \label{sec:sample}

\subsection{CTTS Sample} \label{sec:ctts}

We use the sub-sample of {208 CTTS studied in \citet{Serna_2021}. 
This sub-sample has been confirmed and characterized by \citet{Briceno_2019} and \citet{Hernandez_2014}. It includes low-mass accretors of the on/off cloud population in the Orion OB1 association, which span ages from 1 to 10 Myr (e.g., ONC, $\sigma$ Ori, 25 Ori). Each of these targets has measurements of the {\halpha} equivalent width, {\vsini}, as well as masses and ages derived using the \texttt{MassAge} code \citep[][Hern\'andez et al., in preparation]{Serna_2021}. The list of parameters can be seen in Table \ref{tab:parameters}.\\

\subsection{APOGEE-2 Data and Stellar Parameters} \label{sec:apogee}

The APOGEE-2 northern spectrograph \citep[APOGEE;][]{Majewski2017} is located on the 2.5 m SDSS telescope at the Apache Point Observatory and observed thousands of stars for the APOGEE-2 Young Cluster Survey. This instrument can simultaneously observe up to 300 objects at high resolution ($R\sim22500$) in the H-band (15100–17000{\angstrom}) across a 1.5 deg radius field-of-view.\\

\citet{Kounkel_2018} analyzed data from the APOGEE-2 survey in the Orion Star-Forming Complex and reported nearly 2400 kinematic members using six-dimensional analysis (positions, parallax, and proper motions from \textit{Gaia}-DR2 and radial velocities (RV) from APOGEE-2). Additionally, for each one of these members, \citet{Kounkel_2019} report (effective temperature $T_{eff}$, surface gravity $log(g)$, $v\sin(i)$, RV). However, the authors warn that there are likely systematic features in the parameter space due to theoretical templates not offering a perfect match to the real data. To reduce these issues on the estimated parameters, \citet{Olney_2020}, based on a deep learning analysis, have provided more reliable predictions of $\log g$, $T_{eff}$, and [Fe/H] for low-mass stars. More recently, \citet{Serna_2021} estimated {\vsini} using the Fourier method, an independent method that does not require theoretical templates, and found that measurements are in agreement with the estimations of \citet{Kounkel_2019}, indicating that the systematic features in the parameter space suggested by these authors do not affect their   {\vsini} derivation. Since the Fourier method requires a high signal-to-noise ratio, applying this method to the entire CTTS sample was not possible. Thus, we adopt \citet{Kounkel_2019} measurements for our sample of CTTS.\\

\subsection{LAMOST}
\label{s:lamost_data}
We use the data archive from the Large Sky Area Multi-Object Fiber Spectroscopic Telescope \citep[LAMOST;][]{Luo_2015,Liu_2020}, located on the 4-meter quasi-meridian reflecting Schmidt telescope at the Xinlong station of the National Astronomical Observatory. We downloaded 55 spectra of our CTTS sample at low-resolution (R$\sim$1800) in the wavelength range 3650–9000{\angstrom}, and 7 spectra at medium-resolution (R$\sim7500$) in two bands, which cover the wavelength ranges 4950-5350{\angstrom}, and 6300-6800{\angstrom}. From these spectra, only 24 have the presence of the {\oi} line, which was used for the mass-loss rate estimations \S \ref{sec:mass_loss_rates}. The remaining spectra of the 55 stars without [OI] were not used in the analysis.
The details of the LAMOST data and plots of the {\oi} profile are presented in Appendix \ref{ap:A}.\\

\subsection{X-Shooter and Giraffe/ESO}
\label{s:eso_data}
We download spectra for 25 CTTS using the ESO Archive Science Portal. From this sample, there are 15 stars with resolution $R<18340$ (X-Shooter) and 10 stars with $R=24000$ (Giraffe); both instruments are installed in the ESO Very Large Telescope \citep{Vernet_2011}. We rejected three stars observed with Giraffe that do not exhibit the {\oi} line.
We use the data reduced and flux-calibrated by the ESO calibration pipelines available in the ESO archive. Additionally, we used the \texttt{Molecfit} tool \citep{Kausch_2015} to correct each spectrum from telluric lines. 
Table \ref{tab:obs} shows a summary and observations details.\\


\begin{longrotatetable}
\begin{deluxetable*}{ccccccccc}
\tablecaption{Stellar parameters of the CTTS sample\label{tab:parameters}}

\tablewidth{0pt}
\tablehead{ \colhead{2MASS ID} & \colhead{$v\sin(i)$\tablenotemark{a}} & \colhead{$M_{\ast}$} & \colhead{Age} & \colhead{$P_{rot}$\tablenotemark{b}} & \colhead{$P_{rot}$\tablenotemark{c}} & \colhead{$\log{\dot{M}_{acc}}$} & \colhead{$\log{\dot{M}_{wind}}$} & \colhead{Binary\tablenotemark{a}} \\
\colhead{} & \colhead{(km~s$^{-1}$)}  &
\colhead{($M_{\odot}$)} & \colhead{(Myr)} & \colhead{(days)} & \colhead{(days)} &
\colhead{($\log$$M_{\odot}$ yr$^{-1}$)} & \colhead{($\log$$M_{\odot}$ yr$^{-1}$)} & 
}
\decimalcolnumbers
\startdata
05390878-0231115 & 12.8 $\pm$ 0.6 & 0.3 $\pm$ 0.03 & 2.36 $\pm$ 0.53 & \nodata & \nodata & -9.37 $\pm$ 0.31 & -11.13 $\pm$ 0.83 & 1\\
  05380097-0226079 & 16.2 $\pm$ 0.5 & 0.37 $\pm$ 0.03 & 3.77 $\pm$ 0.85 & \nodata & \nodata & -9.01 $\pm$ 0.3 & -10.43 $\pm$ 0.84 & 1\\
  05344178-0453462 & 10.7 $\pm$ 0.7 & 0.44 $\pm$ 0.04 & 1.73 $\pm$ 0.33 & \nodata & \nodata & -8.95 $\pm$ 0.3 & -10.05 $\pm$ 0.84 & 1\\
  05373094-0223427 & 7.7 $\pm$ 0.6 & 0.36 $\pm$ 0.03 & 6.66 $\pm$ 1.46 & \nodata & \nodata & -8.93 $\pm$ 0.3 & -10.38 $\pm$ 0.84 & 1\\
  05391151-0231065 & 13.7 $\pm$ 0.5 & 0.5 $\pm$ 0.04 & 1.3 $\pm$ 0.21 & \nodata & \nodata & -8.89 $\pm$ 0.3 & -9.76 $\pm$ 0.84 & 1\\
  05332852-0517262 & 8.1 $\pm$ 1.2 & 0.5 $\pm$ 0.04 & 1.56 $\pm$ 0.27 & \nodata & \nodata & -8.7 $\pm$ 0.29 & -10.26 $\pm$ 0.83 & 1\\
  05380826-0235562 & 8.8 $\pm$ 0.5 & 0.4 $\pm$ 0.03 & 1.03 $\pm$ 0.18 & \nodata & \nodata & -8.68 $\pm$ 0.29 & -9.9 $\pm$ 0.84 & 1\\
  05384027-0230185 & 18.3 $\pm$ 1.1 & 0.49 $\pm$ 0.03 & 0.8 $\pm$ 0.11 & 7.69 $\pm$ 0.01 & \nodata & -8.67 $\pm$ 0.29 & -9.37 $\pm$ 0.85 & 1\\
  05402461-0152309 & 14.4 $\pm$ 0.5 & 0.54 $\pm$ 0.04 & 1.9 $\pm$ 0.34 & \nodata & \nodata & -8.66 $\pm$ 0.29 & -10.02 $\pm$ 0.84 & 1\\
  05393938-0217045 & 14.1 $\pm$ 0.5 & 0.6 $\pm$ 0.04 & 0.98 $\pm$ 0.16 & \nodata & 13.52 $\pm$ 0.01 & -8.61 $\pm$ 0.29 & -9.93 $\pm$ 0.83 & 1\\
  05343395-0534512 & 22.3 $\pm$ 0.5 & 0.54 $\pm$ 0.04 & 0.82 $\pm$ 0.12 & \nodata & \nodata & -8.56 $\pm$ 0.29 & -9.03 $\pm$ 0.85 & 1\\
  05395362-0233426 & 0.1 $\pm$ 0.6 & 0.35 $\pm$ 0.03 & 2.68 $\pm$ 0.55 & \nodata & \nodata & -8.56 $\pm$ 0.29 & -10.75 $\pm$ 0.83 & 1\\
  05324196-0539239 & 60.7 $\pm$ 1.7 & 0.75 $\pm$ 0.07 & 0.41 $\pm$ 0.07 & 1.7 $\pm$ 0.01 & 1.69 $\pm$ 0.01 & -8.56 $\pm$ 0.29 & -9.29 $\pm$ 0.84 & -1\\
  05400195-0221325 & 8.8 $\pm$ 0.5 & 0.36 $\pm$ 0.03 & 0.96 $\pm$ 0.15 & \nodata & \nodata & -8.54 $\pm$ 0.28 & -10.71 $\pm$ 0.83 & 1\\
  05391883-0230531 & 47.9 $\pm$ 2.6 & 1.1 $\pm$ 0.18 & 2.45 $\pm$ 1.33 & \nodata & 1.83 $\pm$ 0.01 & -8.53 $\pm$ 0.28 & -9.79 $\pm$ 0.83 & -1\\
  05380674-0230227 & 15.3 $\pm$ 2.1 & 0.52 $\pm$ 0.03 & 0.77 $\pm$ 0.11 & \nodata & \nodata & -8.51 $\pm$ 0.28 & -9.05 $\pm$ 0.85 & 1\\
  05401274-0228199 & 50.9 $\pm$ 1.4 & 0.45 $\pm$ 0.03 & 0.52 $\pm$ 0.05 & 1.53 $\pm$ 0.01 & \nodata & -8.49 $\pm$ 0.28 & -10.86 $\pm$ 0.81 & -1\\
  05354600-0057522 & 10.3 $\pm$ 0.7 & 0.65 $\pm$ 0.05 & 1.75 $\pm$ 0.32 & \nodata & \nodata & -8.49 $\pm$ 0.28 & -10.01 $\pm$ 0.83 & 1\tablenotemark{*}\\
  05341420-0542210 & 5.4 $\pm$ 1.1 & 0.62 $\pm$ 0.05 & 2.53 $\pm$ 0.52 & \nodata & \nodata & -8.42 $\pm$ 0.28 & -9.12 $\pm$ 0.85 & 1\\
  05394017-0220480 & 16.1 $\pm$ 0.8 & 0.77 $\pm$ 0.05 & 1.6 $\pm$ 0.27 & 5.8 $\pm$ 0.01 & 5.64 $\pm$ 0.01 & -8.4 $\pm$ 0.28 & -9.73 $\pm$ 0.84 & 1\\
\enddata
\tablenotetext{a}{\citet{Kounkel_2018,Kounkel_2019}: 0- Undeconvolvable cross-correlation function (CCF); 1- Only a single component in the CCF; 2- Multiple components in the CCF; -1- Spotted pairs or SB2 Uncertain}
\tablenotetext{b}{TESS Sector 6}
\tablenotetext{c}{TESS Sector 32}
\tablenotetext{*}{\citet{Tokovinin_2020} visual binaries}
\tablecomments{Only a portion of the table is shown here. The full version is available in electronic form.}
\end{deluxetable*}
\end{longrotatetable}

\begin{deluxetable*}{lccccccc}[t]
\tablecaption{Summary of ESO archival spectroscopy \label{tab:obs}}
\tablehead{
\colhead{Target} 	&
\colhead{Instrument\tablenotemark{*}} 	&
\colhead{Spectral} &
\colhead{Obs. Date} &
\colhead{Exp. time} &
\colhead{SNR} &
\colhead{ESO Program} \\
\colhead{} 		&
\colhead{} 		&
\colhead{Resolution} 		&
\colhead{(UT)} 	&
\colhead{(s)} &
\colhead{}      &
\colhead{ID}
}
\startdata
\multicolumn{7}{c}{}\\
SO 73 & XSHOOTER  & 8935     & 2019-11-18    &  2400  & 106   & 0104.C-0454(A) \\
SO 299 & GIRAFFE & 24000     & 2009-02-04    &  2775  & 22   & 082.C-0313(B) \\
SO 341 & XSHOOTER  & 1029     & 2020-02-07    &  90  & 28   & 0104.C-0454(A) \\
SO 362 & XSHOOTER  & 1512     & 2020-02-07    &  200  & 41   & 0104.C-0454(A) \\
SO 662 & XSHOOTER  & 1512     & 2020-02-07    &  60  & 38   & 0104.C-0454(A) \\
SO 908 & GIRAFFE & 24000     & 2010-01-03    &  2775  & 39   & 084.C-0282(A) \\
SO 927 & GIRAFFE & 24000     & 2010-01-03    &  2775  & 62   & 0103.C-0887(B) \\
SO 984 & XSHOOTER  & 18340     & 2019-11-17    &  420  & 83   & 0104.C-0454(A) \\
SO 1036 & XSHOOTER 	& 1512 	& 2019-10-11 	& 60 	&  37 	& 0104.C-0454(A) \\
SO 1075 & GIRAFFE & 24000     & 2010-01-03    &  2775  & 26   & 084.C-0282(A) \\
SO 1152 & XSHOOTER  & 18340     & 2019-10-12    &  800  & 73   & 0104.C-0454(A) \\
SO 1153 & XSHOOTER   & 1512     & 2021-02-13    & 60  & 37   & 106.20Z8.002 \\
SO 1156 & XSHOOTER   & 1512     & 2019-10-28    & 60  &  44   & 0104.C-0454(A) \\
SO 1260 & GIRAFFE & 24000     & 2010-01-03    &  2775  & 40   & 084.C-0282(A) \\
SO 1267 & XSHOOTER   & 1512     & 2019-11-17    & 60  & 33   & 0104.C-0454(A) \\
SO 1327 & GIRAFFE & 24000     & 2010-01-03    &  2775  & 33   & 084.C-0282(A) \\
SO 1368 & GIRAFFE & 24000     & 2010-01-03    &  2775  & 79   & 0103.C-0887(B) \\
SO 1361 & XSHOOTER  & 1512     & 2019-12-09    &  60  & 40   & 0104.C-0454(A) \\
CVSO 58 & XSHOOTER   & 1512     & 2020-12-02    & 65  &  27   & 106.20Z8.002 \\
CVSO 90	& XSHOOTER  & 18340     & 2020-12-04    &  740  & 80   & 106.20Z8.002 \\
CVSO 107 & XSHOOTER  & 18340     & 2020-12-04    &  740  & 80   & 106.20Z8.002 \\
CVSO 146 & XSHOOTER  & 1512     & 2020-12-09    &  40  & 42   & 106.20Z8.002 \\
CVSO 1876 & GIRAFFE & 24000     & 2010-01-03    &  2775  & 72   & 084.C-0282(A) \\
CVSO 1885 & GIRAFFE & 24000     & 2010-01-03    &  2775  & 49   & 084.C-0282(A) \\
V605 Ori B & GIRAFFE & 24000     & 2010-01-03    &  2775  & 39   & 084.C-0282(A) \\
\enddata
\tablenotetext{*}{The spectral coverage for the X-Shooter data 533.7-1020 nm and Giraffe 611.38-640.37 nm.}
\end{deluxetable*}

\subsection{TESS Data} \label{sec:TESS}

From the initial sample of 208 CTTS, we have 194 stars observed by TESS at 10 min and 30 min cadences for at least $\sim$25 days. We build each light curve (LC) using the \href{https://www.tessextractor.app}{\texttt{TESSExtractor}\footnote{\url{https://www.tessextractor.app}}} application \citep{Serna_2021}, which uses simple aperture photometry (SAP) to extract the stellar fluxes from the Full-frame images (FFI) provided by TESScut \citep{TESSCut}. Also, we use the Cotrending basis vectors (CBVs) from TESS and the task \texttt{kepcotrend} of the PyKE package \citep{PyKE} to correct the LCs from any systematic effect on the data. Once LCs were adequately processed, we performed a visual inspection and flagged the sources that exhibited periodic or quasi-periodic events as candidates for their posterior period analysis. Figure \ref{fig:LC} shows an example of the graphical product generated by the {\texttt{TESSExtractor}} application. This product includes the light curve, the phase-folded curve, and the periodogram used to obtain rotational periods.
For more details, see section \ref{sec:periods}.

\begin{figure}[ht!]
    \centering
    \includegraphics[width=0.5\textwidth]{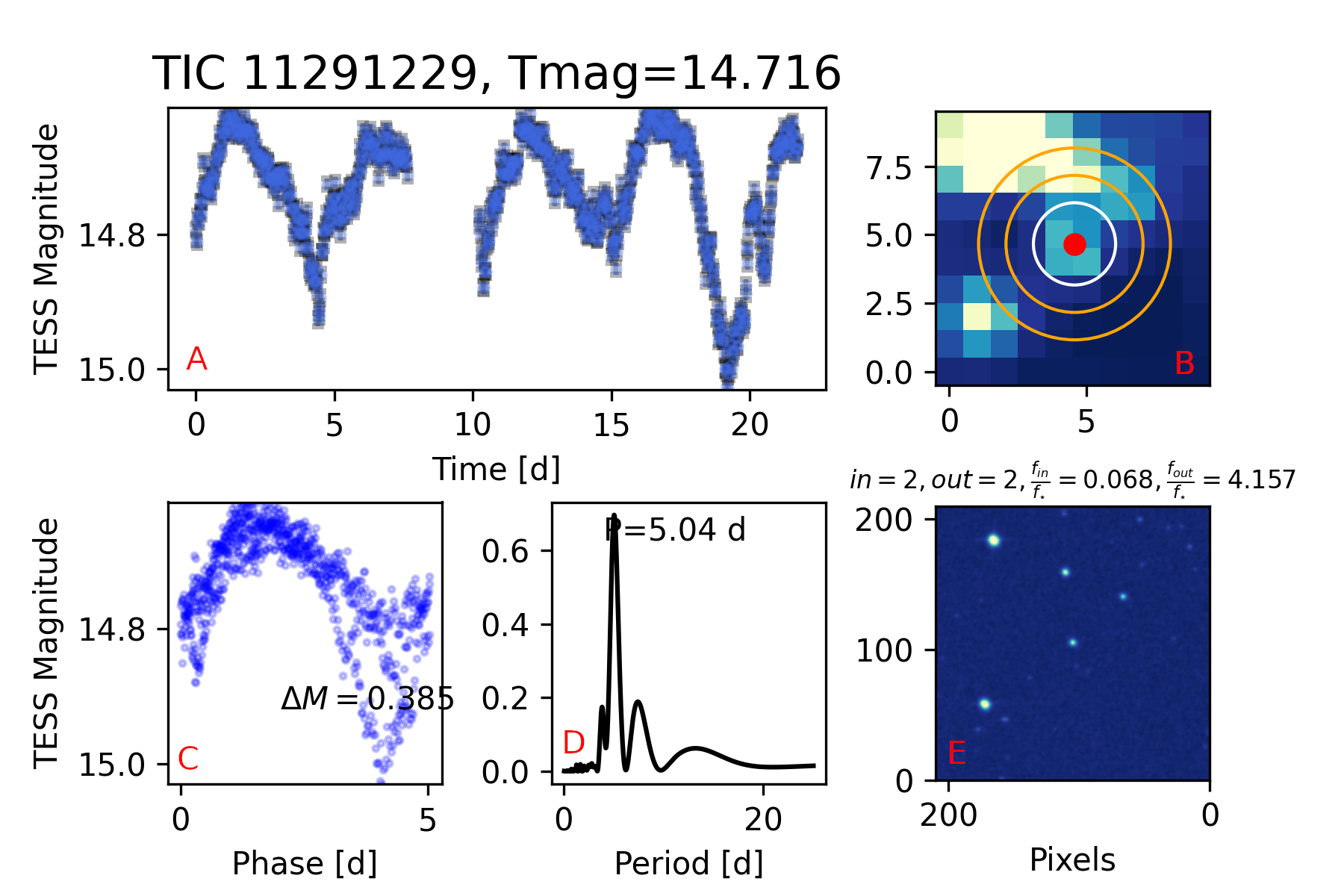}
    \caption{TESS light curve and analysis for the target TIC 11291229. \textit{Top}: (A) Light curve in TESS magnitudes, the error bars are contained in the marker symbols. The label on top refers to the star identification and the mean TESS magnitude. (B) Field of view (210 x 210 sq arcsec) corresponding to a TESS image of 10x10 pixels. The white circle shows the photometric aperture, and the orange circle shows the sky annulus. The red dot marks the centroid of the star. \textit{Bottom}: (C) Phase-folded light curve to the estimated best period. The legend shows the amplitude. (D) Lomb-Scargle periodogram, with the estimated period. (E) 210 x 210 pixels Digital Sky Survey (DSS2) thumbnail, same field of view as (B).
    } 
    \label{fig:LC}
\end{figure}

\section{Analysis and Results} \label{sec:results}

\subsection{Stellar Parameters}

\subsubsection{Rotation Period} \label{sec:periods}

Several variability studies in young stellar objects (YSOs) have revealed that most CTTS exhibit complex flux variations on timescales of hours, days, and even years \citep{Bouvier1993,Cody2010,Roggero_2021}. In some cases, brightness fluctuations are observed at levels of 1\% to 10\%, with fading or brightening events and quasi-periodic structures in most instances \citep{Cody2014}.
It has been proposed that these flux variations are generated by various mechanisms, such as the rotation of hot and cold starspots, accretion bursts \citep[e.g.,][]{Espaillat_2021}, occultations by dusty disk structures \citep[e.g.,][]{Ansdell2020}, and accretion streams falling at random locations on the star \citep{Kurosawa2013}.
These processes may be acting at once in CTTS, challenging the estimations of {\prot} for most stars. However, according to the DL model, the magnetic connection between the star and its inner disk results in synchronization between the stellar angular velocity and the Keplerian velocity at the corotation radius \citep{Artemenko2012}, particularly for the first 3 Myr of the CTTS evolution \citep{Matt_2010,Matt2012}. 
Then, assuming a Keplerian rotation regime, any periodic signal produced at the corotation zone could proxy the stellar rotation at the stellar surface. Therefore, under this assumption, we have estimated rotation periods for those candidates that show an apparent periodicity in the TESS data (\S\ref{sec:TESS}). \\

From the sample of LCs in \S \ref{sec:TESS}, only 51 stars exhibit periodic behavior to study rotation. 
We have selected light curves with a false alarm probability (FAP) below 0.01\%, indicating that the period measurements are highly reliable. We use the task \texttt{statistics.false\_alarm\_probability} from the astropy package \citep{2013A&A...558A..33A}.
We use the Lomb-Scargle periodogram \citep{Lomb_1976, Scargle_1982} to estimate the rotational period from the LC. We use a resolution of 1000 steps within the interval $0.01<P<25$ days. The best period was selected as the highest peak in the periodogram (see lower middle panel of Figure \ref{fig:LC}). These periods have been confirmed by checking the periodic pattern using the phase-folded LC at the estimated period, as shown in Figure \ref{fig:LC}. The {\prot} are listed in Table \ref{tab:parameters}. We plot the histogram of the periods in Figure \ref{fig:periods}. The range of periods estimated in this section will be used as input in the models \S\ref{sec:grid}, see Table \ref{tab:models}.\\

\begin{figure}
    \centering
    \includegraphics[scale=0.35]{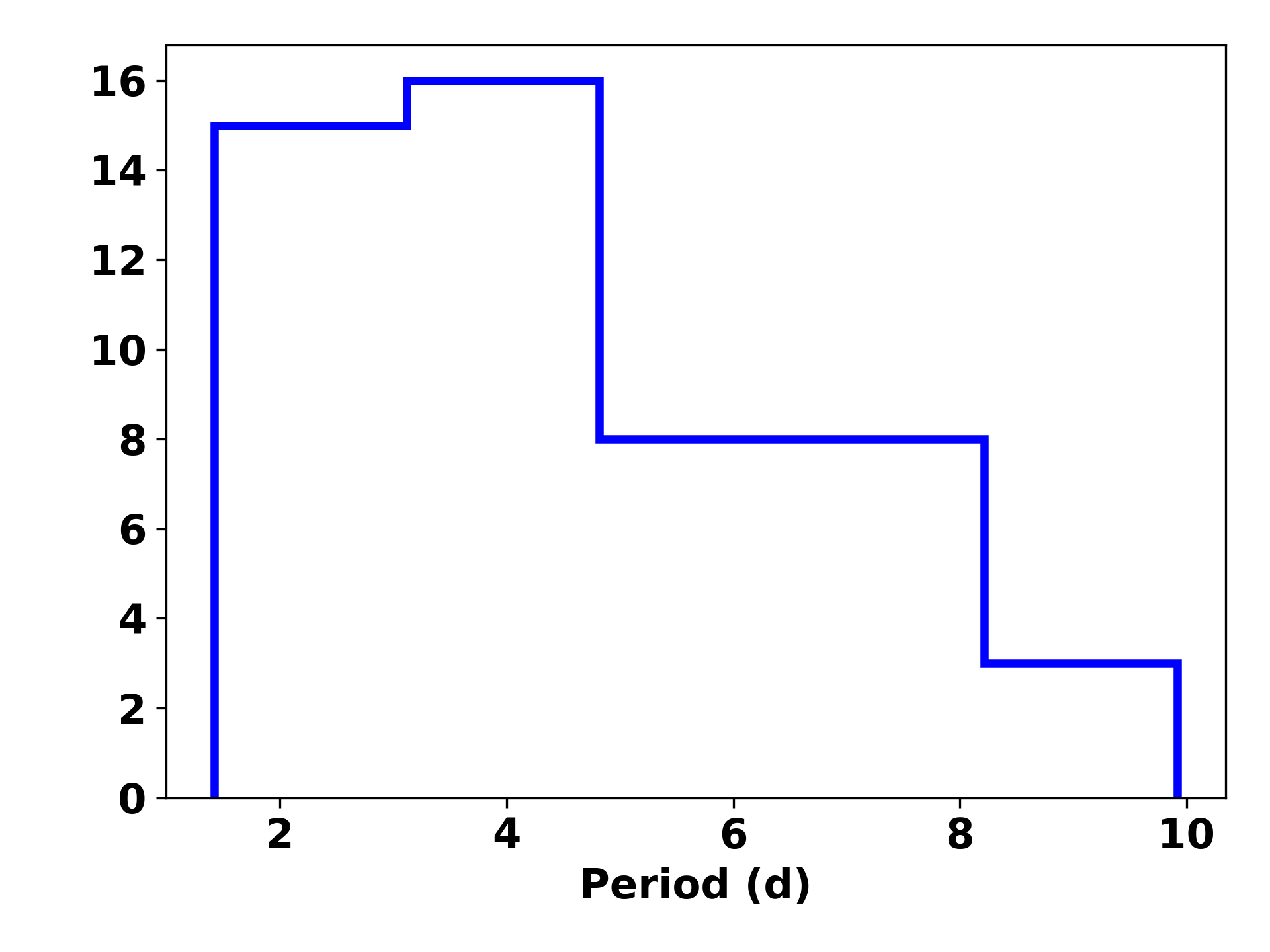}
    \caption{Rotational periods for the CTTS sample obtained through TESS light-curve analysis. The bin size used in the histogram follows Scott's rule \citep{Scott_1979}.}
    \label{fig:periods}
\end{figure}

\subsubsection{Mass Accretion Rate}
\label{sec:mass_acc_rates}

We determined mass accretion rates for the 208 stars of the sample in \S \ref{sec:ctts} based on the H$\alpha$ equivalent width ($EWH_{\alpha}$), spectral types, and reddening $A_{V}$ reported by \citet[Hern\'andez et al., in preparation]{Briceno_2019,Hernandez_2014}. Also, we use the \textit{Gaia} DR3 parallax measurements corrected by systematics and the \textit{Gaia} $G$, $G_{BP}$, and $G_{RP}$ photometry \citep{GAIADR3,Lindegren2021}. Using the relationship between the Johnson-Cousins photometric system and the \textit{Gaia} photometry\footnote{from Table 5.8 of the \textit{Gaia} data release 2 documentation \url{https://gea.esac.esa.int/archive/documentation/GDR2/}}, we obtain the $V$ magnitude. Given the spectral type and using the V magnitude corrected by extinction, the $I_{c,0}$ magnitude is obtained from the intrinsic $[V - I_{c}]_{0}$ color from \citet{Pecaut2013}. To obtain the continuum flux at the H$\alpha$ line ($F_{cont}$), we estimate the fluxes at the $V$(0.55 $\mu m$) and $I_{c}$ (0.79$\mu m$) bands and interpolate the flux at 0.6563 $\mu m$.
Using the H$\alpha$ equivalent width and distance estimated by the inverse relation with the parallax, we estimate the luminosity of the H$\alpha$ as follows:

\begin{equation}\label{eq:Luminosity}
    L_{H\alpha}=4\pi d^{2}\times(EWH_{\alpha}) \times F_{cont}
\end{equation}

Subsequently, we obtain the accretion luminosity using the following relation \citep{Ingleby_2013}:

\begin{equation}\label{eq:acclum}
    \log(L_{acc})=1.0(\pm0.2)\log(L_{H\alpha}) + 1.3(\pm 0.7).
\end{equation}

\noindent Finally, the mass accretion rate is obtained from $L_{acc}=GM_{\ast}\dot{M}_{acc}/R_{\ast}$, where $M_{\ast}$ and $R_{\ast}$ are determined using the MIST evolutionary models \citep{Dotter2016}. The mass accretion rates distribution is shown in Figure \ref{fig_Mdot_all_sample}, and corresponding values are included in Table \ref{tab:parameters}.

\begin{figure}
    \centering
    \includegraphics[scale=0.35]{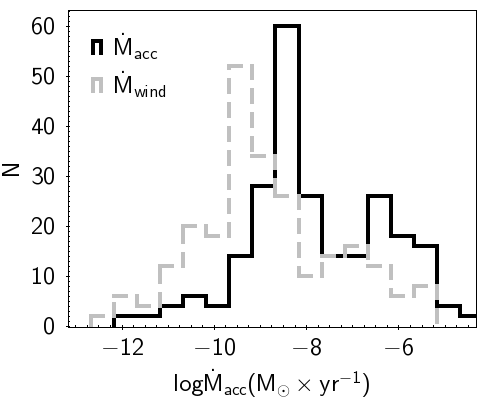}
    \caption{Mass accretion and mass loss rates for the sample of CTTS.}
    \label{fig_Mdot_all_sample}
\end{figure}

\subsubsection{Mass loss Rate Determination}
\label{sec:mass_loss_rates}

We computed mass loss rates for a sub-sample of the CTTS described in \S \ref{sec:ctts} via the forbidden emission line of {\oi}, which has been widely used as a tracer of bi-polar jets, stellar and disk winds \citep[e.g.,][]{Cabrit1990,Edwards2003,Watson_2016,Pascucci2022}. 
Specifically, we compute mass loss rates for 24 CTTS with LAMOST spectra (\S \ref{s:lamost_data}), 15 CTTS with X-Shooter spectra, and 7 CTTS with Giraffe spectra (\S \ref{s:eso_data}).
Luminosities of [OI] emission lines are computed from equivalent widths and dereddened fluxes at wavelengths near 6300{\angstrom} and used as indicators of the amount of mass in the outflow \citep{Hartigan1995}. 
Forbidden lines typically exhibit two components: a low-velocity component (LVC), which is symmetric and slightly blue-shifted, and a high-velocity component (HVC), with its peak either blue or red-shifted and separated from the LVC just a few tens of {\kms} \citep[e.g.,][]{Hartigan1995,Natta2014,simon2016,Banzatti2019}.\\

Following the physical treatment to the mass-loss rate in the Appendix A1 of \citet{Hartigan1995} and references therein, we use
the luminosity of the {\oi}, identified in the optical spectra of our CTTS to estimate the mass-loss rate ($\dot{M}_{wind}=MV_{\perp}/l_{\perp}$), where M is the mass in the flow whose dependence with the {\oi} luminosity was taken from \citet{Hartigan1995}, the $V_{\perp}$ is the component of the velocity of the projected wind in the plane of the sky, and $l_{\perp}$ corresponds to the size of the slit projected in the sky. In what follows, we assume $V_{\perp}=150$ {\kms} as an average for all the stars \citep{Hartigan1995}, and based on the average distance to Orion ($d\sim400$pc), we adopt $l_{\perp}=6\times 10^{15}$ $cm$ for X-shooter and Giraffe observations. For the case of the LAMOST spectra, the slit at the output end of the fibers is assumed to be equal to 2/3 of the fiber width leading to marginal differences relative to $l_{\perp}$. Thus, the correlation of \citet{Hartigan1995} may be written as follows:

\begin{equation}
\label{eq:wind}
    \log_{10}\Big(\dot{M}_{wind}\Big)=-4.65+\log_{10}\Big(\frac{L_{\lambda 6300}}{L_{\odot}} \Big)
\end{equation}

Before analyzing any optical forbidden line profiles, we first remove any telluric and photospheric absorption contaminating the region of interest in the spectra. We used the \texttt{Molecfit} tool \citep{smette2015}, which corrects for telluric absorption lines based on synthetic modeling of the Earth's atmospheric transmission. To remove telluric lines, each spectrum was normalized and corrected by radial velocity (RV), using the RV reported in \citet{Kounkel_2018, Hernandez_2014}. Based on the effective temperature and $\log(g)$ of the star, we selected templates from the PHOENIX spectral library \citep{Husser2013}.
In the specific case of the X-Shooter data, we use the XSL DR2 \citep{Gonneau_2020} to select a proper template for each star.
Once spectra were corrected by RV, all the templates were broadened by their reported {\vsini} using the rotational kernel implemented in the \texttt{PyAstronomy} package \citep{pyastronomy}. Then, the broadened templates were subtracted from each observed spectrum. \\

In Figures \ref{fig:OI_1}, and \ref{fig:OI_2} we show the line profiles.
In order to get the luminosity of the line {\oi} ($L_{\lambda6300}$), we have used the distances of \textit{Gaia} DR3, the equivalent width of {\oi} (EW[OI]) obtained with the task \textit{splot} of \texttt{IRAF}, a continuum flux estimation around the {\oi}, and the equation (\ref{eq:Luminosity}).
In particular, for LAMOST and Giraffe spectra, the flux at the continuum was estimated interpolating two bands around 6300{\angstrom} as described in section \ref{sec:mass_acc_rates}. For the X-shooter spectra, we directly measure the flux because the spectra have been properly flux calibrated. Subsequently, using the equation (\ref{eq:wind}), we estimate the mass-loss rate $\dot{M}_{wind}$ of our sample and their uncertainty following the error propagation procedure.\\

This analysis remains valid as long as we have the presence of HVC in the {\oi} line profiles. In some cases, our spectral resolution is insufficient to distinguish the HVC. We warn the reader to be cautious with results because the mass-loss rates for some sources could be overestimated. However, our results provide a helpful statistical estimation of the mass loss rate for this study. We summarize our results in Table \ref{tab:parameters}.\\

Using the mass accretion rates estimated in \S \ref{sec:mass_acc_rates} and the mass-loss rates of our sample, we roughly estimate the branching ratio $\chi$ of these systems.
We plot in Figure \ref{b} the accretion and mass-loss rates for our sample compared to other estimations of CTTS in the literature. Considering the limited mass and age intervals covered by our sample, accretion rate values range between $10^{-9}-10^{-8}$ {\msunyr} and mass-loss rates between $10^{-10}-10^{-8}$ {\msunyr} as shown in Figure \ref{fig_Mdot_all_sample}. To have a more comprehensive view of a diverse population of CTTS, we included reported measurements for CTTS in Taurus \citep{Herzceg_2008}, Lupus \citep{Natta2014}, active CTTS systems from \citet{Hartigan1995,Gullbring1998}, and CTTS in nine molecular cloud complexes studied by \citet{Watson_2016} using Spitzer data.
According to Figure \ref{b}, most of our CTTS remain below the APSW boundary (long-dashed black line), confirming that APSW is a suitable scenario to explain the observed rotation rates.\\

Although age information is not available in Figure \ref{b}, some studies have suggested evidence of temporal evolution between the {\mdot} and {\mwind} relationship. For instance, \citet{Watson_2016} mostly includes objects younger than CTTS (e.g., Class 0 and Class I) and demonstrated that different classes of young objects, used as an age proxy, are naturally separated into different plot regions \citep[see Figure 7 of][]{Watson_2016}. Our study supports that evidence, providing more CTTS to the sample and showing even better how CTTS are well positioned in the plot.\\

\begin{figure*}[ht!]
    \centering
    \includegraphics[width=0.8\textwidth]{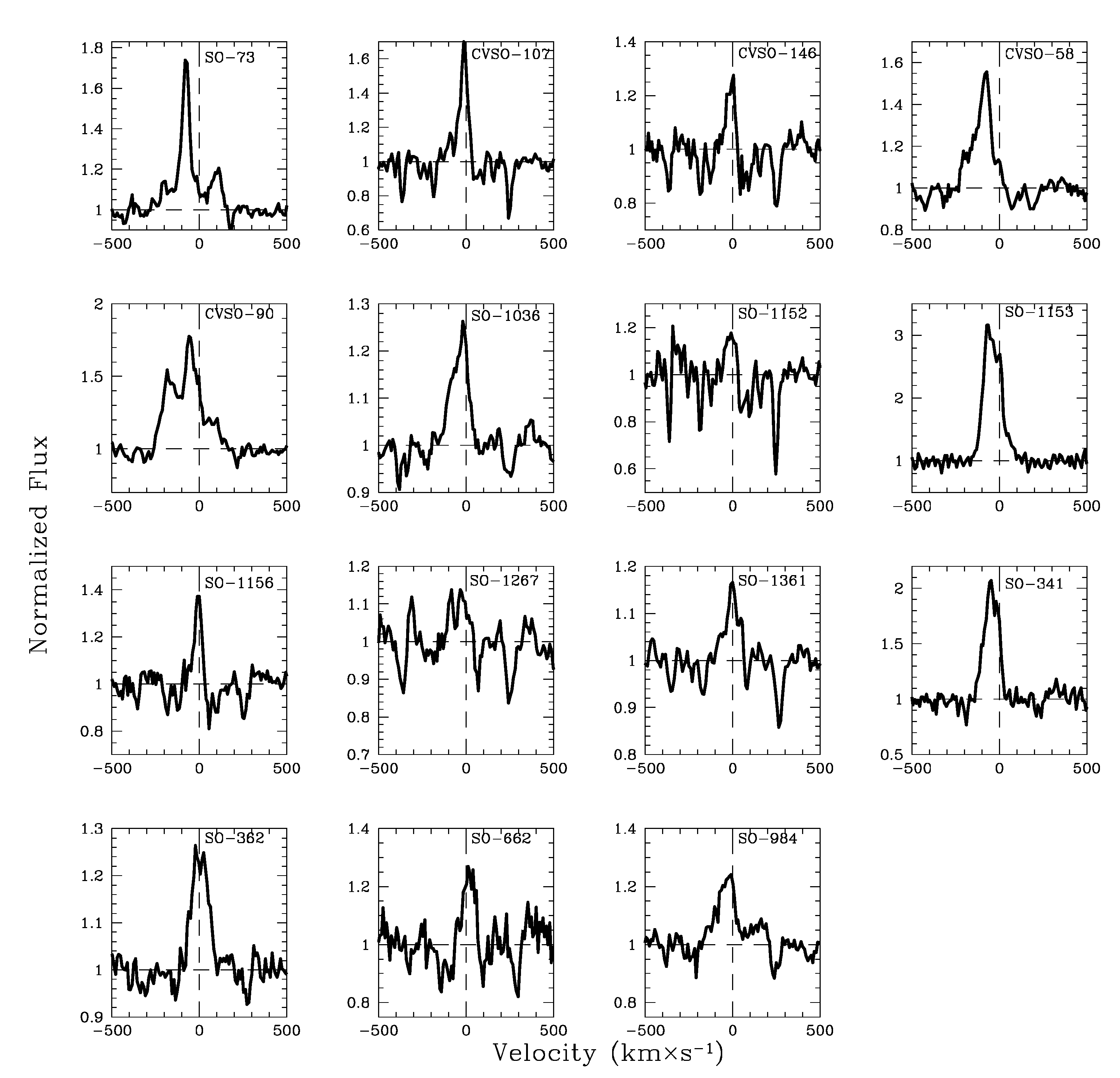}
    \caption{{\oi} residual line profiles for 15 CTTS found in the X-Shooter data archive. The stellar rest velocity is indicated with a vertical dashed line. Residual emission lines were obtained by subtracting photospheric contributions using Phoenix templates and posterior removing telluric lines.}
    \label{fig:OI_1}
\end{figure*}

\begin{figure*}[ht!]
    \centering
    \includegraphics[width=0.8\textwidth]{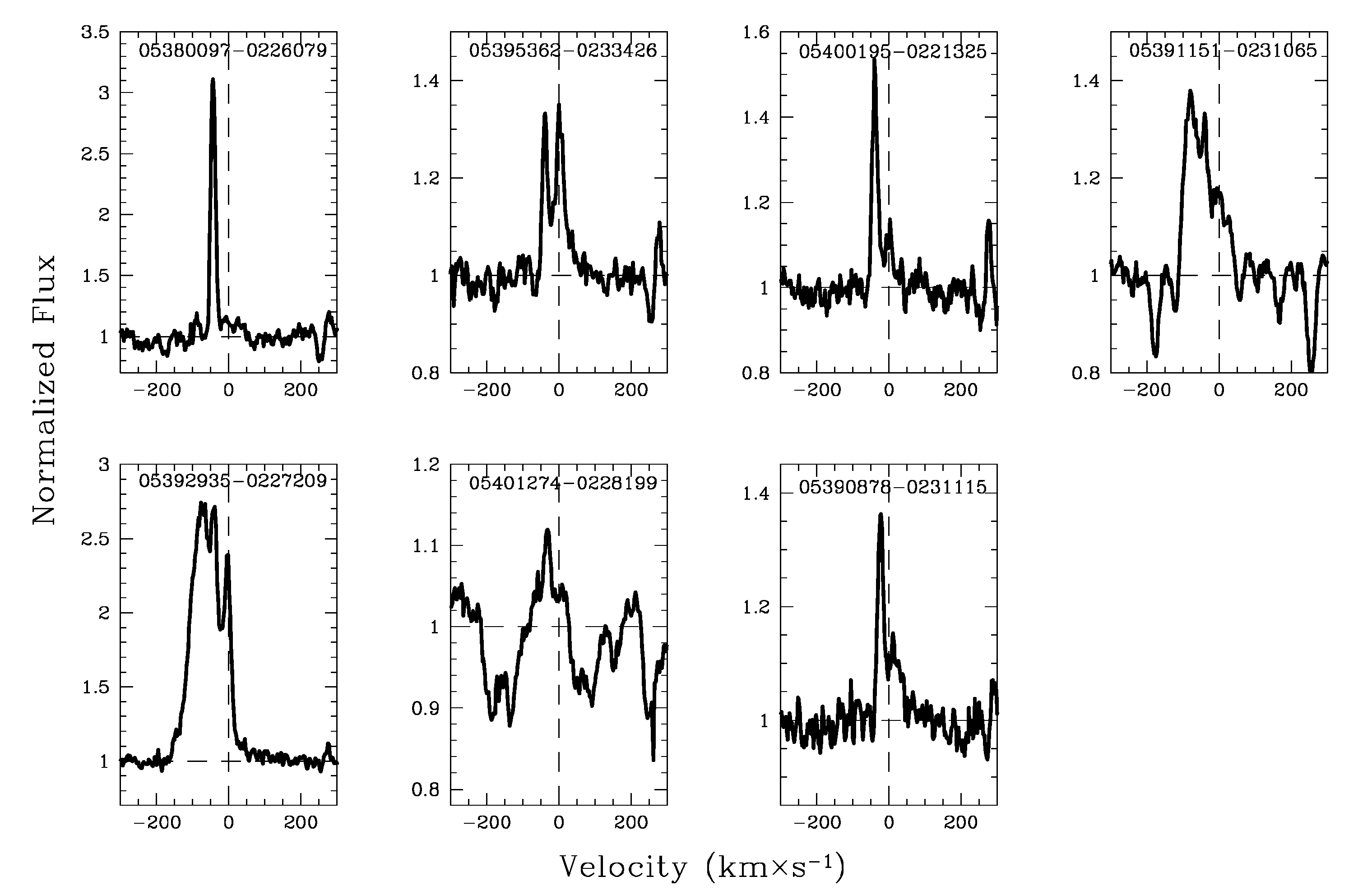}
    \caption{{\oi} residual lines profiles for 7 CTTS found in data archive of Giraffe.}
    \label{fig:OI_2}
\end{figure*}

\begin{figure}[ht!]
    \centering
    \includegraphics[width=0.5\textwidth]{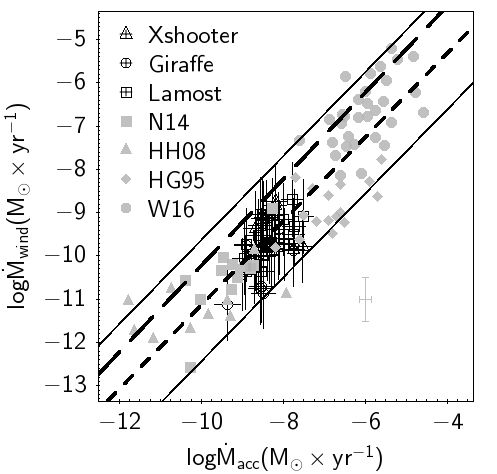}
    \caption{Accretion and mass-loss rates for our sample of CTTS (black crosses) and other studies (gray symbols). Complementary data shown in gray correspond to low mass accretors in Taurus from \citet{Herzceg_2008}, (triangles) active CTTS systems from \citet{Hartigan1995,Gullbring1998} (diamonds) and selected CTTS from \citet{Natta2014} (squares) and \citet{Watson_2016} (circles). The long-dashed line indicates the upper limit given by APSW models of \citet{Matt2012}, i.e. $\dot{M}_{wind}=\chi\times \dot{M}_{acc}^\eta$, with $\eta=1$, and $\chi$=0.6. The short-dashed line corresponds to the linear regression fit using all the data, which leads to $\eta=(0.97\pm0.05)$ and $\chi=0.04^{+0.05}_{-0.02}$. Upper and lower solid lines represent the 3$\sigma$ level for the whole sample.}
    \label{b}
\end{figure}

\subsubsection{Magnetic field topology and evolution}

Observations indicate that the surface magnetic fields of young stars have high-order multipolar components rather than magnetic dipoles. While stars more massive than 0.5{\msun} exhibit toroidal and non-axisymmetric poloidal components, low-mass stars below 0.5{\msun} show strong poloidal large-scale magnetic fields, which are mainly axisymmetric \citep{Donati_2019}, where most CTTS have typical intensities of the order of 1-3 kG \citep{Johns_Krull_2007}, and significant star spot coverages \citep{Somers2020, Cao_2022}.
These magnetic fields are expected to impact CTTS, their accretion disks, and planets \citep{Strugarek_2015}.\\

\citet{Schatzman_1962} was the first to suggest that open magnetic field lines or the breaking of closed lines could trigger mass loss phenomenons, which could, in the long run, be responsible for taking a large amount of angular momentum from the system.
For this reason, the geometry of the magnetic field is an essential element in describing stellar winds. For example, \citet{Reville_2015} and \citet{Garraffo_2015} found that complexity in the geometry of the magnetic field can dramatically decrease the loss rate of angular momentum to a few orders of magnitude. This is mainly because higher orders of magnetic moments would facilitate the generation of a magnetosphere with many closed lines that would suppress mass loss. \citet{Viard_2017,Folsom_2016} observed that the topology of the magnetic field changes strongly as the star ages because as the radiative core becomes bigger, the dipole components decrease, and the magnetic field becomes more and more complex.
\citet{Folsom_2016} and \citet{vidotto_2014} found empirical relationships between the large-scale component of the magnetic field strength versus age for pre-main to the main sequence.\\

Although there are many efforts to measure magnetic fields in T Tauri stars, nowadays, there is no conceptual model that allows us to describe magnetic fields accurately. Some authors approximate the magnetic field topologies, mixing different geometries. For example, \citet{Finley_2017} suggest that combinations of dipolar and quadrupolar fields change the relative orientation of the stellar wind with respect to any planetary or disk magnetic field. In a general view, these components are important in the morphology of the wind. Also, they confirm that the original prescription from \citet{Matt2012}, which is the formulation of the present work, uses pure dipolar magnetic fields and remains robust in most cases, even for significantly nondipolar fields.\\

\subsection{Model Assumptions}
\label{sec:spinmodel}

We developed stellar spin multi-parametric models for young low-mass and solar-type stars magnetically linked to a surrounding gaseous accreting disk. We assume that stars can produce their stellar magnetic fields when they rotate, through dynamo processes in the convective regions of the stellar interior. Due to the size of the convective region varying with the spectral type, we use non-rotational, PMS evolutionary models of \citet{Baraffe_2015} that provide a robust treatment of the internal stellar structure. In the next section, we describe in detail the main assumptions of our parametric spin evolution model.

For simplicity, we assume a dipolar field with strength $B_{\ast}$ at the equator of the stellar surface and co-rotating with the star. At larger stellar radii, this dipole enables angular momentum transport with the disk and magnetosphere. We assume values range between 500-3500 G in agreement with previous studies \citep{Johns_Krull_2007} and recent surveys such as \citet{Lavail_2017,Lavail_2019}. The main assumptions for constructing our grid of rotational models are the following:

\begin{itemize}
\item \textbf{A solid body approach}: 
We consider a simple approach in our models, assuming that the stars rotate as a solid body with a core and envelope coupled to each other. We neglect any difference in angular momentum between the stellar surface and its stellar interior. Rotation is thus entirely controlled by the external torques. 
This approach is sufficient for fully convective stars on their Hayashi tracks, although our simplified treatment should be revised for stars with radiative zones in their interiors using a more realistic treatment, similar to the considered redistribution of angular momentum between the two layers \citep{MacGregor1991}. We do not expect a strong impact on the rotation rates predicted by our models in the CTTS phase ($<$10 Myr) because the expected time scale for the core-envelope decoupling interaction starts around $\sim$10-30 Myr \citep{Gallet2013,Gallet_2015}.

\end{itemize}

\begin{itemize}
\item \textbf{Accretion rate onto the star}:  We assume an exponential decay for the accretion rate as a function of age ($\dot{M}_{acc}=\dot{M}^{in}_{acc}\,e^{-\frac{(t-t_0)}{\tau_{a}}}$), starting from an initial value  $\dot{M}^{in}_{acc}$ at $t_0=0.5$ Myr and with a characteristic accretion timescale of $\tau_{a}=2.1$ Myr reported by \citet{Briceno_2019}. 
Since the exponential relation for the accretor fraction evolution \citep{Briceno_2019} represents a probability that stars in a stellar group have accretion rates above the threshold detection limit, the $\tau_{a}$ obtained from this relation is a statistical proxy for the timescale of the general accretion rate evolution behavior in stellar groups.
The accretion process adds angular momentum that is transferred from the disk truncation radius ($R_t$) toward the stellar surface at a rate of:
\begin{equation}
\label{eq:acctorque}
    \tau_{acc}=\dot{M}_{acc}\sqrt{GM_{\ast}R_{t}}
\end{equation}

$R_t$ denotes the distance at which the magnetic field fully governs the stress in the internal disk. At this location, disk material and the stellar surface share identical rotation. We assumed $R_t$ inside disk corotation, which is computed using the Newton-Raphson method and the equation 15 of \citet{Matt_2005a}.

\item \textbf{Stellar winds}:  We consider the APSW an important source of angular momentum loss in CTTS \citep{Matt_2008a}. Previous studies have shown that open field regions in the magnetic field lead to the formation of ASPW that efficiently removes substantial amounts of angular momentum from the star \citep[e.g.,][]{Hartmann1989, Uzdensky2004, Pantolmos_2020,Matt2012,Pinzon2021}. It is assumed that a fraction $\chi$ of the mass in the accretion flow is transferred to a stellar wind, i.e., $\dot{M}_{wind}=\chi\,\dot{M}_{acc}$, providing an efficient angular momentum loss mechanism. The formulation is inspired by the analytic work of \citet{Weber1967}, in which the net torque from a one-dimensional wind is as follows:

\begin{equation}
\label{eq:windtorque}
    \tau_{wind}=-\dot{M}_{wind}\,\Omega_{\ast}\,r^{2}_{A}
\end{equation}
\noindent where {\mwind} is the integrated wind mass-loss rate, $\Omega_{\ast}$ is the angular velocity of the star, and $r_{A}$ is the Alfv\'en radius. 
In our one-dimensional approach, we used the semi-analytical results obtained from numerical MHD simulations of \citet{Matt_2008a}:
 
\begin{equation}
\label{eq:alven}
    \frac{r_{A}}{R_{\ast}}=K\Bigg(\frac{B^{2}_{\ast} R^{2}_{\ast}}{\dot{M}_{wind}v_{esc}}\Bigg)^{m}
\end{equation}

\noindent where $K\approx2.11$ and $m\approx0.223$ are dimensionless constants and $v_{esc}=(2GM_*/R_*)^{1/2}$ is the escape velocity from the star.\\
\end{itemize}

\begin{itemize}
\item{\textbf{Star-disk interaction}:} Along with angular momentum losses through disk winds, CTTS also spin down due to their magnetic interaction with the Keplerian rotating disk \citep[e.g.,][]{Ghosh_1978, Koenigl1991,CollierCameron1993}. In a scenario of large Reynolds numbers of the star-disk interaction (low diffusion parameter $\beta$), the strong coupling between disk matter and stellar magnetic field lines leads to a twisting of the vertical field component $B_z$ and thus to a spin-down torque given by:
\end{itemize}
\begin{equation}
\label{eq:DLtorque}
    \tau_{DL}=\frac{B^{2}_{\ast}R^{6}_{\ast}}{3\beta R^{3}_{co}}\Bigg[2\Big(\frac{R_{co}}{R_{out}}\Big)^{\frac{3}{2}}-\Big(\frac{R_{co}}{R_{out}}\Big)^{3}-2\Big(\frac{R_{co}}{R_{t}}\Big)^{\frac{3}{2}}+\Big(\frac{R_{co}}{R_{t}}\Big)^{3}\Bigg]
\end{equation}
here $R_{out}=R_{co}\big(1+\beta\gamma_{c}\big)^{\frac{2}{3}}$ defines the extent of the connected magnetic region. $R_{co}$ is the co-rotation radius, $R_t$ the disk truncation radius, and $\gamma_{c}=B_{\phi}/Bz$ characterizes the proportion of magnetic field twisted at the disk's surface at each radial location. We use $\gamma_c=1$ and $\beta=10^{-2}$ as fixed parameters in all our simulations \citep{Matt_2010}.\\

\subsubsection{Grid of Rotational Models}
\label{sec:grid}

The evolution of the stellar rotation rate ($\Omega_{\ast}$) is computed by solving the following differential equation:
\begin{equation} \label{eq:diffeq}
    \frac{d\Omega_{\ast}}{dt}=\frac{\tau_{\ast}}{I_{\ast}}-\frac{\Omega_{\ast}}{I_{\ast}}\frac{dI_{\ast}}{dt}
\end{equation}

\noindent where $\tau_{\ast}$ is the sum of the individual torques due to the accretion, winds, and star–disk interaction as described above, and $I_*=k^2 M_*R_*^2$ is the stellar moment of inertia and $k^2$ is the gyration radius obtained from the stellar evolutionary models of \citet{Baraffe_2015}.  We note that the accretion rate is incorporated into the calculation of $dI_{\ast}/dt$ and used to update $M_*$.\\

\begin{figure*}[htp!]
    \centering
    \includegraphics[scale=0.5]{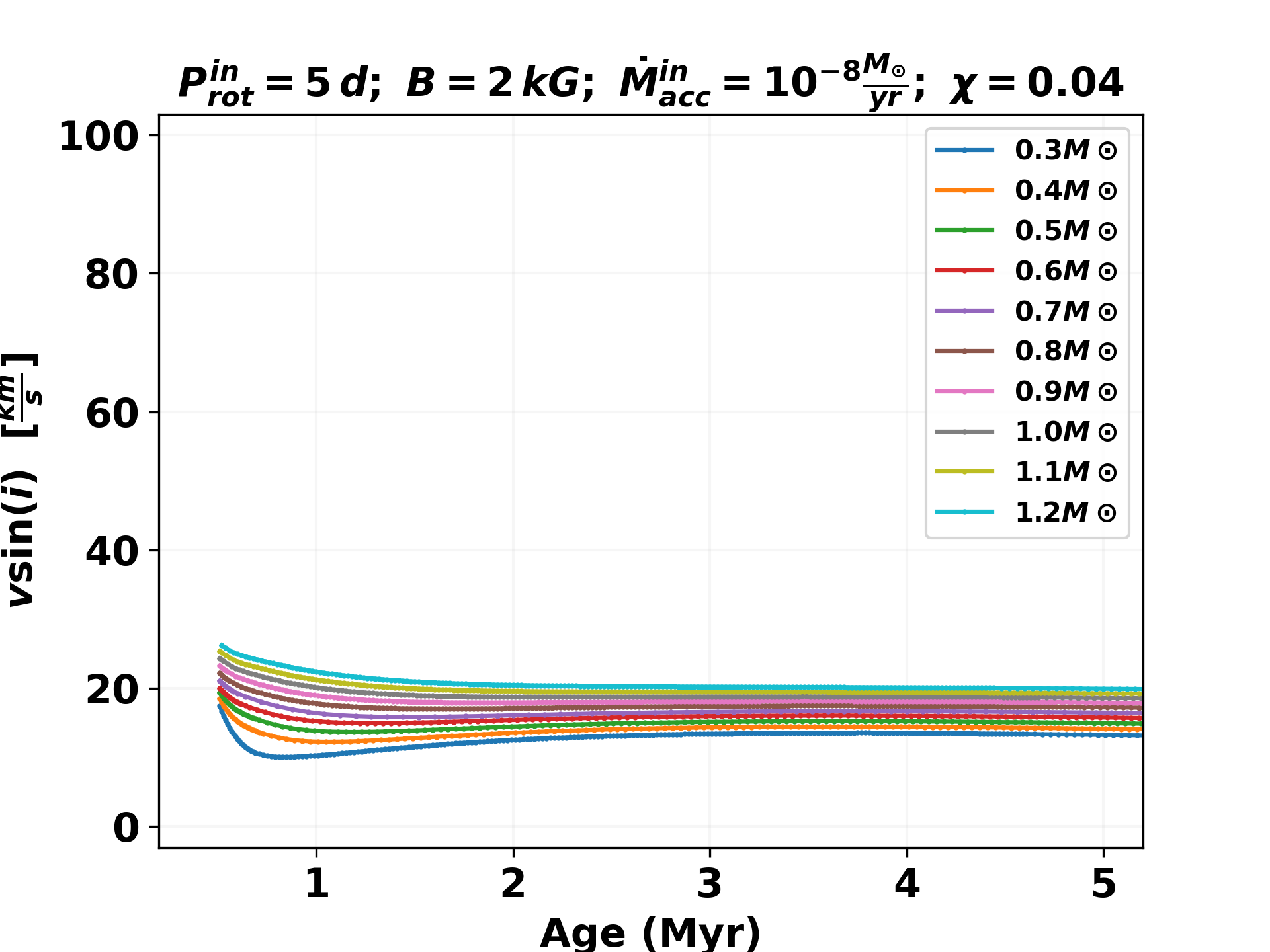}
    \includegraphics[scale=0.5]{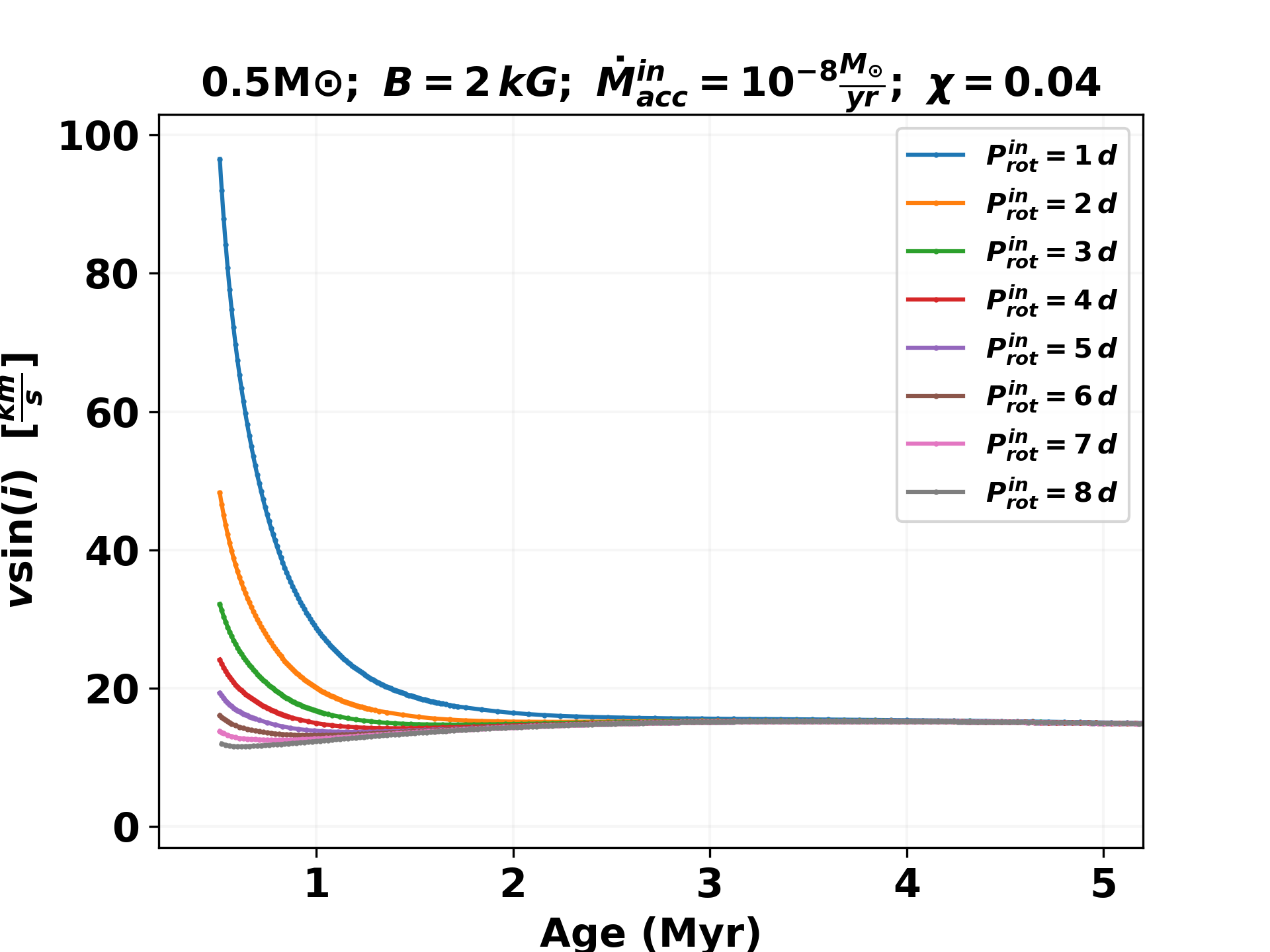}
    \includegraphics[scale=0.5]{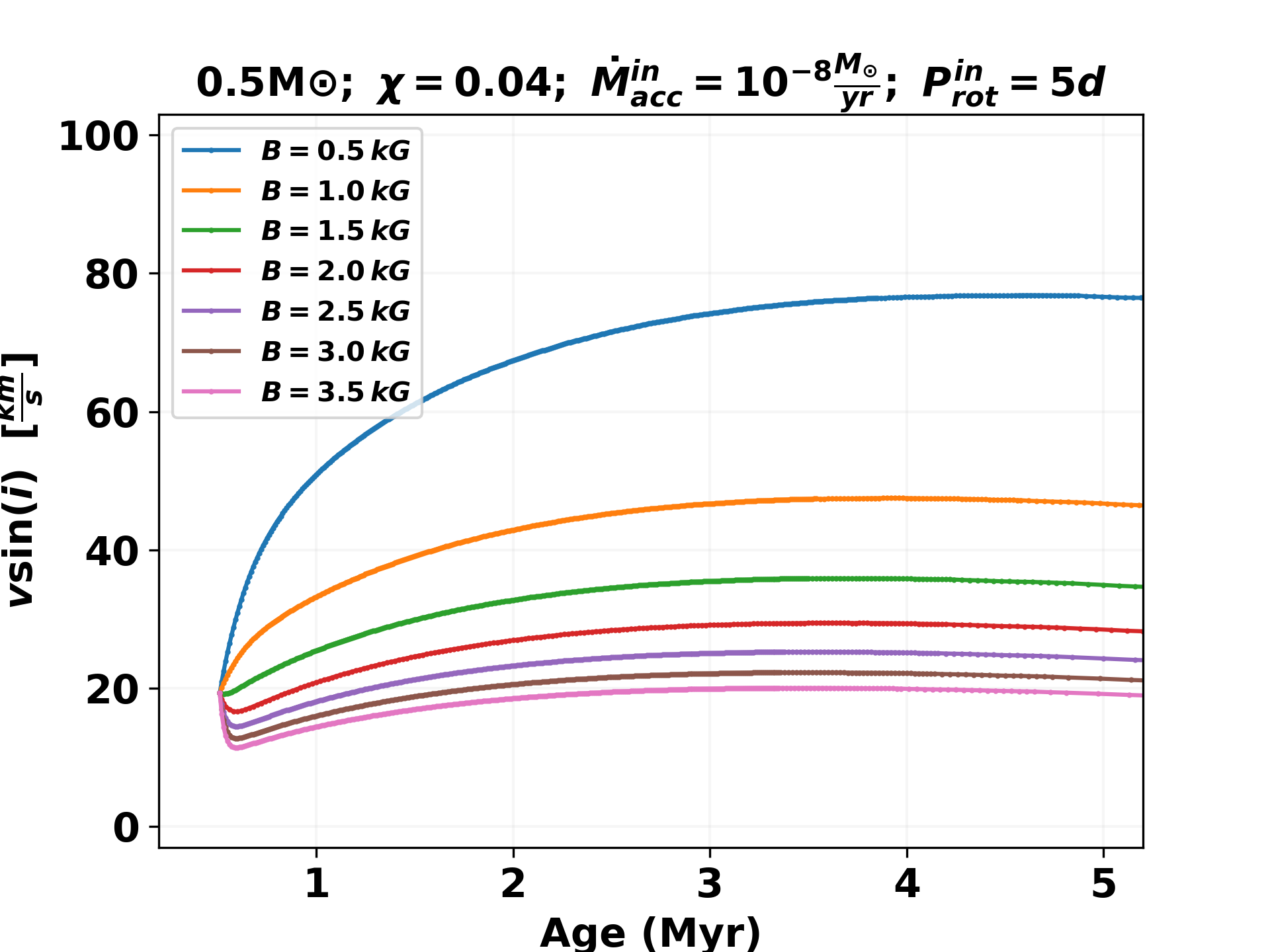}
    \includegraphics[scale=0.5]{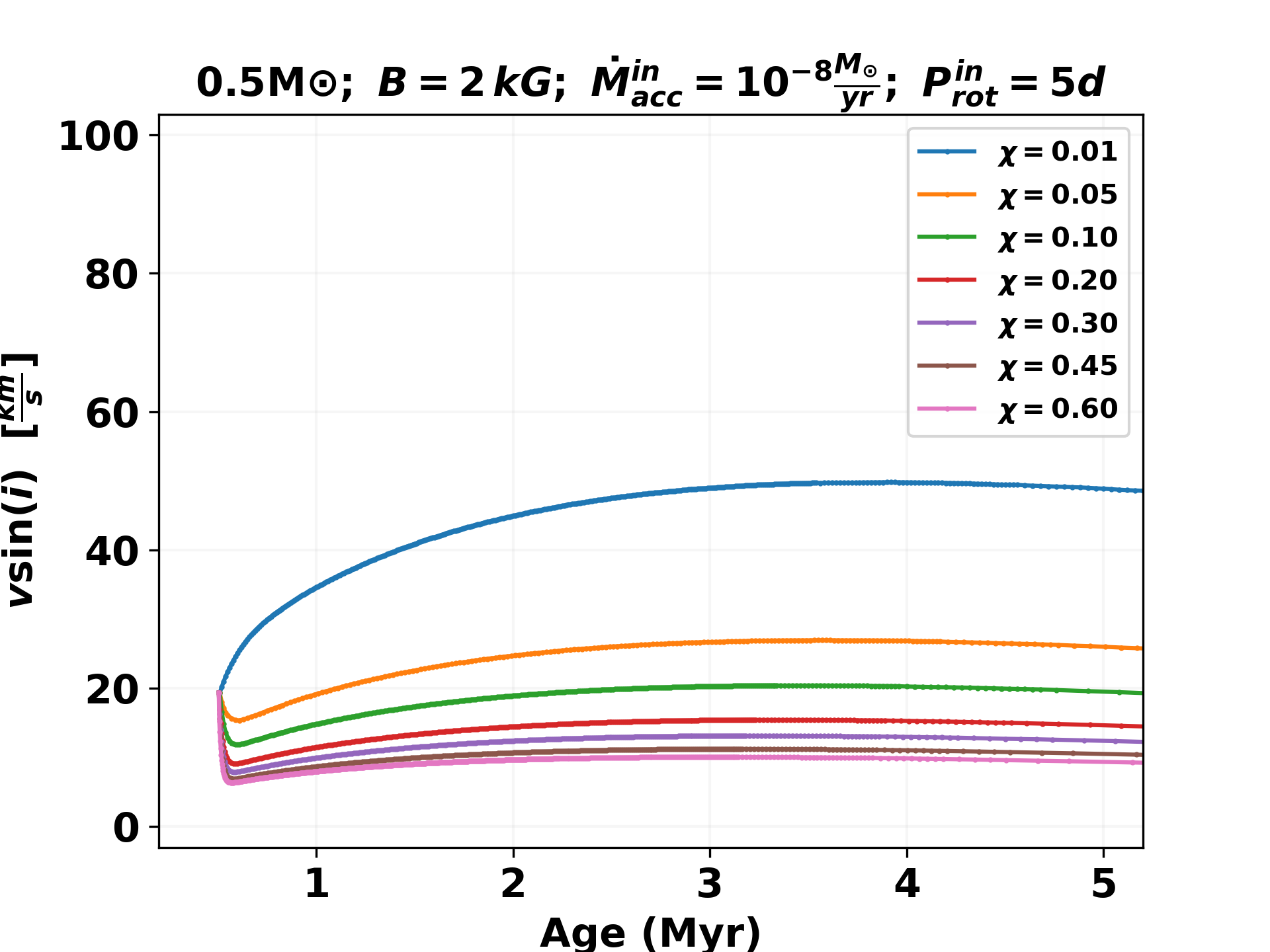}
    \includegraphics[scale=0.5]{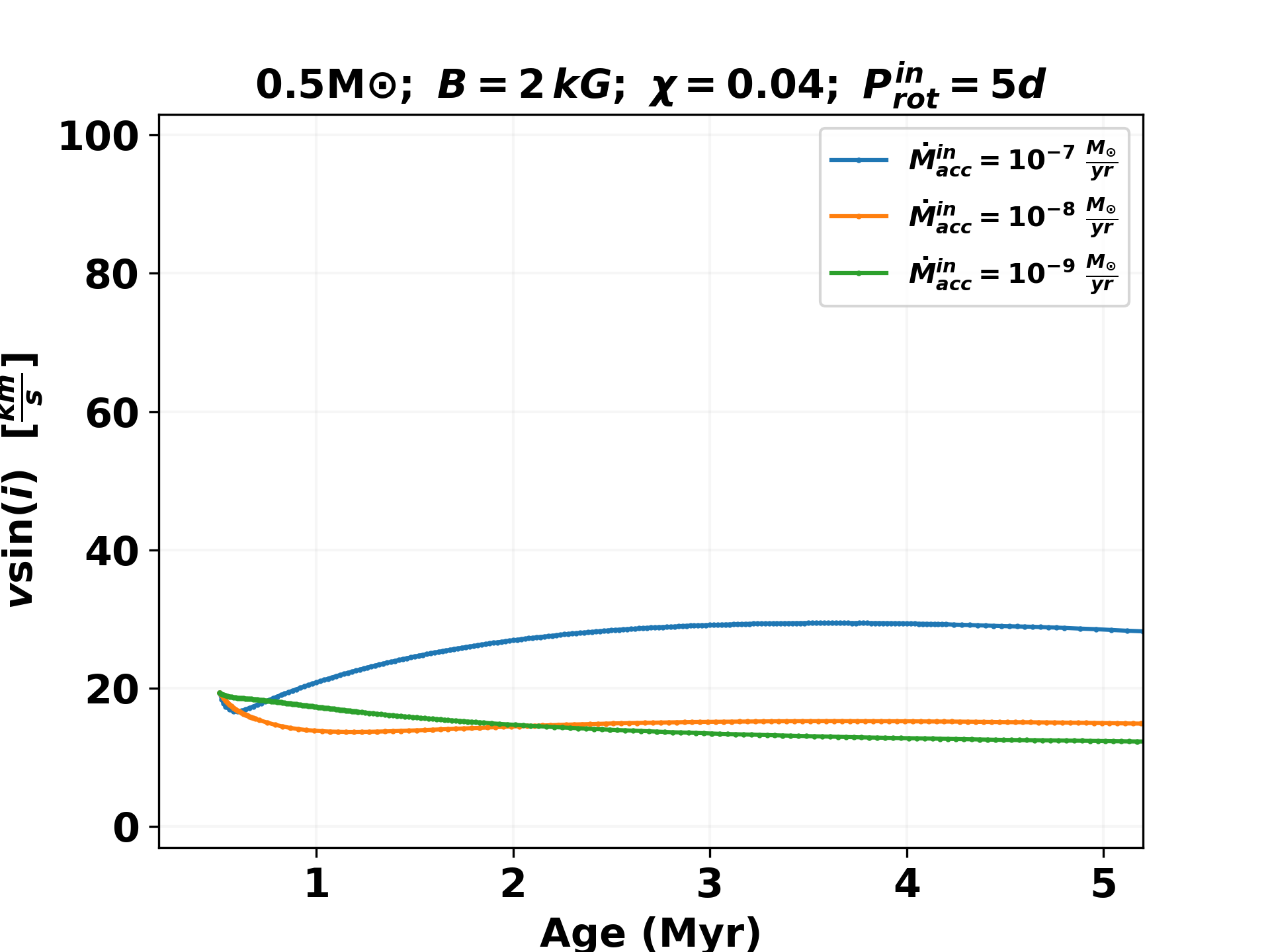}

    \caption{Spin evolutionary tracks represented as $v\sin(i)=\frac{\pi}{4}v_{rot}$ as a function of age for individual cases. The title of each panel shows the fixed parameters. Meanwhile, the legend colors represent different cases for a respective variable in the models.}
    \label{fig:track}
\end{figure*}

Using the fourth-order Runge-Kutta method with adaptive step size, we find the solution to the equation (\ref{eq:diffeq}) \footnote{The models are available on GitHub in the following link: \href{https://github.com/javiserna/Rotational-models-of-CTTS}{Rotational-models-of-CTTS}}. At each time step of the simulation, we compute the next time step, ensuring changes over the $\Omega_{\ast}$ solution remain lower than $1\%$ per step.\\

We use the set of parameters [$M_{\ast}$, $B_{\ast}$, $\chi$, $P^{in}_{rot}$, $\dot{M}^{in}_{acc}$] based on the values of Table \ref{tab:models}. We use all possible combinations of the parameters, without repetitions, and within the value ranges and steps to build a grid with $2.232\times10^{6}$ individual models. 
For each combination of the five parameters in Table \ref{tab:models}, the angular momentum conservation, stellar evolutionary models, and the assumed exponential decay for the accretion rate as a function of age regulate the temporal variation of the synthetic rotational values (e.g., angular velocity, {\vsini}) estimated in our grid.
Simulations start at 0.5 Myr and end at 15 Myr, the age at which there are some long-lived CTTS. Nevertheless, statistically, the typical age when the gas component of the disk has been dissipated is around 5 Myr \citep{Carpenter_2006,Hernandez_2008}. Given a simulation, both the angular velocity (stellar period) and the mass accretion rate evolve with time. \\

\subsection{{\vsini} synthetic distributions}

\begin{figure*}[htp!]
    \centering
    \includegraphics[scale=0.5]{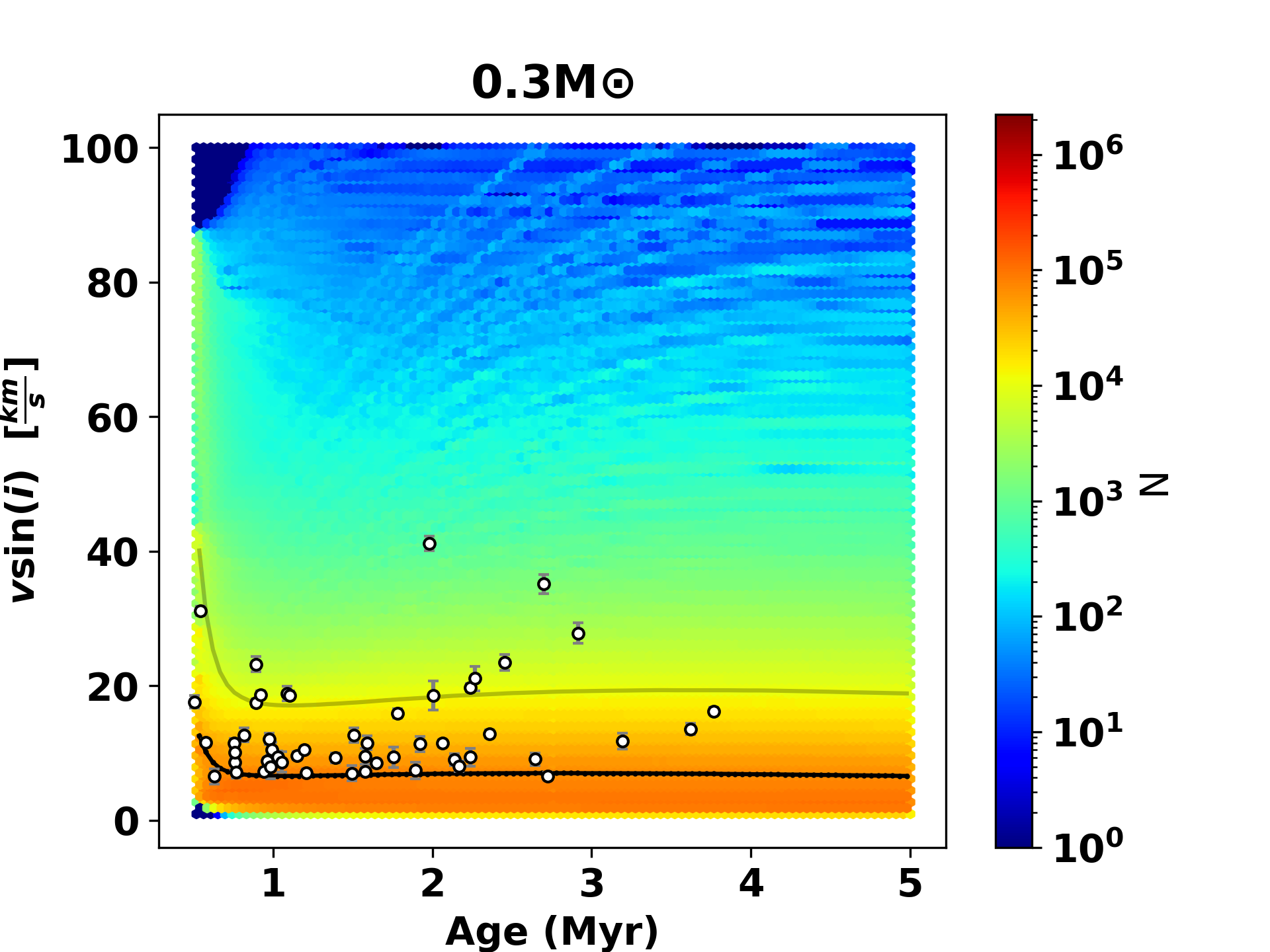}
    \includegraphics[scale=0.5]{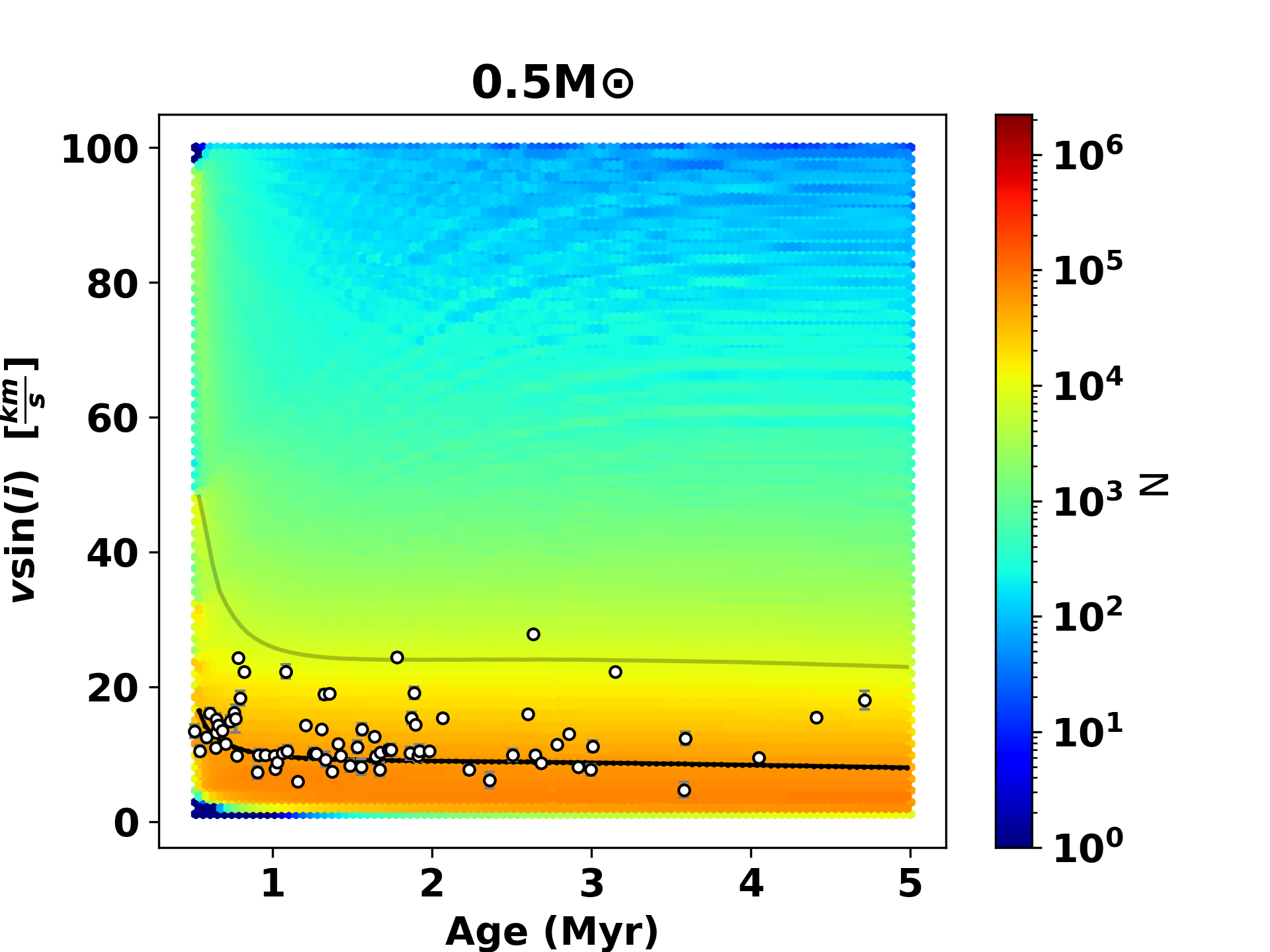}
    \includegraphics[scale=0.5]{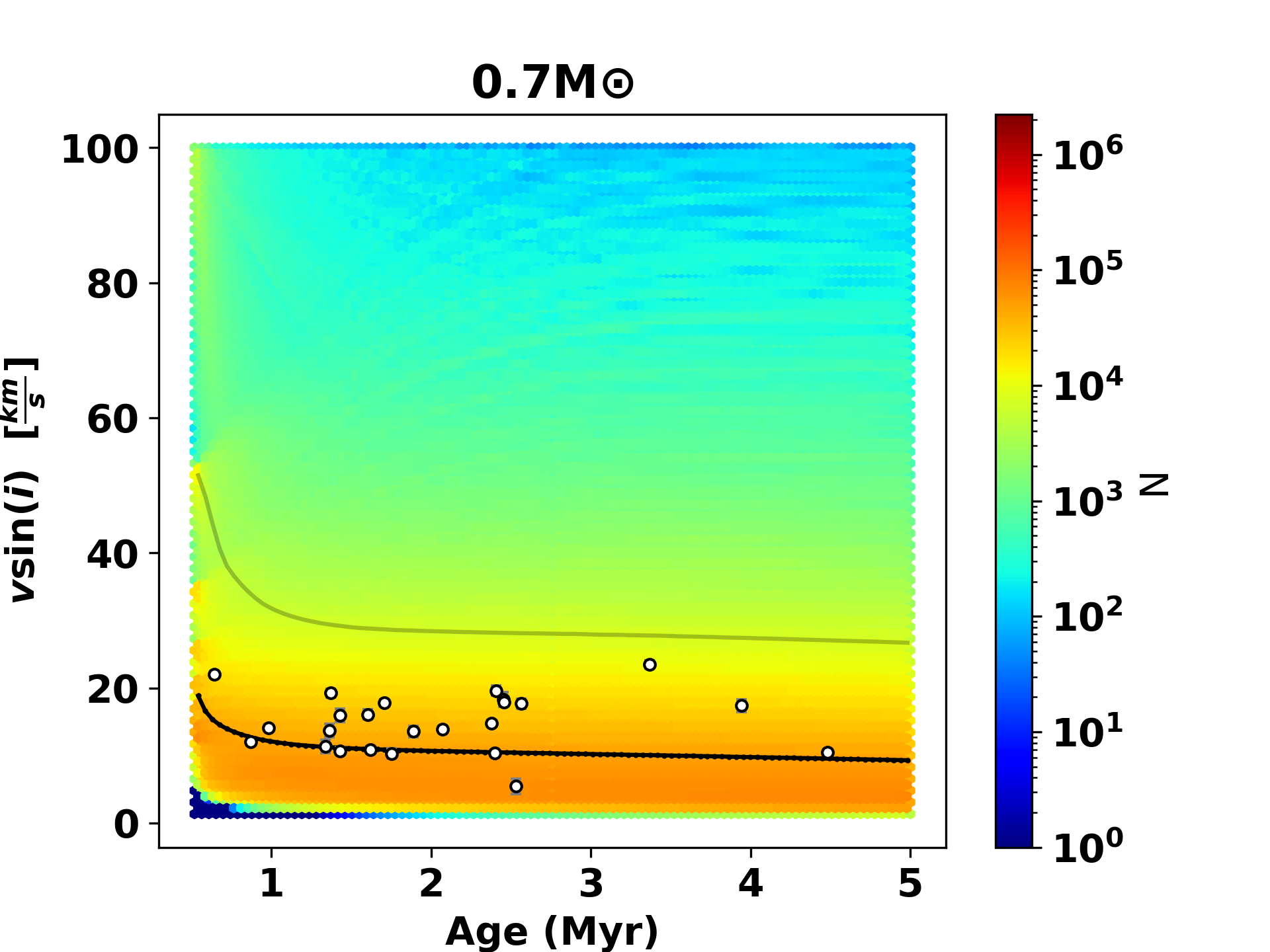}
    \includegraphics[scale=0.5]{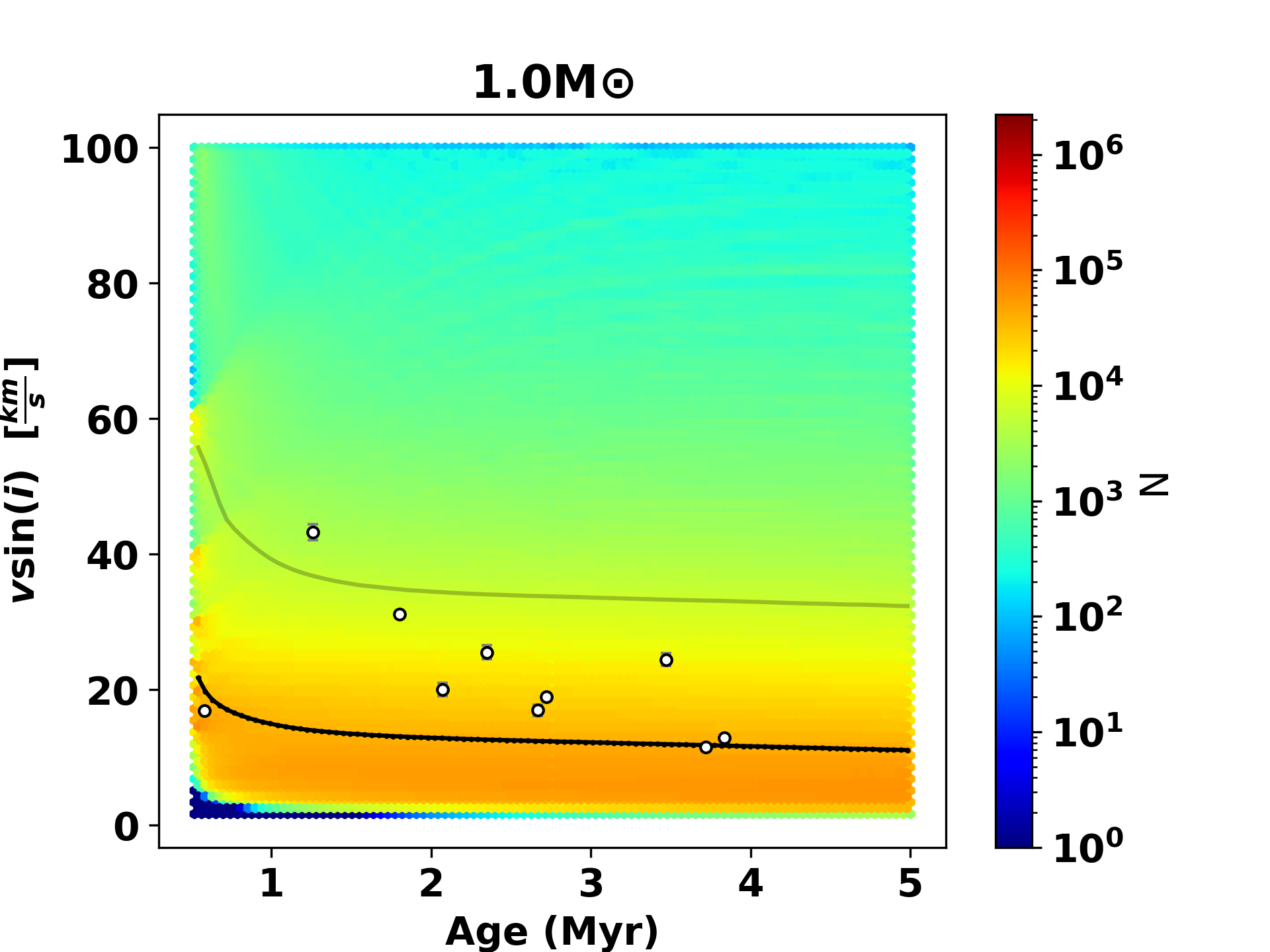}

    \caption{Grid of models illustrated as a density plot of $v\sin(i)=\frac{\pi}{4}v_{rot}$ as a function of the age, in different values of mass. The black line shows the grid's median {\vsini}, while the gray line contains 90\% of the models below it. The color scale represents the number of models per hex bin pixel. The white dots represent the {\vsini} and age for the CTTS sample.}
    \label{fig:grid}
\end{figure*}

The rotation rates given by our grid of models are related to the projected rotational velocities through the previous knowledge of spin-axis orientation. 
By assuming that the spin-axes of objects are randomly oriented, and utilizing the inclination average method proposed by \citet{Chandrasekhar_1950}, the projected rotational velocity can be expressed as $v\sin(i)=\frac{\pi}{4}R_{\ast}\Omega_{\ast}$, where $R_{\ast}$ is the stellar radius and $\Omega_{\ast}$ is the angular velocity of the star.\\

In Figure \ref{fig:track}, we show individual evolutionary tracks of {\vsini} for specific cases of the parameter set. While in Figure \ref{fig:grid}, we display the complete grid of models for masses of 0.3, 0.5, 0.8, and 1.0$M_{\odot}$. In each panel, the color scale represents the number of models enclosed in each hexagonal bin pixel of the {\vsini} versus age diagram. We have included the $90^{th}$ percentile and the median per pixel indicated with grey and black lines, respectively. \\

The panels in Figure \ref{fig:grid} show a decreasing trend of the median of {\vsini} with age. Despite the wide range of initial conditions considered in Table \ref{tab:models}, most solutions lie on the slow rotator regime with rotation below 10$\%$ of the break-up limit as expected for CTTS. Synthetic velocities increase with stellar mass, as confirmed by the gradual growth of the $90^{th}$ percentile of the {\vsini} distributions, plotted with a gray line at each panel of Figure \ref{fig:grid}. We also note that rotational equilibrium is reached very rapidly, particularly for low-mass stars ($<$1 Myr). For comparison purposes, we have added the rotation rates measured for our sample of CTTS. With a few exceptions, especially for 0.3{\msun}, the bulk of {\vsini} values remains below the $90^{th}$ percentile of our simulations, reaching a rotational locked state by the end of the Hayashi track ($<$3 Myr) in agreement with previous studies \citep{Gallet2013}.\\

\begin{table}
\centering
\caption{Input parameters of the models, their ranges, and steps. \label{tab:models}}
\begin{tabular}{cccc} 
 \hline
 Parameters & Min & Max & Step \\
 \hline\hline
 $P^{in}_{rot}$ (days) & 1 & 8 & 1 \\
 $\dot{M}^{in}_{acc}$ $(M_{\odot}~yr^{-1})$ & $10^{-10}$ & $10^{-7}$ & $10^{-0.2}$ \\
 \hline
 $M_{\ast}$ $(M_{\odot})$ & 0.3 & 1.2 & 0.1 \\
 $B_{\ast}$ (G) & 500 & 3500 & 100 \\
 $\chi$ & 0.01 & 0.60 & 0.01 \\
 \hline
\end{tabular}
\end{table}

\subsection{Spin-torque equilibrium}

The total torque on the system is the sum of the accretion, winds, and star-disk interaction torques:

\begin{equation}
    \tau_{\ast}=\tau_{acc}+\tau_{wind}+\tau_{DL}
\end{equation}
Values of $\tau_{\ast}=0$ are also known as the torque equilibrium state. In the APSW scenario, stellar rotation evolves rapidly ($<$1 Myr) toward an equilibrium state in which angular momentum transferred toward the star by the disk accretion is balanced with that transferred outward via winds and magnetic star-disk interaction \citep{Matt2005,Matt2012}. For bonafide samples of CTTS with reliable masses, ages, accretion, and rotation rates, the determination of this equilibrium state permits indirectly inferring the response of the wind to different magnetic strengths and topologies and allows for searching statistical trends between the branching ratio and the stellar magnetic field.\\

Panel A of Figure \ref{fig:torque} reveals that the torque equilibrium state, on average, is reached at 1.5 Myr. On the other side, panels B and C of Figure \ref{fig:torque} also indicate that an equilibrium state can be achieved with higher values of $B_{\ast}$ and $\chi$ at ages earlier than 1.5 Myr. The majority of the models are placed on the equilibrium zone, especially stars older than 1.5 Myrs, to ensure that equilibrium is rapidly reached, a fact consistent with previous studies \citep{Yi1994,Armitage1996,Matt2012}.\\

Interestingly, stellar winds can achieve equilibrium spin rates during the Hayashi track when magnetic field strengths and branching ratios are large. In \S\ref{sec:discussion}, we discuss this result based on statistical techniques in depth. \\

\begin{figure*}[htp!]
    \centering
    \includegraphics[scale=0.5]{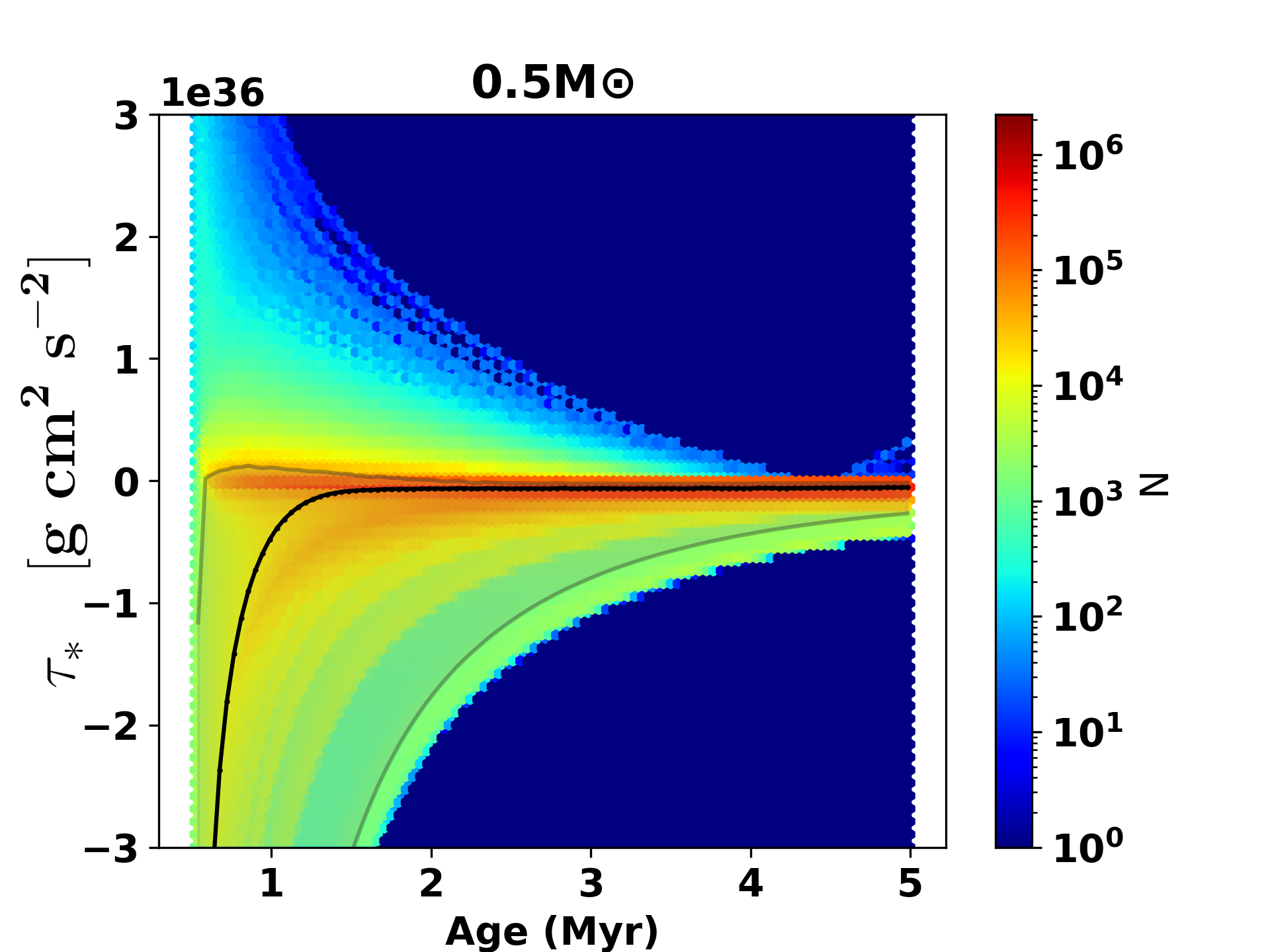}
    \includegraphics[scale=0.5]{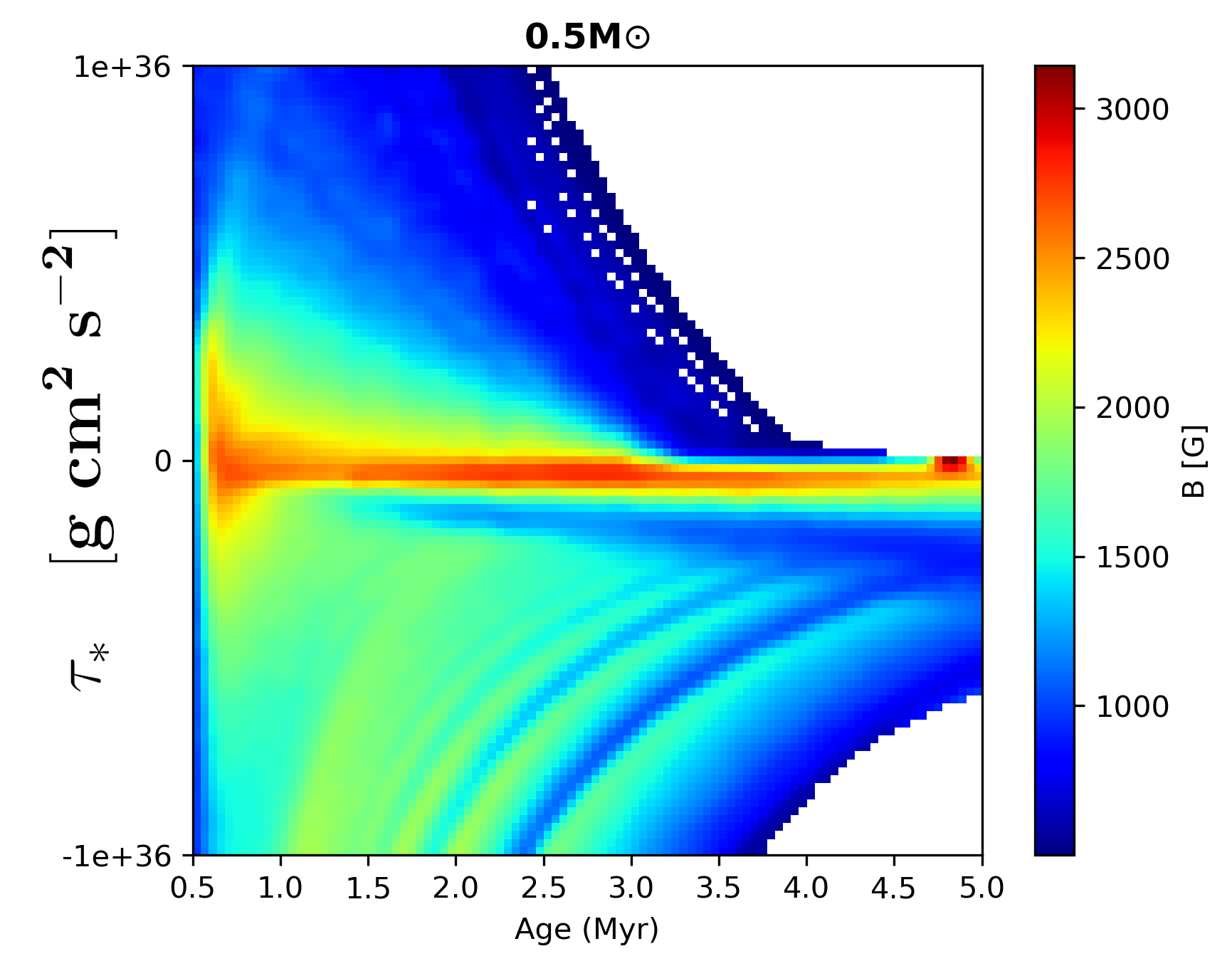}
    \includegraphics[scale=0.5]{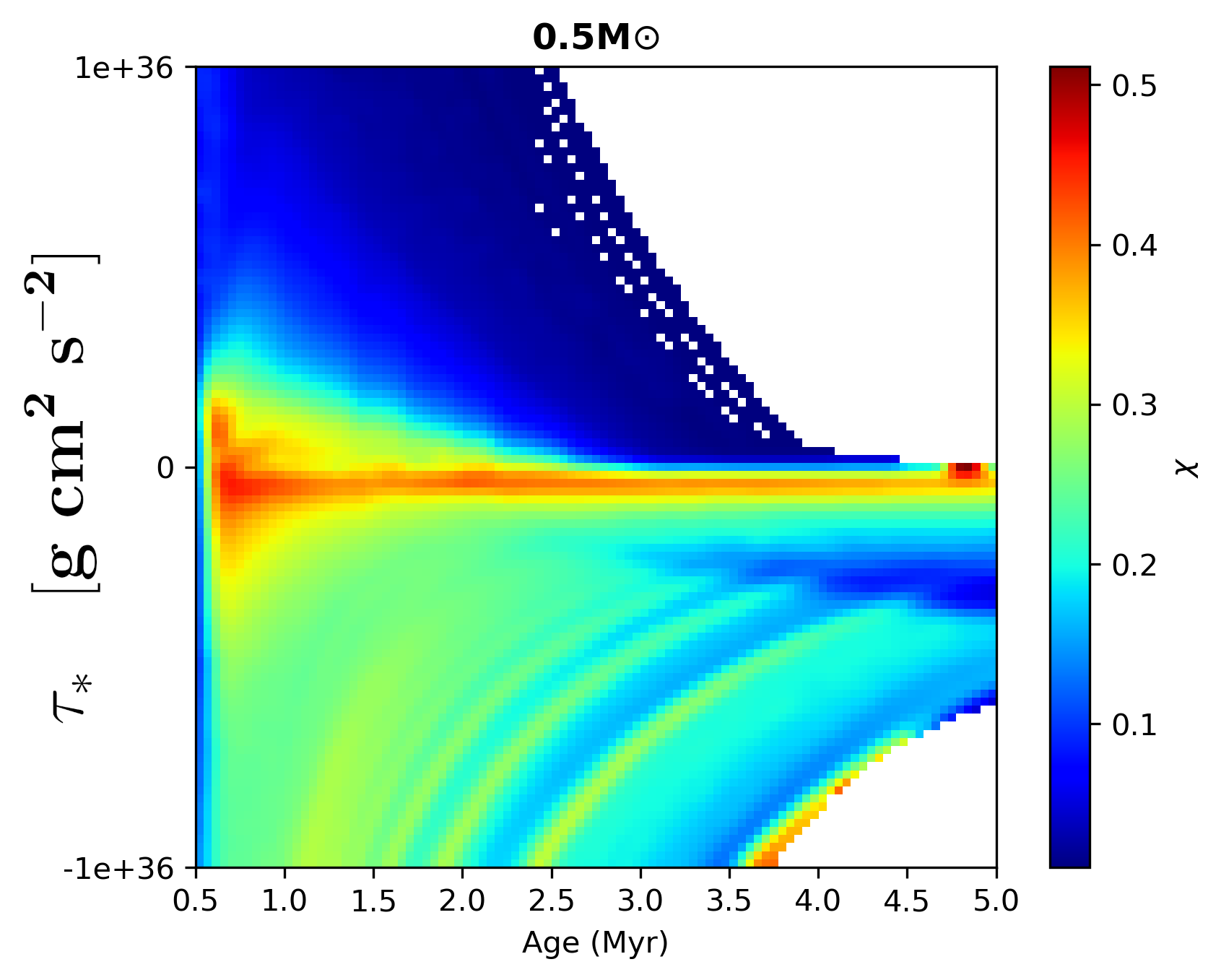}
    \caption{Rotational grid for the case $0.5$ $M_{\odot}$ obtained using the set of input parameters in Table \ref{tab:models}. (1) The distribution of the total torque applied onto the stars as a function of age. The color bar represents the number of models per bin. The median of the models is represented as the black line. The grey zone contains 90$\%$ of the models. (2) Distribution of magnetic stellar field strengths versus age. The color bar illustrates the median value per bin. (3) Distribution of branching ratio versus age. Similarly to (2), the color bar represents the median values per bin.}
    \label{fig:torque}
\end{figure*}

\subsection{Observations and models comparison}
\label{sec:comparison_obs}

With the aim of conducting careful comparisons between observations and models, we implement Bayesian techniques. We use our sample of CTTS (\S \ref{sec:ctts}), together with the models described in \S \ref{sec:spinmodel}, to explore the interplay between stellar magnetic field strength and the branching ratio $\chi$ in young accreting systems during the first Myr of their stellar evolution. 
To perform the comparison, we use the observed rotation rates compiled in Table \ref{tab:parameters} and the rotation rates provided by the models, varying the parameters in the ranges shown in Table \ref{tab:models}.
Below we describe the use of the Bayesian technique in more detail.\\

\subsubsection{Stellar magnetic field and wind mass-loss rate evolution: forward modeling}
\label{sec:abc}

We carry out a statistical study to understand the changes in the physical parameters related to the evolution of the rotation velocities of young stars. In order to perform an analysis as robust as possible, we split our sample of CTTS described in \S \ref{sec:ctts} into four age bins as follows: 
bin 1: 0-1 \text{Myr} (62 stars), bin 2: 1-2 \text{Myr} (68 stars), bin 3: 2-3 \text{Myr} (40 stars), and bin 4: 3-13 \text{Myr} (30 stars). We use the compiled {\vsini} measurements (see \S \ref{sec:apogee}) to build the corresponding distribution of rotational velocities for each age bin. These distributions are shown in Figure \ref{fig_vsini_dist}. \\

\begin{figure}[ht!]
    \centering
    \includegraphics[scale=0.25]{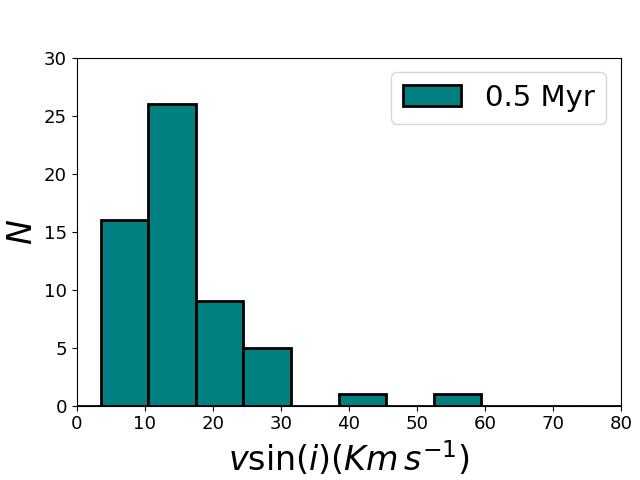}
    \includegraphics[scale=0.25]{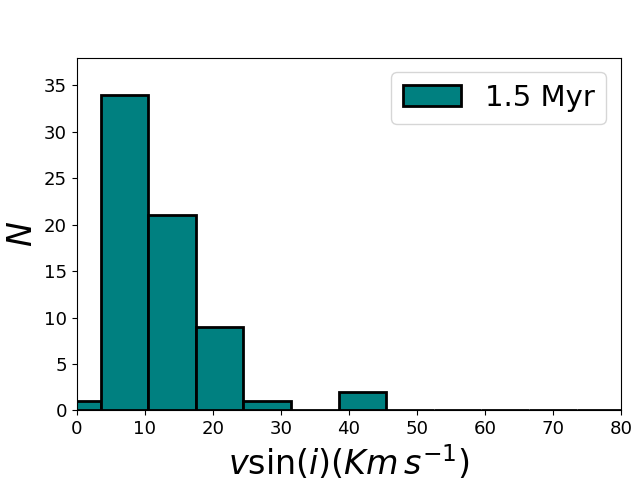}
    \includegraphics[scale=0.25]{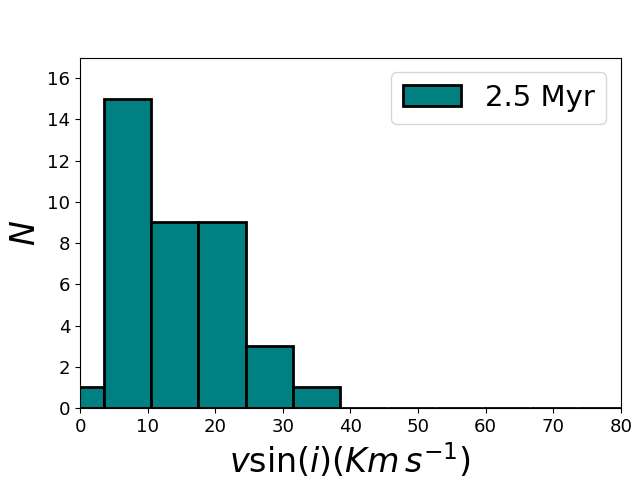}
    \includegraphics[scale=0.25]{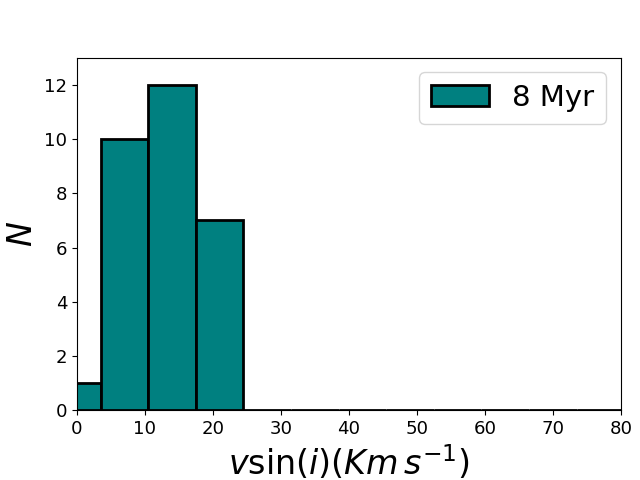}
    \caption{Rotational velocity distributions for the age bins described in \S \ref{sec:abc}.}
    \label{fig_vsini_dist}
\end{figure}

First, we selected from the grid model generated in section \ref{sec:spinmodel} an ensemble of models for each age bin. Thus, we have four ensembles of models, one per each CTTS sample.
Then, we reproduce the observed {\vsini} histograms for each age bin using the Approximate Bayesian Computation (ABC) method, a statistical approach that compares models with observed data \citep[e.g.,][]{Marjoram2015,Manzo_2020}. This is achieved by randomly selecting {\vsini} values from the respective ensemble of models previously selected. For each {\vsini} distribution of Figure \ref{fig_vsini_dist}, we keep and stack the values that fall within the range of the observations and discard those that fall out of this range. We continue selecting theoretical {\vsini} values randomly until we complete one realization, meaning that the observed histogram is exactly reproduced \citep[e.g.][]{Turner2012}. 
Due to the limitations of the ABC method, for each realization, only one parameter ($B_{\ast}$ or $\chi$) can be chosen to obtain the posterior distribution. If $B_{\ast}$ is chosen, a fixed value of $\chi=0.3$ is used (Case A), representing the minimum value for spin equilibrium \citep{Ireland_2020}. If $\chi$ is chosen, $B_{\ast}$ is fixed at 2000 G (Case B), the average magnetic field strength from CTTS measurements \citep{Johns_Krull_2007}. This analysis is repeated 100 times for each age bin to build statistically robust distributions of the chosen parameter.

\begin{figure}[htpb]
    \centering
    \includegraphics[scale=0.25]{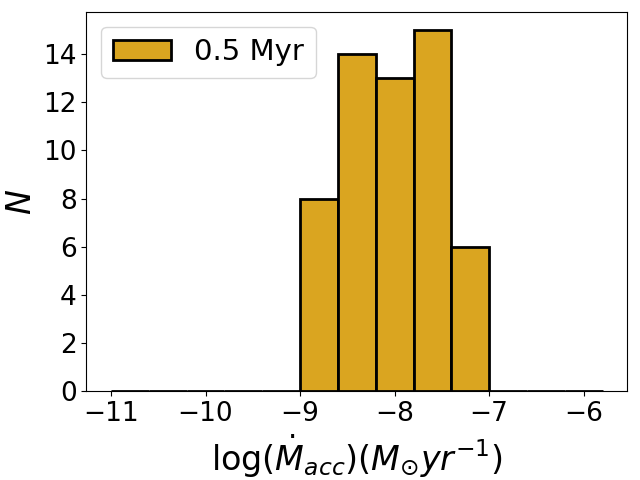}
    \includegraphics[scale=0.25]{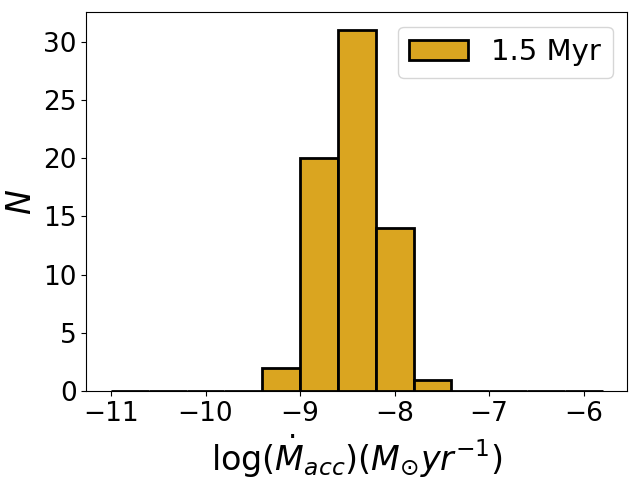}
    \includegraphics[scale=0.25]{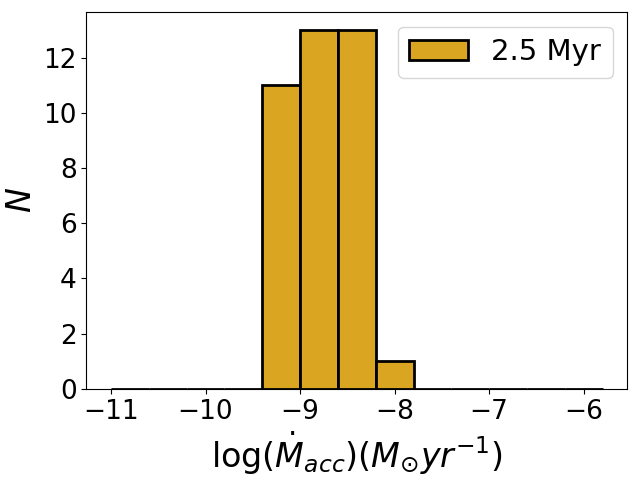}
    \includegraphics[scale=0.25]{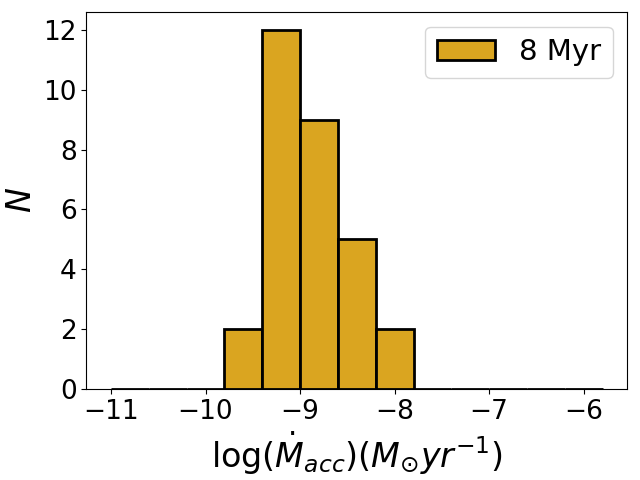}
    \caption{Mass accretion rate distributions for the age bins defined in the Bayesian analysis.}
    \label{fig:mdot_dist}
\end{figure}

In the ABC analysis, we use $M_{\ast}=0.5M_{\odot}$ since it is the median stellar mass from the sample of stars used in each bin. We decided to use a representative value of stellar mass for this analysis since the mass range of the CTTS sample is relatively narrow (e.g., an interquartile range of $\sim$0.2 $M_\odot$), and we do not have enough CTTS to additionally split them into several mass bins. We also have included the observed distribution of mass accretion rates discussed in \S\ref{sec:mass_acc_rates} as a constraint. For this purpose, we separate the sample of stars with accretion measurements into the same age bins defined in this section and use the corresponding $\dot{M_{acc}}$ distribution at each age bin as a prior distribution. Figure  \ref{fig:mdot_dist} shows the distributions of mass accretion rates for each bin. It can be seen that, as the age increases, the distributions move towards lower values of accretion, as expected \citep[e.g.][see \S \ref{sec:mass_acc_rates} for more discussion]{Hartmann_2016}. The randomly selected $\dot{M_{acc}}$ values used in the ABC analysis follow the corresponding cumulative distribution of accretion rates for each age bin. Finally, we obtained the posterior distribution of  $\chi$ and $B_{\ast}$ (see Figure \ref{fig_parameters_dist}).
Using the 100 realizations of the ABC techniques, we construct the histograms using the mean and the standard deviation for each bin of $\chi$ and $B_{\ast}$, respectively.\\

\begin{figure}[htpb]
    \centering
    \includegraphics[scale=0.25]{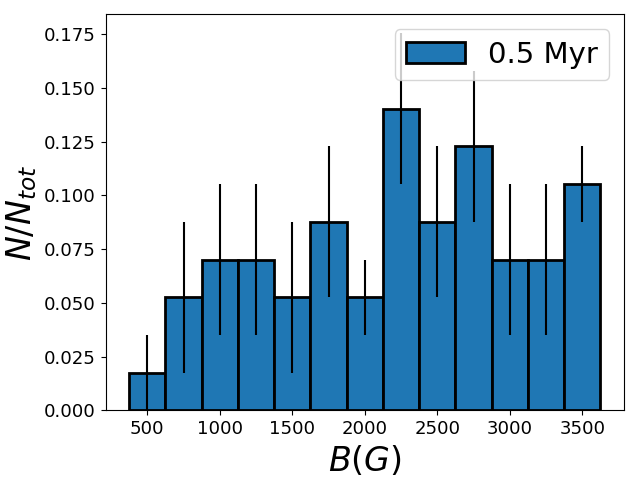}
    \includegraphics[scale=0.25]{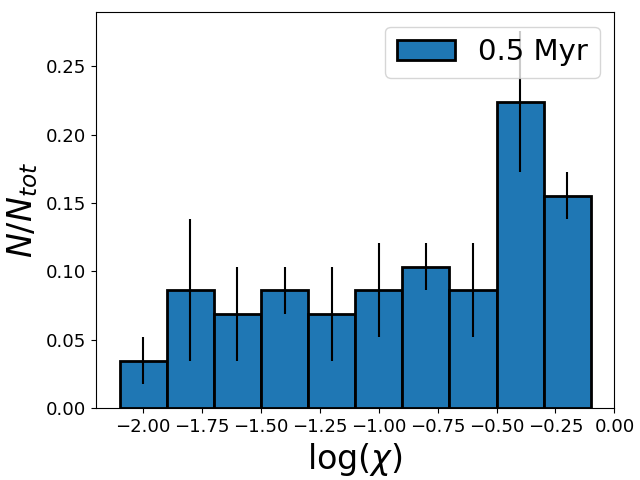}
    \includegraphics[scale=0.25]{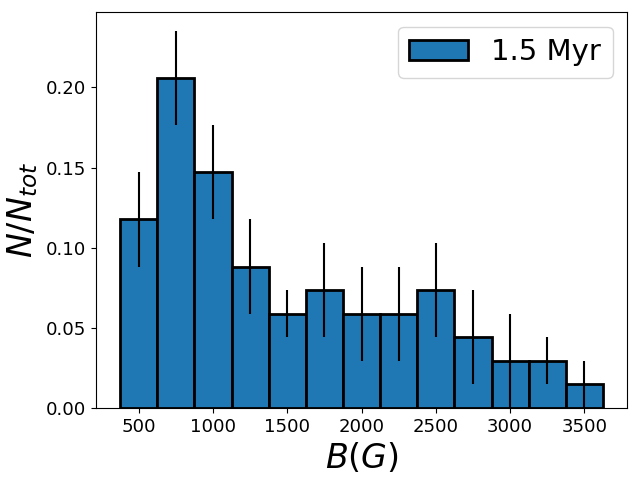}
    \includegraphics[scale=0.25]{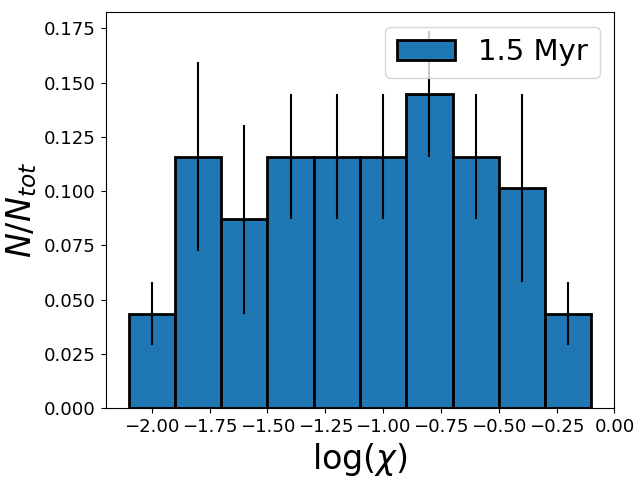}
    \includegraphics[scale=0.25]{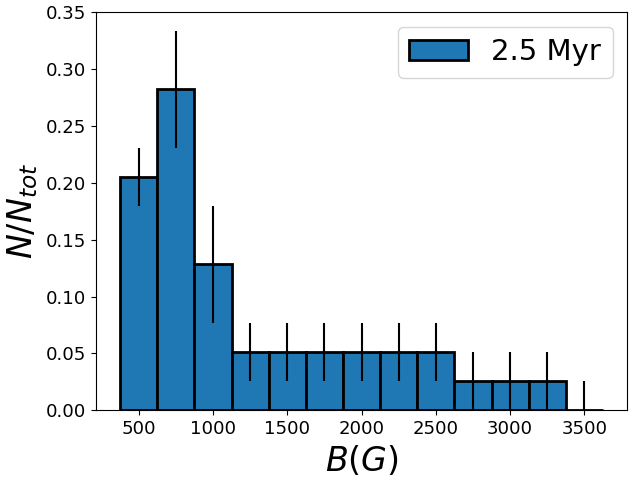}
    \includegraphics[scale=0.25]{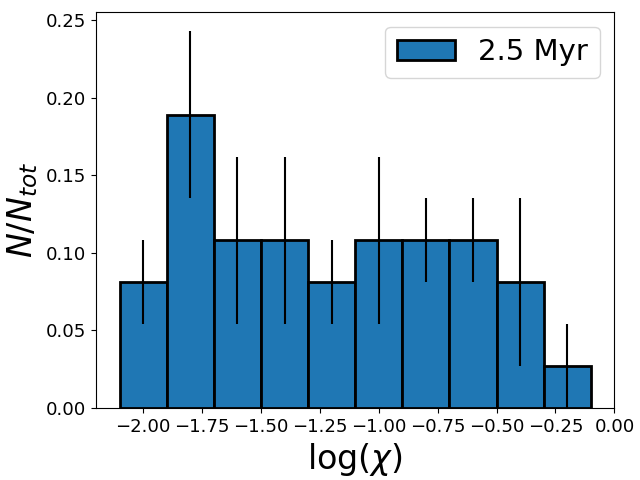}
    \includegraphics[scale=0.25]{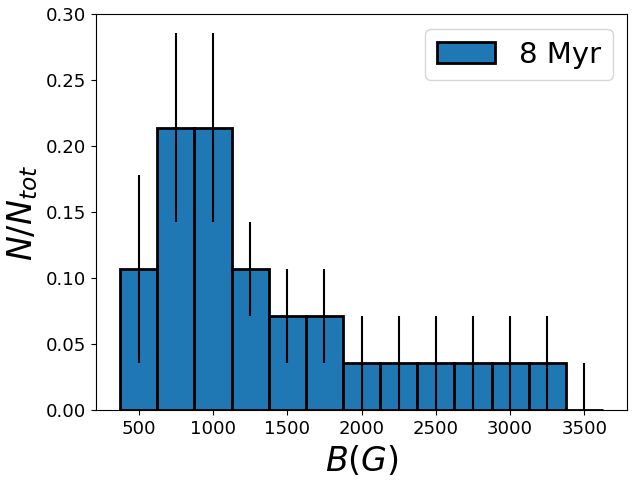}
    \includegraphics[scale=0.25]{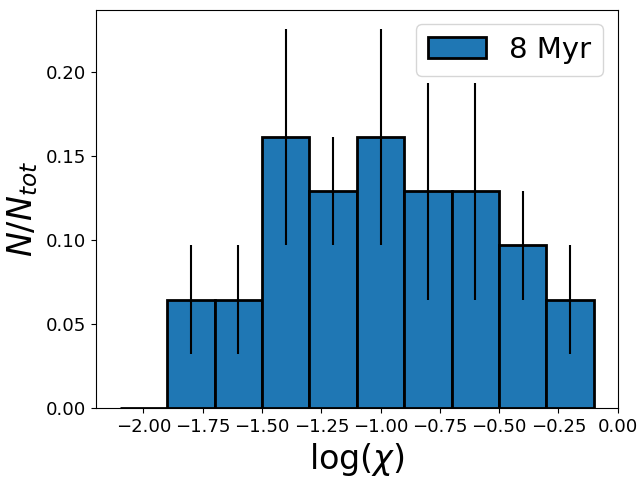}
    \caption{Evolution of $B_{\ast}$ and $\chi$ in young stars revealed by the ABC analysis. The left panel shows the posterior distributions of magnetic field strength $B_{\ast}$ for different age bins, obtained assuming a fixed $\chi=0.3$ (Case A). The right panel shows the posterior distributions of $\log(\chi)$ for different age bins, obtained by fixing $B_{\ast}=2000$ G. The results suggest that one of the two parameters, either $B_{\ast}$ or $\chi$, likely decreases with age, while other remain fixed play a significant role in the rotational evolution of young stars.}
    \label{fig_parameters_dist}
\end{figure}

Figure \ref{fig_parameters_dist} shows the resulting distribution for (1) $B_{\ast}$ (left column: Case A) and (2) $\log(\chi)$ (right column: Case B) obtained from our Bayesian analysis. In these plots, the age increases downwards.

Based on this comparison, we obtain that $B_{\ast}$ distributions with systematically smaller values can reproduce better the observed {\vsini} distributions when we move to older bin ages. Specifically,
for the stellar magnetic field, the distribution at bin 1 shows that median value is $\sim2100$ G, and the median in bins 2, 3, and 4 moves toward lower values. This suggests that $B_{\ast}$ decreases as the stars evolve during the pre-main sequence. Similarly, for the branching ratio, the distribution in bin 1 shows a median value at $\chi=0.15$, and the median values of the 2 and 3 distributions move to lower branching ratios than the first bin. At bin 4, the peak appears slightly shifted toward larger $\chi$ values ($\chi\sim 0.04$) with respect to bin 3. These results are shown more clearly in Figure \ref{fig:cumulative_distl}, in which we plot the cumulative distributions of the data from Figure \ref{fig_parameters_dist}.\\

\begin{figure}[ht!]
    \centering
    \includegraphics[scale=0.40]{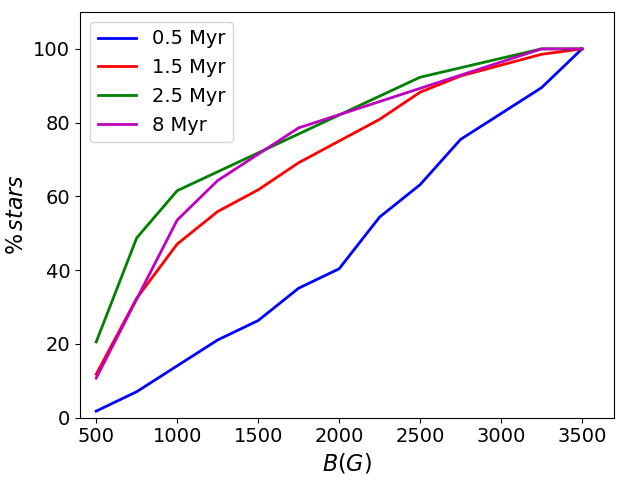}
    \includegraphics[scale=0.40]{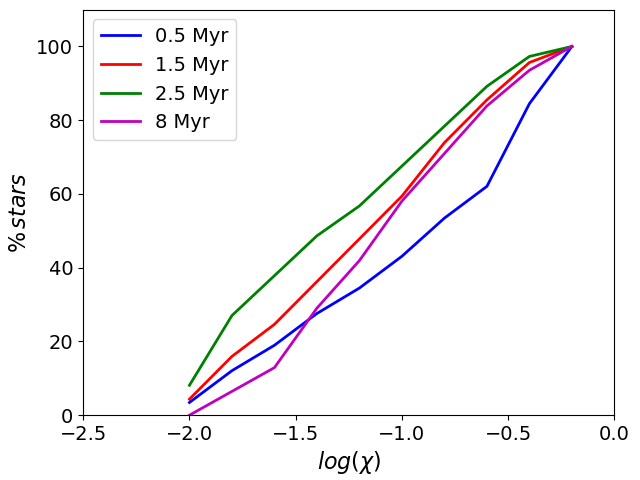}
    \caption{Cumulative distributions for $B_{\ast}$ (upper panel) and $\log(\chi)$ (lower panel), for bin 1 (blue), bin 2 (red), bin 3 (green), and bin 4 (magenta).}
    \label{fig:cumulative_distl}
\end{figure}

The histograms in Figure \ref{fig_parameters_dist} suggest a gradual decrease in the branching ratio and magnetic field strength for the first three age bins. The slight increase in the branching ratio ($\chi)$ from bin 3 to bin 4 is unexpected. It must be noticed that the oldest bin has an age spread of 3-13 Myr and is the smallest sample of the different bins. If the sample in bin 4 is not a representative distribution for $v\sin(i)$ in this age interval, the results for $\chi$ from the ABC method could be biased towards values that describe this incomplete sample.\\

In order to test the robustness of our results with the ABC technique, particularly the impact of low numbers in the sample used in bin 4, we performed an additional Bayesian test using synthetic prior distributions for $v\sin(i)$. We generated random Gaussian distributions for each age bin, using the corresponding median value and the median absolute deviation (MAD) of the observed $v\sin(i)$ histograms shown in Figure \ref{fig_vsini_dist}. We demonstrate in Appendix \ref{ap:B} that the general trends seen in Figure \ref{fig_parameters_dist} hold, and they are independent of the number of stars in each bin. Further details of this test can be found in Appendix \ref{ap:B}.
Although our results suggest that the observed increase of $B_{\ast}$ and $\chi$ toward bin 4 (Figure \ref{fig_parameters_dist})  is not a statistical fluctuation caused by low numbers, this result should be taken carefully since:
(1) The sample of CTTS in the last bin could not be as representative as in the younger bins. In particular, given the variable nature of TTS, slow accretors or CTTS approaching the end of their accretion phase \citep{Briceno_2019, Thanathibodee_2022} could not be included in the last bins;
(2) A wide age range (3-13 Myr) is considered in the last bin, in which the stars significantly change their properties during the pre-main sequence evolution. However, torque equilibrium in the system is reached earlier than 3 Myr, and models beyond this age do not change significantly (see figure \ref{fig:grid});
(3) We assume that the parameters $\gamma_{c}$ and $\beta$ are constant. These parameters are related to the magnetic connection between the star and the disk, which might change, together with the complexity of the magnetic field, during the evolution of the star; (4) The models ignore the changes in the inner stellar structure, particularly, the core-envelope decoupling is not considered in the models, which might impact the angular momentum evolution at bin 4. Some of these effects are not so important for very young low-mass stars, however, solar-type stars are expected to develop a radiative core at the end of the Hayashi track \citep{Iben_1965}. Exploring these effects' impact is out of this paper's scope.\\

We applied a Kolmogorov-Smirnov (K-S) test, which is used to determine if two samples are drawn from the same distribution, to the $B_{\ast}$ and $\log(\chi)$ posterior distributions obtained from the ABC analysis. The $p$ values are shown in Table \ref{tab:ks}. For bins 1 and 2, we find the low $p$ values of $0.05\%$ and $5\%$, for $B_{\ast}$ and $\log(\chi)$, respectively, suggest evolution of both parameters, as discussed above. In contrast, for $B_{\ast}$, the high $p$ value between bin 3 and bin 4 shows that there is barely an evolution, which suggests that most of the stars have reached a state in which the strength of their magnetic field has decreased to lower values. For $\log(\chi)$, the difference between bins 3 and 4 seems more clear; the possible explanations for these differences are also discussed above.\\

\begin{table}
\centering
\caption{K-S test and $p$ values \label{tab:ks}.}
\begin{tabular}{cccc} 
 \hline\hline
 Bin pairs & parameter & K-S & p-value \\
 \hline
 1-2 & $B_{\ast}$ & 0.35 & 0.0005 \\
     & $\log(\chi)$ & 0.23 & 0.05 \\
 \hline
 2-3 & $B_{\ast}$ &  0.16 & 0.45 \\
     & $\log(\chi)$ & 0.13 & 0.73 \\
 \hline
 3-4 & $B_{\ast}$ & 0.16 & 0.68 \\
     & $\log(\chi)$ & 0.24 & 0.20 \\
 \hline
\end{tabular}
\end{table} 

Our results suggest that the median of $B_{\ast}$ and $\log(\chi)$ might be decreasing as the age of the bins increases, this implies that both parameters could evolve during the first million years. These results remain consistent under testing with back-forward models. For additional information, please refer to the model insights in Appendix \ref{ap:C}. Additional measurements of magnetic fields and stellar winds are needed to confirm the observed trends.

\section{Discussion} \label{sec:discussion}

\subsection{Binarity}

We make a quality astrometric selection of the CTTS using the Re-normalized unit weighted error (RUWE) reported by \textit{Gaia} DR3. For the selection, we use RUWE values less than 1.4, suggested by \citet{Lindegreen2020a}. However, \citet{Stassun_2021} and \citet{Kounkel_2021} confirmed that RUWE values from 1.0 to 1.4 may contain unresolved binary systems.\\

One of the main consequences of having binary stars in our sample will be the inconsistency of the physical parameters (e.g., Rotation rates, masses, ages, etc.). In that sense, we have cleared our sample of possible close companions systems using the spectroscopic binaries reported by \citep{Kounkel_2019} and visual binaries reported by \citet{Tokovinin_2020}.
Also, we did not find discrepancies in rotation rates for CTTS groups by mass and age that could indicate the presence of binaries.
As a double-check test, we inspected each LC available in TESS (\S\ref{sec:periods}), and as a result, we did not detect signatures of eclipsing binaries in the sample.\\

The RUWE and the LCs inspection can not reject all multiple systems in our sample, particularly unresolved systems. However, we expect that binary systems do not dominate our sample and, thus, do not affect our results.\\

\subsection{Clues about the branching ratio}

Several studies have explored the branching ratio parameter from a numerical perspective to explain the slow rotations in CTTS. At the same time, only some efforts have focused on the observational viewpoint.
For example, \citet{Hartmann1989,Matt_2005b,Matt2012} have suggested that if there is a stellar wind with a sustained mass-loss rate of about $0.1$ of the accretion rate, the wind can carry away enough angular momentum to achieve a stellar-spin equilibrium. \citet{Matt_2008a} suggested a hard upper limit of $\chi\leq0.6$. \citet{Cranmer_2008}, by his part, predicts a maximum of $0.014$. \citet{Gallet2019} studied two cases where they argued that a branching ratio $\chi=0.01$ would favor the stellar wind theory. However, simulations would require stronger dipolar magnetic fields to reproduce the rotation rates, while $\chi=0.1$ requires weaker dipolar fields that are commonly observed.
\citet{Ireland_2020} suggest that $\chi>0.3$ is required to achieve spin-equilibrium in their simulated cases. Particularly, for the BP Tau star, they predict a $\chi\approx0.25$ to achieve an equilibrium between the wind and accretion torques.
\citet{Pantolmos_2020} predict that winds should eject more than $0.1$ of the mass-accretion rate to counteract the stellar spin-up due to accretion.
\citet{Gehrig_2022} reproduce the observed rotational period of most young stars with values less than $0.05$.
Independently of their numerical approaches and strategies to describe the star-disk interactions and winds, nowadays, no consensus has been reached.
Observational analyses of \citet{Hartigan1995} have found that $\chi$ can reach up to $0.2$, but most values are below $0.02$ in all ages between 1 and 10 Myr.
Measurements provided by this work could help to contribute to a general perspective of the expected $\chi$ in CTTS. \\

\begin{figure}[htpb]
    \centering
    \includegraphics[scale=0.5]{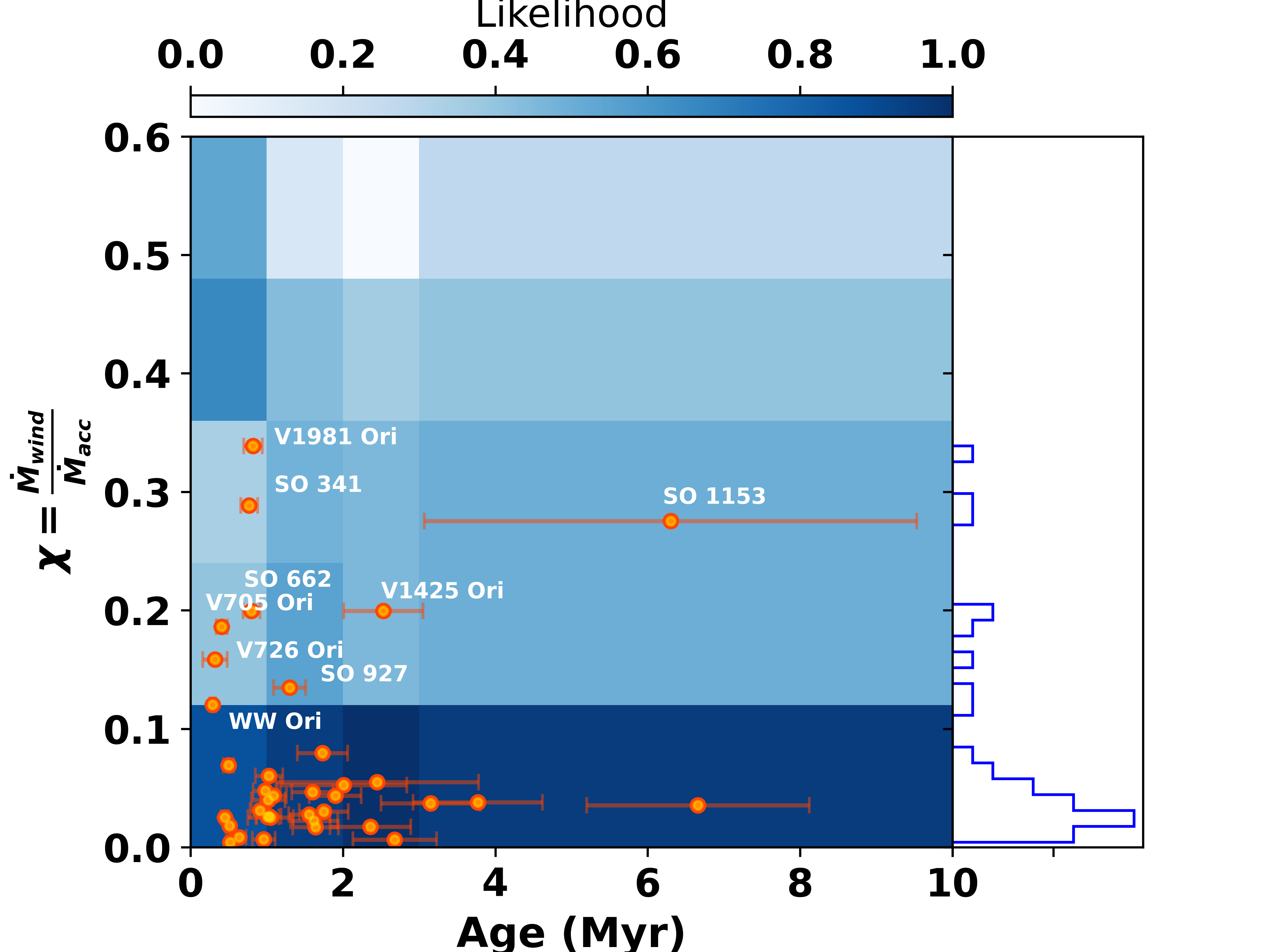}
    \caption{Branching ratio for measured Classical T Tauri stars versus age (left panel). The background map illustrates the density of the posterior distribution of $\chi$ given by the ABC analysis. We display their name for those stars with $\chi>0.1$.
    The distribution of the branching ratio represented with the histogram on the right side exhibits a pronounced peak at $\chi=0.04$ as expected from data in Figure \ref{b}.}
    \label{fig:chi evolution}
\end{figure}

We have 37 stars with available $\chi$ measurements that simultaneously hold an estimation of age (Table \ref{tab:parameters}). These stars were utilized in Figure \ref{fig:chi evolution} to investigate the relationship between $\chi$ and age (left panel). As seen in the histogram (right panel), branching ratio values tend to be below $0.1$, and the most likely values are around $0.04-0.05$. Even though few data are displayed in Figure \ref{fig:chi evolution}, within the uncertainties, our data suggest that strong stellar wind with a mass loss rate greater than $0.1$ of the accretion rate is possible at earlier ages ($<3$ Myr).\\

The ABC analysis in \S\ref{sec:abc} plus measurements reported here could suggest a possible trend for the branching ratio with age. In the future, supplementary surveys of accretion and mass-loss rates in CTTS \citep[e.g.,][]{Kounkel_2023} will improve the sampling and allow us to understand better the nature and strength of the APSW mechanism at the early stages of the stars.\\

\section{Summary and Conclusions}
\label{sec:conclusions}

In this work, we developed one-dimensional numerical models of the spin evolution of CTTS, which allowed us to build a comprehensive grid of specific cases for CTTS. We build approximately two million simulations, varying physical parameters such as magnetic field, wind, accretion, initial rotation periods, and mass.

\noindent Our models allow for the first time to study how the magnetic field strength ($B_{\ast}$) and branching ratio ($\chi$) change in relation to rotation rates and ages. The results are broadly consistent with the APSW picture and predict a steady evolution of $B_{\ast}$ and $\chi$ during the CTTS phase. Below we list the results of this work:

\begin{itemize}

\item Ninety percent of our model's grid is compliant with {\vsini} rotation rates below 50{\kms}.

\item Using the observed rotation and accretion rates as prior distributions, we performed a Bayesian analysis that suggests a decreasing trend of $\chi$ and $B_{\ast}$ intensities with age.

\item The models indicate that strong dipolar magnetic field components and higher branching ratios would achieve an effective equilibrium state of torque at the first stages of the evolution in CTTS ($<1.5$ Myr). However, at least for the branching ratio, from the observational measurements and the posterior distributions of the  ABC analysis, we speculate that $\chi<0.1$ would be more likely than $\chi>0.1$.

\item Models suggest a branching ratio decreasing with age, in particular, during the first 2 Myr when a pronounced peak for $\chi$=0.04 is observed. This result is in agreement with our measurements using a diverse sample of CTTS (Figure \ref{b}) suggesting that small $\chi$ values up to 10\% of the APSW upper limit are capable of extracting enough angular momentum from the CTTS system.

\end{itemize}

\section*{acknowledgments}

\noindent A special thanks to Lee Hartmann, Jerome Bouvier, Luis Aguilar, and the anonymous referee for their helpful opinions and recommendations for improving the article. J.S. acknowledges from the CONACYT by fellowship support in the Posgrado en Astrof\'{i}sica graduate program at Instituto de Astronom\'{i}a UNAM.
J.H. acknowledges support from the National Research Council of México (CONACyT) project No. 86372 and the PAPIIT UNAM projects IA102921, IA102319, and CG-101723. E.M.M. acknowledges the support from a CONACyT postdoctoral grant.
K.M. is funded by the European Union (ERC, WANDA, 101039452). Views and opinions expressed are however those of the author(s) only and do not necessarily reflect those of the European Union or the European Research Council Executive Agency. Neither the European Union nor the granting authority can be held responsible for them.
C.R-Z. acknowledges support from project CONACyT CB2018 A1-S9754 and the PAPIIT UNAM project IN112620. A.B. acknowledges partial funding by the Deutsche Forschungsgemeinschaft Excellence Strategy - EXC 2094 - 390783311 and the ANID BASAL project FB210003.\\

This paper includes data collected with the TESS mission, obtained from the MAST data archive at the Space Telescope Science Institute (STScI). Funding for the TESS mission is provided by the NASA Explorer Program. Funding for the Sloan Digital Sky Survey IV has been provided by the Alfred P. Sloan Foundation, the U.S. Department of Energy Office of Science, and the Participating Institutions. SDSS-IV acknowledges support and resources from the Center for High-Performance Computing at the University of Utah. The SDSS 
website is www.sdss.org. SDSS-IV is managed by the Astrophysical Research Consortium for the Participating Institutions of the SDSS Collaboration including the Brazilian Participation Group, the Carnegie Institution for Science, Carnegie Mellon University, Center for 
Astrophysics | Harvard \& Smithsonian, the Chilean Participation 
Group, the French Participation Group, Instituto de Astrof\'isica de Canarias, The Johns Hopkins University, Kavli Institute for the Physics and Mathematics of the Universe (IPMU) / University of 
Tokyo, the Korean Participation Group, Lawrence Berkeley National Laboratory, Leibniz Institut f\"ur Astrophysik Potsdam (AIP),  Max-Planck-Institut f\"ur Astronomie (MPIA Heidelberg), 
Max-Planck-Institut f\"ur Astrophysik (MPA Garching), 
Max-Planck-Institut f\"ur Extraterrestrische Physik (MPE), 
National Astronomical Observatories of China, New Mexico State University, New York University, University of Notre Dame, Observat\'ario Nacional / MCTI, The Ohio State University, Pennsylvania State University, Shanghai Astronomical Observatory, United Kingdom Participation Group, Universidad Nacional Aut\'onoma de M\'exico, University of Arizona, University of Colorado Boulder, University of Oxford, University of 
Portsmouth, University of Utah, University of Virginia, University of Washington, University of Wisconsin, Vanderbilt University, 
and Yale University. Guoshoujing Telescope (the Large Sky Area Multi-Object Fiber Spectroscopic Telescope LAMOST) is a Major National Scientific Project built by the Chinese Academy of Sciences. Funding for the project has been provided by the National Development and Reform Commission. LAMOST is operated and managed by the National Astronomical Observatories, Chinese Academy of Sciences.

\vspace{5mm}
\facilities{TESS, APOGEE-2, \textit{Gaia}-DR3, X-Shooter, GIRAFFE, and LAMOST.}

\software{Numpy \citep{Harris2020}, Matplotlib \citep{Hunter2007}, Astropy \citep{2013A&A...558A..33A}, Molecfit \citep{Kausch_2015}.}

\appendix
\section{LAMOST data archive Summary}
\label{ap:A}

In this Appendix, we compiled the low-resolution plots and table of sources with {\oi} line available in our study of CTTS. 

\begin{figure*}[ht]
    \centering
    \includegraphics[width=\textwidth]{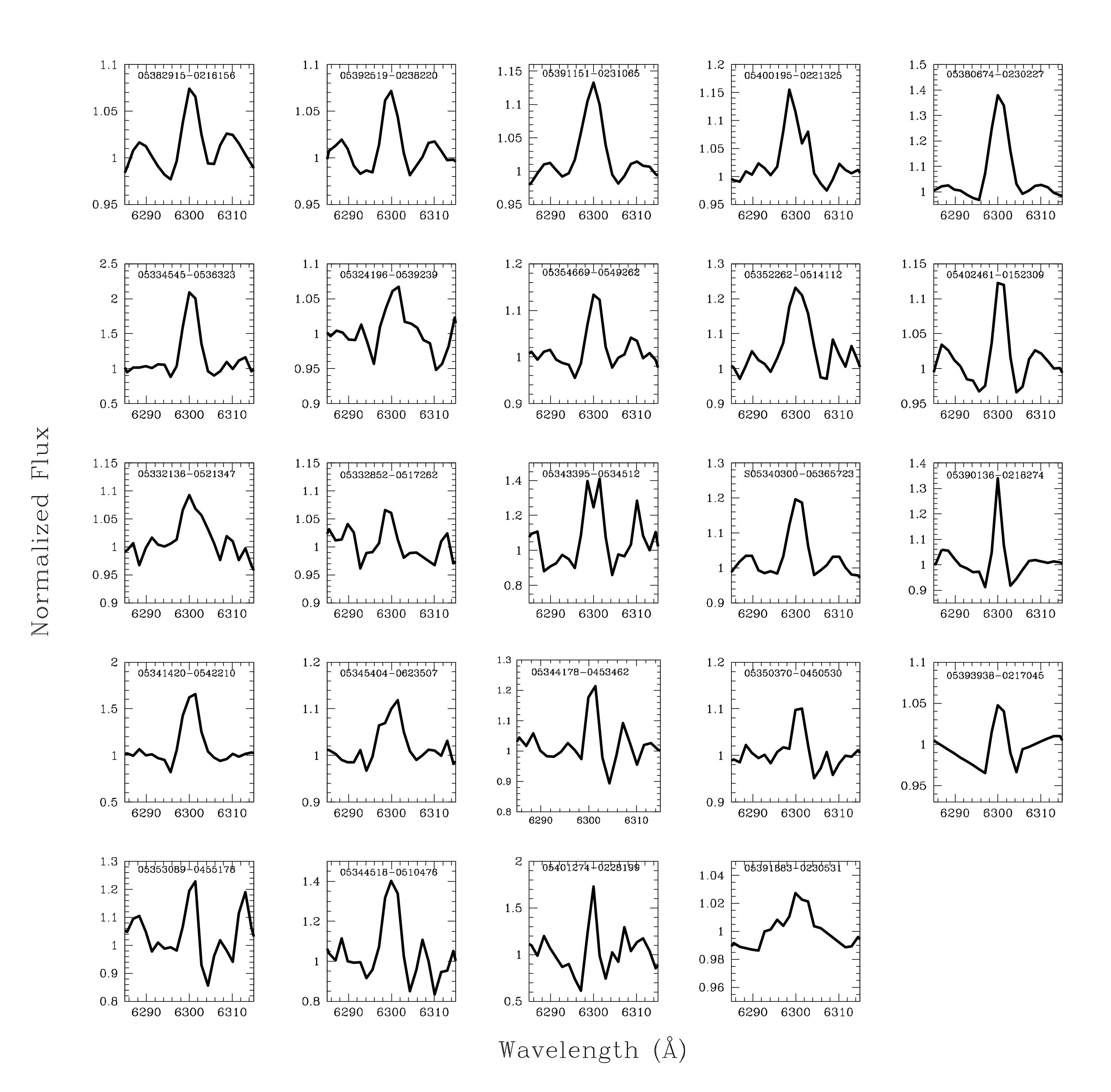}
    \caption{{\oi} Residual lines profiles for 24 CTTS found in data archive of LAMOST.}
    \label{fig:OI_3}
\end{figure*}


\begin{table}[ht]
\label{tab:obs2}
\centering
\caption{LAMOST Summary of archival data.}
\begin{tabular}{cccccc} 
 \hline
 2MASS & Spectral Resolution & Obs. Date & Exp. time & SNR\tablenotemark{a} & Plan ID \\
      ID  &                     & (UT)      & (s)       &     &  \\
 \hline\hline
05334545-0536323 & 1800 & 2013-11-20T17:26:00 & 1800.0 & 4.09 & VB081S05V1 \\
05324196-0539239 & 1800 & 2013-11-20T17:26:00 & 1800.0 & 24.66 & VB081S05V1 \\
05354669-0549262 & 1800 & 2013-11-20T17:26:00 & 1800.0 & 16.0 & VB081S05V1 \\
05352262-0514112 & 1800 & 2013-11-20T17:10:00 & 1800.0 & 13.83 & VB081S05V1 \\
05332136-0521347 & 1800 & 2013-11-20T17:10:00 & 1800.0 & 20.7 & VB081S05V1 \\
05332852-0517262 & 1800 & 2013-11-20T17:10:00 & 1800.0 & 16.21 & VB081S05V1 \\
05343395-0534512 & 1800 & 2013-11-20T18:24:00 & 1800.0 & 3.59 & VB081S05V2 \\
05340300-0536572 & 1800 & 2013-11-20T18:24:00 & 1800.0 & 38.85 & VB081S05V2 \\
05341420-0542210 & 1800 & 2013-11-20T18:24:00 & 1800.0 & 7.1 & VB081S05V2 \\
05345404-0623507 & 1800 & 2013-11-20T18:24:00 & 1800.0 & 19.61 & VB081S05V2 \\
05344178-0453462 & 1800 & 2013-11-20T18:08:00 & 1800.0 & 13.3 & VB081S05V2 \\
05350370-0450530 & 1800 & 2013-11-20T18:08:00 & 1800.0 & 31.88 & VB081S05V2 \\
05353089-0455178 & 1800 & 2013-11-20T18:08:00 & 1800.0 & 11.25 & VB081S05V2 \\
05344518-0510476 & 1800 & 2013-11-20T18:08:00 & 1800.0 & 4.96 & VB081S05V2 \\
05401274-0228199 & 1800 & 2015-03-01T12:13:00 & 1800.0 & 2.78 & HD053813S011009V01 \\
05391883-0230531 & 1800 & 2016-12-15T17:01:00 & 1800.0 & 13.19 & HD054602S041606V01 \\
05402461-0152309 & 1800 & 2017-01-01T14:26:00 & 4500.0 & 101.73 & GAC085S03B1 \\
05390136-0218274 & 1800 & 2017-01-01T14:26:00 & 4500.0 & 121.31 & GAC085S03B1 \\
05393938-0217045 & 1800 & 2017-01-01T14:26:00 & 4500.0 & 180.07 & GAC085S03B1 \\
05382915-0216156 & 1800 & 2017-01-01T14:26:00 & 4500.0 & 224.63 & GAC085S03B1 \\
05380674-0230227 & 1800 & 2017-01-01T14:26:00 & 4500.0 & 119.59 & GAC085S03B1 \\
05400195-0221325 & 1800 & 2017-01-01T14:26:00 & 4500.0 & 49.09 & GAC085S03B1 \\
05391151-0231065 & 1800 & 2017-01-01T14:26:00 & 4500.0 & 136.68 & GAC085S03B1 \\
05392519-0238220 & 1800 & 2017-01-01T14:26:00 & 4500.0 & 233.1 & GAC085S03B1 \\
 \hline
\end{tabular}
\tablenotetext{a}{The reported SNR was estimated at the r' Sloan band in which the effective central wavelength corresponds to 620.4 nm.}
\end{table}

\clearpage
\section{ABC analysis additional test}
\label{ap:B}

We performed an additional test using the ABC technique to explore the robustness of the results discussed in \S \ref{sec:abc}. In particular, to explore if the increase of $\log(\chi)$ from bin 3 to bin 4 is an effect of low numbers in the last bin. For this purpose, for each $v \sin(i)$ distribution shown in Figure \ref{fig_vsini_dist}, we used the corresponding median value and the median absolute deviation to build synthetic random Gaussian distributions using 200 points. The median and MAD values are (13.6, 3.41), (10.48, 2.17), (13.01, 4.98), and (12.89, 5.13) {\kms}, for bin 1, bin 2, bin 3, and bin 4, respectively. The artificial Gaussian distributions are shown in Figure \ref{fig_vsini_dist_test}.

\begin{figure}[htpb!]
    \centering
    \includegraphics[scale=0.25]{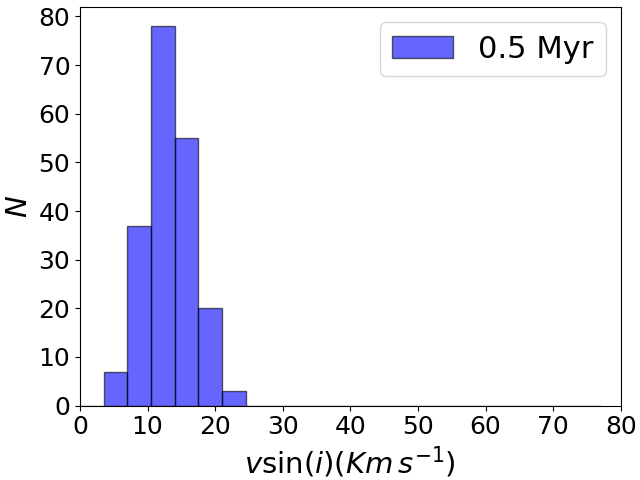}
    \includegraphics[scale=0.25]{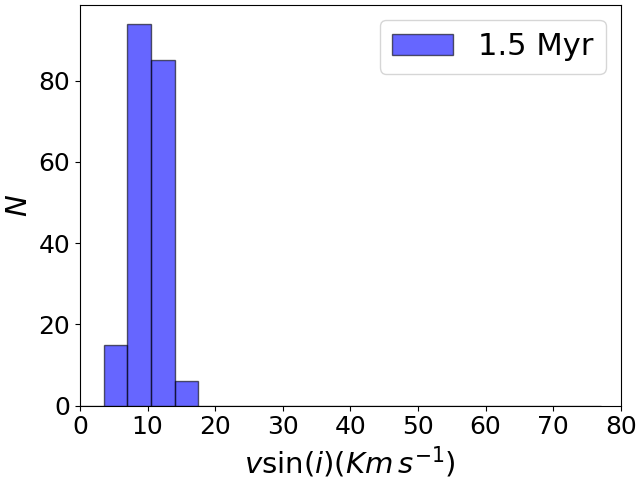}
    \includegraphics[scale=0.25]{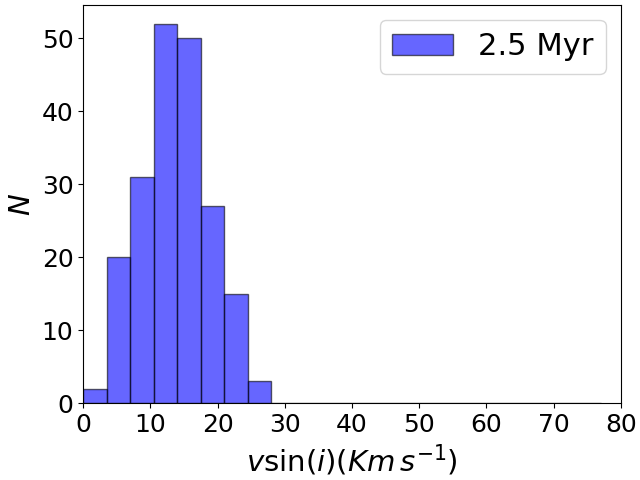}
    \includegraphics[scale=0.25]{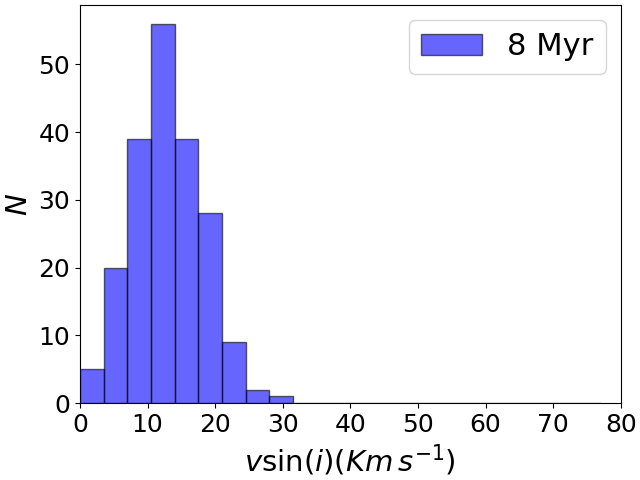}
    \caption{Synthetic distribution of {\vsini} at each age bin.}
    \label{fig_vsini_dist_test}
\end{figure}

In this ABC test, we performed 100 realizations to fill the synthetic $v\sin(i)$ histograms shown in Figure \ref{fig_vsini_dist_test}. The obtained posterior distributions are shown in Figure \ref{fig_parameters_dist_test}. For $\log(\chi)$, we also find a gradual transition from high to low branching ratios from bin 1 to bin 3, similar to what is seen in Figure \ref{fig_parameters_dist}. In addition, the increase in values of $\log(\chi)$ from bin 3 to bin 4 also is present. For the magnetic field strength $B_{\ast}$, the posterior distributions are the same as the corresponding distributions from Figure \ref{fig_parameters_dist}. Since we use more complete $v\sin(i)$ samples to be forward modeled, the standard deviation, represented with vertical bars in each bin value of $\log(\chi)$ and $B_{\ast}$, is smaller compared with the results seen in Figure \ref{fig_parameters_dist}. This test confirms that the observed increase, in particular for $\log(\chi)$ from bin 3 to bin 4. is not a consequence of low numbers in the $v\sin(i)$ sample used in bin 4. The interpretation of this result, as well as possible explanations, are discussed in \S\ref{sec:abc}.

\begin{figure}[htpb!]
    \centering
    \includegraphics[scale=0.25]{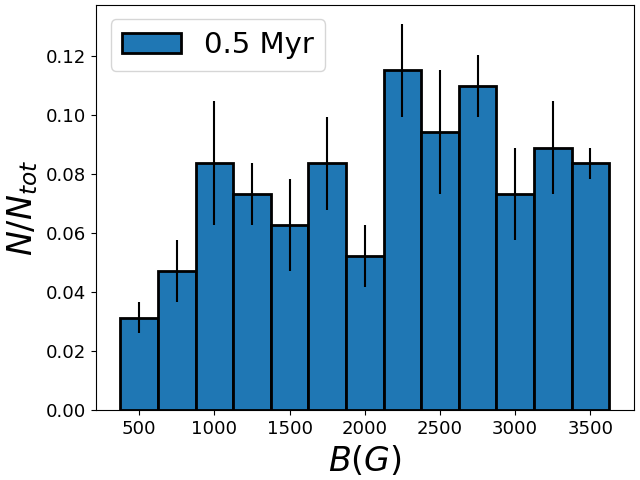}
    \includegraphics[scale=0.25]{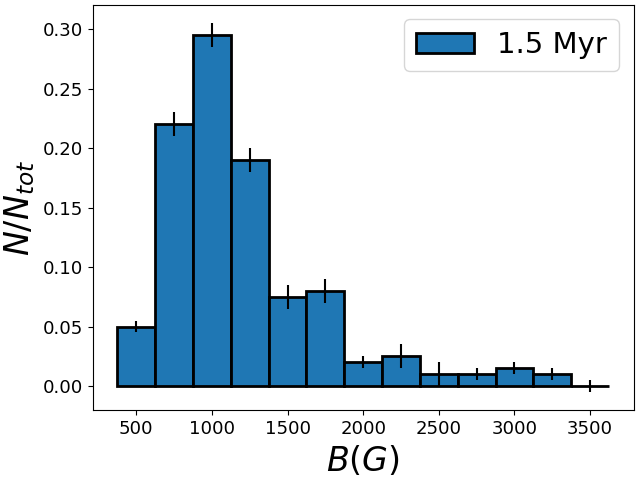}
    \includegraphics[scale=0.25]{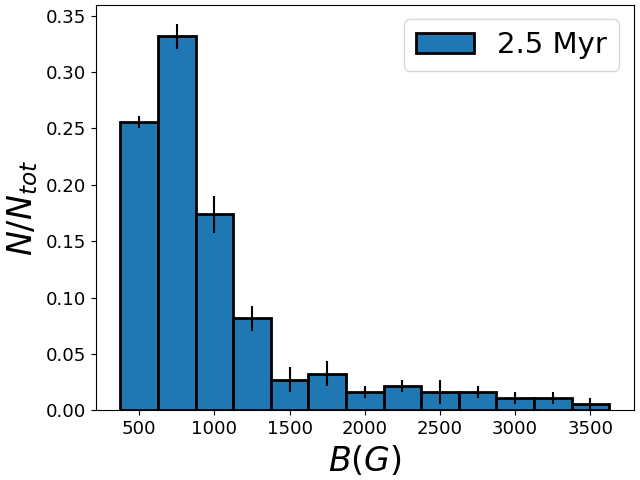}
    \includegraphics[scale=0.25]{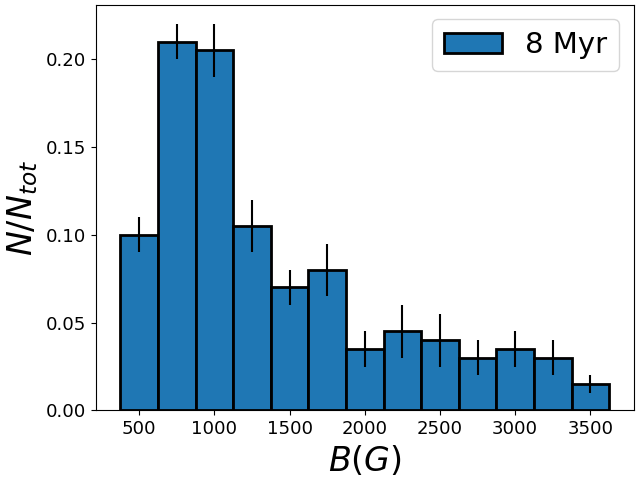}
    \includegraphics[scale=0.25]{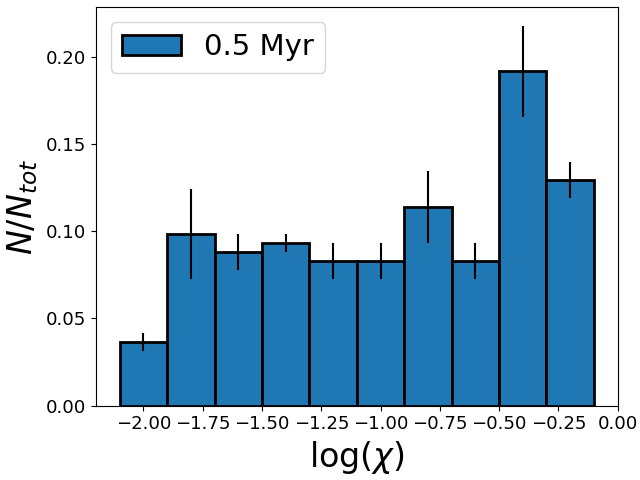}
    \includegraphics[scale=0.25]{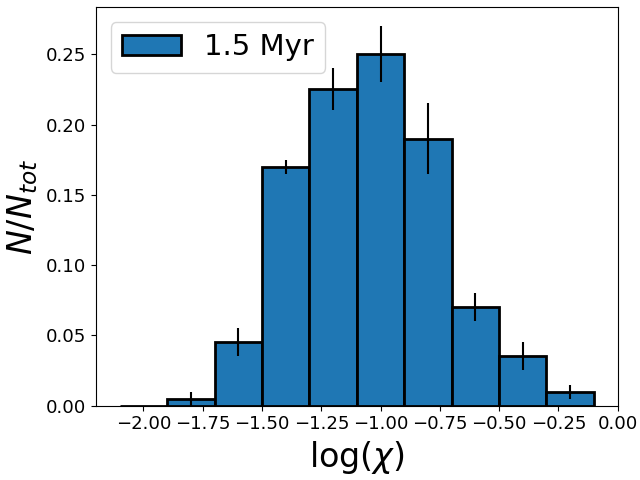}
    \includegraphics[scale=0.25]{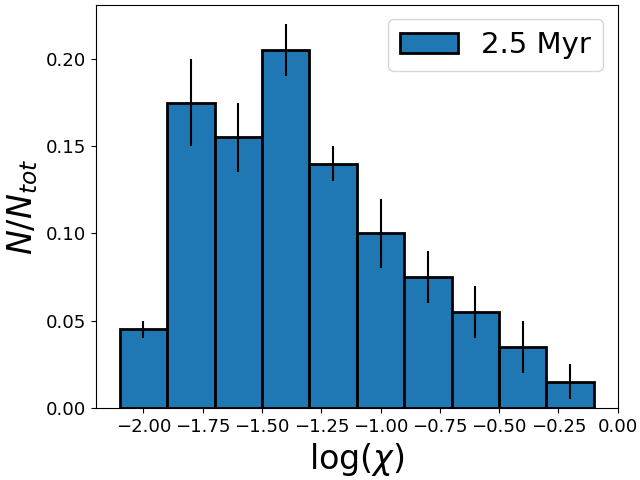}
    \includegraphics[scale=0.25]{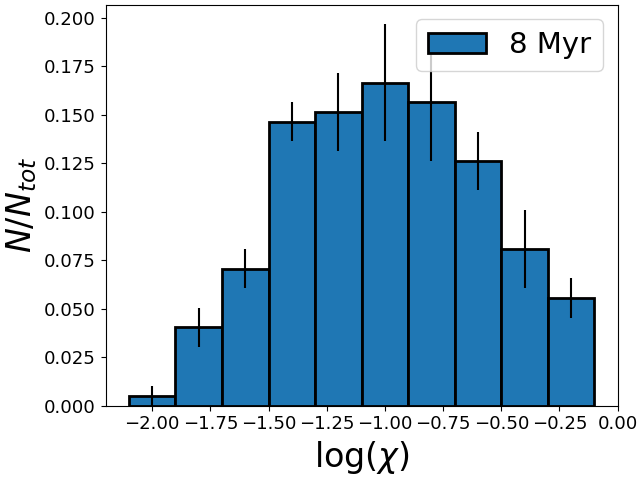}
    \caption{Distributions of parameters at each age bin for the test.}
    \label{fig_parameters_dist_test}
\end{figure}

\section{Auto-consistency of ABC Results (ABC Back-forward Tests)}
\label{ap:C}

To validate the findings from the Approximate Bayesian Computation (ABC) analysis (Section \ref{sec:abc}) and ensure the self-consistency of the method, we have implemented a series of "back-forward tests". These tests aim to demonstrate the ability of the posterior model parameters derived from ABC to reproduce approximately the observed $v\sin(i)$ distributions. The following are the key steps of back-forward testing:

\begin{enumerate}
    \item Model input: Feed the spin models in each bin age with posterior distributions of $B_{\ast}$ or $\chi$, obtained from the ABC analysis and employ identical values for other parameters as used in the ABC procedure of Section \ref{sec:abc}.
    \item Model output: Generate model-predicted $v\sin(i)$ distributions for each age bin.
    \item Comparison with observations and quantifying similarity: Compare the model-generated $v\sin(i)$ distributions with the observed data and assess the similarity between distributions using the histogram intersection algorithm.
\end{enumerate}

\subsection{Spin models input parameters}\label{ap:c1}

Based on the equation \ref{eq:diffeq}, we developed a script that uses a set of input parameters [$M_{\ast}$, $B_{\ast}$, $\chi$, $P^{in}_{rot}$, $\dot{M}^{in}_{acc}$] to produce the $v\sin(i)$ value for the corresponding age, as output. The analysis, consistent with the ABC approach (Section \ref{sec:abc}), is divided into two cases:\\

\textbf{Case A: Magnetic field decreases with age}
\begin{itemize}
    \item Fixed parameters for each age bin: $M_{\ast}=0.5M_{\odot}$, $\chi=0.3$. 
    \item Free parameters randomly chosen from specified ranges: \subitem $P^{in}_{rot}:(1,2,3,4,5,6,7,8)$ days
    \subitem $\dot{M}^{in}_{acc}$: Values following the probability function in Figure \ref{fig_Mdot_all_sample}
    \subitem $B_{\ast}$: Values following the probability function in the upper panel of Figure \ref{fig_parameters_dist_test}.
\end{itemize}

\textbf{Case B: Branching ratio decreases with age}
\begin{itemize}
    \item Fixed parameters for each age bin: $M_{\ast}=0.5M_{\odot}$, $B_{\ast}=2000$ G. 
    \item Free parameters randomly chosen from specified ranges: \subitem $P^{in}_{rot}$ and $\dot{M}^{in}_{acc}$: Same as case A.
    \subitem $\chi$: Values following the probability function in the lower panel of Figure \ref{fig_parameters_dist_test}.
\end{itemize}

For comparison purposes, we also have considered hypothetical cases, for instance, magnetic field increases with age, magnetic field remains constant with age, branching ratio increases with age, branching ratio remains constant with age, and magnetic field with branching ratio simultaneously decreases with age.\\

\textbf{Magnetic field increases with age}
\begin{itemize}
    \item Same fixed parameters to the case A:
    \item Free parameters randomly chosen from specified ranges: \subitem $P^{in}_{rot}$ and $\dot{M}^{in}_{acc}$: Same as case A.
    \subitem $B_{\ast}$: Values follow a series of Gaussian distributions with a standard deviation of 200 G and mean values [1000, 1500, 2000, 2500] G from age bin 1 to age bin 4, respectively.
\end{itemize}

\textbf{Magnetic field remains constant with age}
\begin{itemize}
    \item Same fixed parameters to the case A.
    \subitem $B_{\ast}$: follows a Gaussian distribution with a standard deviation of 200 G and a mean value of 2500 G for all bin ages.
    \item Free parameters randomly chosen from specified ranges:
    \subitem $P^{in}_{rot}$ and $\dot{M}^{in}_{acc}$: Same as case A.
\end{itemize}

\textbf{Branching ratio increases with age}
\begin{itemize}
    \item Same fixed parameters to the case B:
    \item Free parameters randomly chosen from specified ranges: \subitem $P^{in}_{rot}$ and $\dot{M}^{in}_{acc}$: Same as case A.
    \subitem $\chi$: Values follow a series of Gaussian distributions with a standard deviation of 0.05 and mean values [0.10, 0.20, 0.30, 0.50] from age bin 1 to age bin 4, respectively.
\end{itemize}

\textbf{Branching ratio remains constant with age}
\begin{itemize}
    \item Same fixed parameters to the case B.
    \subitem $\chi$: follows a Gaussian distribution with a standard deviation of 0.05 and a mean value of 0.50 for all bin ages.
    \item Free parameters randomly chosen from specified ranges:
    \subitem $P^{in}_{rot}$ and $\dot{M}^{in}_{acc}$: Same as case A.
\end{itemize}

\textbf{Magnetic field and branching ratio simultaneously decreases with age}
\begin{itemize}
    \item Fixed parameters for each age bin: $M_{\ast}=0.5M_{\odot}$. 
    \item Free parameters randomly chosen from specified ranges: 
    \subitem $P^{in}_{rot}$ and $\dot{M}^{in}_{acc}$: Same as case A.
    \subitem $B_{\ast}$ and $\chi$: Values following the probability function in the upper panel and lower panel of Figure \ref{fig_parameters_dist_test}, respectively.
\end{itemize}

\subsection{Comparison with observations}\label{ap:c2}

For each case and each bin age, we obtain a distribution of $v\sin(i)$ generated by spin-models. As an example, we have illustrated the resulting distributions of $v\sin(i)$ generated by spin-models for the model case $\chi$ decreasing with the age at bin 3 in comparison to its observations (Left panel of Figure \ref{similarity})\footnote{For additional cases, we refer the reader to our GitHub repository \url{https://github.com/javiserna/Rotational-Evolution-of-Classical-T-Tauri-Stars-Models-and-Observations-Tests-}}. Employing the histogram intersection algorithm \citep{Swain_1991}, we estimate the intercepted area between the histograms, using this value as a metric for assessing the similarity between distributions. A metric of 1 indicates identical distributions, while a metric of 0 signifies no correlation. Likewise, in the right panel of Figure \ref{similarity}, we show the similarity index bin by bin for all considered cases (Appendix \ref{ap:c1}). The lack of similarity observed for slow rotators is primarily attributed to the contamination of stars with masses below the mass value utilized in our models or similarly below the typical mass for the whole sample. These slower rotators in the initial $v\sin(i)$ container correspond to stars in the age bin 1. Since the size of the star, which is directly related to $v\sin(i)$, depends on the stellar mass for a given age and the mass value used in the theoretical cases is systematically greater than the mass of the observed sample in bin 1, the models do not reproduce the slower rotators in the histograms. Although smaller values of stellar mass could fill the slower rotator container, changing the values of stellar masses would not be directly comparable to the cases addressed in Section \ref{sec:abc}.\\

\begin{figure}[htpb!]
    \centering
    \includegraphics[scale=0.45]{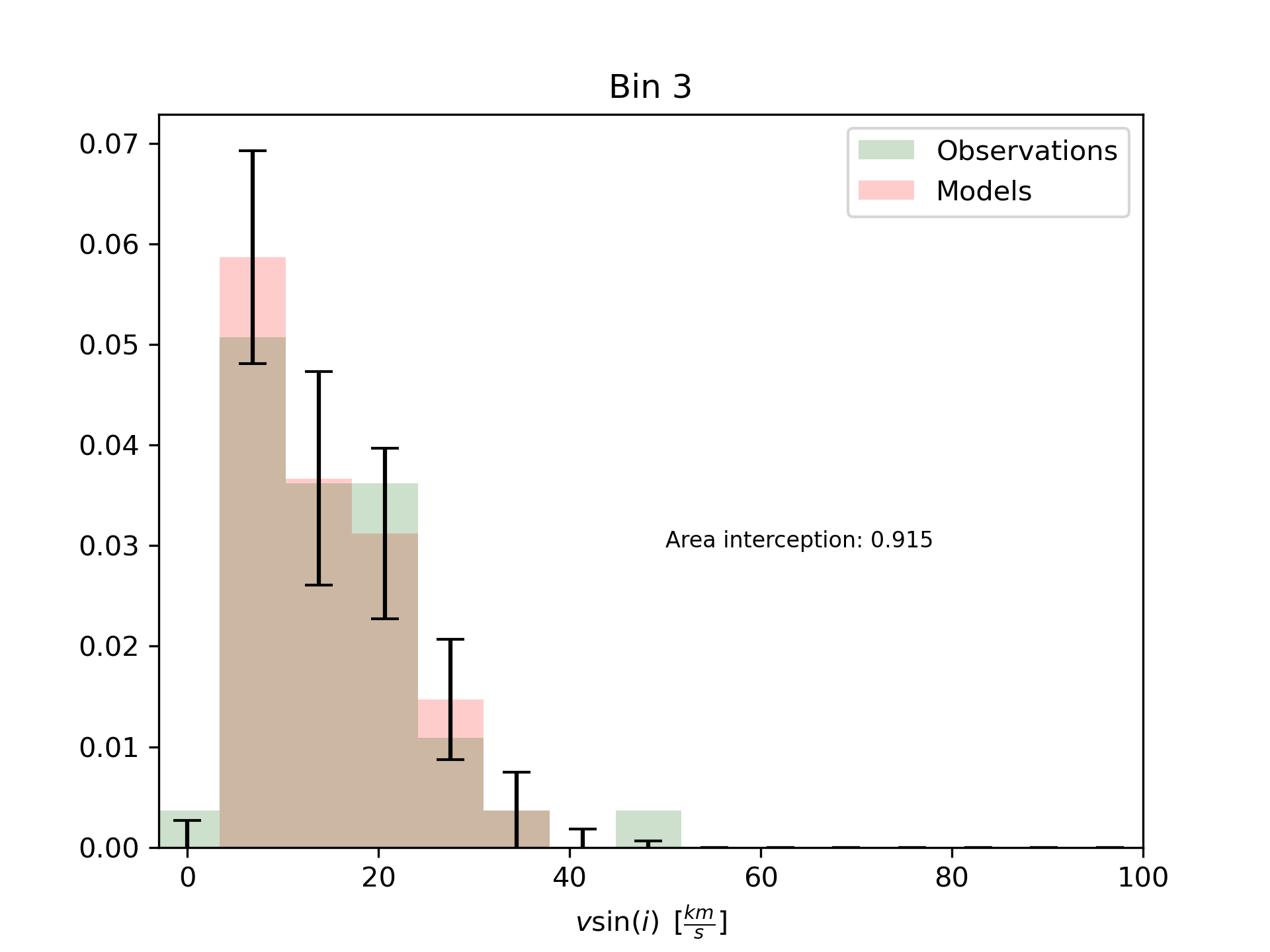}
    \includegraphics[scale=0.45]{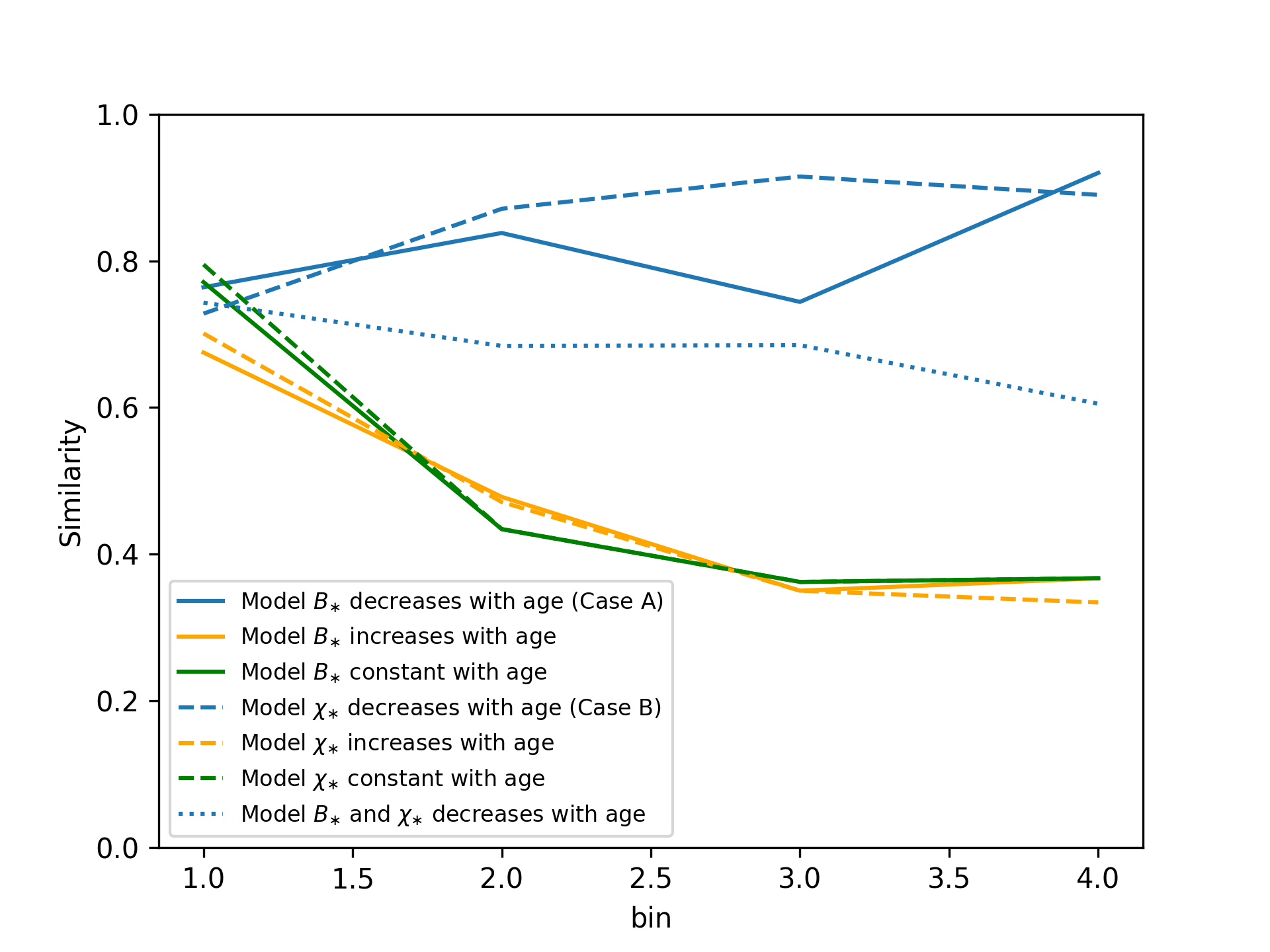}
    \caption{Similarity index between back-forward models and observations. The left panel is a comparison between observed and back-forward $v\sin(i)$ distributions. The back-forward distribution is the median of 100 runs, and the error bars represent the standard deviation. The observed $v\sin(i)$ distribution come from Figure \ref{fig_vsini_dist}. All histograms are properly normalized to the area of each distribution. The right panel shows the comparisons of all considered models.}
    \label{similarity}
\end{figure}

Finally, these tests support that the decrease of magnetic field and branching ratio with time shown in Figure \ref{fig_parameters_dist} effectively account for $v\sin(i)$ observations of Figure \ref{fig_vsini_dist}. Validating the ABC analyses. The right panel of Figure \ref{similarity}, indicates to us that distributions of $v\sin(i)$ observed and predicted by the back-forward test in case A and B, tend to be similar above 80\%. Also, demonstrates that a model with a magnetic field and branching ratio simultaneously decreasing with age compares better with observations than a model with these constant parameters.

\clearpage
\bibliography{main}

\begin{thebibliography}{}
\expandafter\ifx\csname natexlab\endcsname\relax\def\natexlab#1{#1}\fi
\providecommand{\url}[1]{\href{#1}{#1}}
\providecommand{\dodoi}[1]{doi:~\href{http://doi.org/#1}{\nolinkurl{#1}}}
\providecommand{\doeprint}[1]{\href{http://ascl.net/#1}{\nolinkurl{http://ascl.net/#1}}}
\providecommand{\doarXiv}[1]{\href{https://arxiv.org/abs/#1}{\nolinkurl{https://arxiv.org/abs/#1}}}

\bibitem[{{Alencar} \& {Basri}(2000)}]{Alencar2000}
{Alencar}, S. H.~P., \& {Basri}, G. 2000, \aj, 119, 1881,
  \dodoi{10.1086/301300}

\bibitem[{Ansdell {et~al.}(2019)Ansdell, Gaidos, Hedges, Tazzari, Kraus, Wyatt,
  Kennedy, Williams, Mann, Angelo, Dûchene, Mamajek, Carpenter, Esplin, \&
  Rizzuto}]{Ansdell2020}
Ansdell, M., Gaidos, E., Hedges, C., {et~al.} 2019, Monthly Notices of the
  Royal Astronomical Society, 492, 572, \dodoi{10.1093/mnras/stz3361}

\bibitem[{{Armitage} \& {Clarke}(1996)}]{Armitage1996}
{Armitage}, P.~J., \& {Clarke}, C.~J. 1996, \mnras, 280, 458,
  \dodoi{10.1093/mnras/280.2.458}

\bibitem[{{Artemenko} {et~al.}(2012){Artemenko}, {Grankin}, \&
  {Petrov}}]{Artemenko2012}
{Artemenko}, S.~A., {Grankin}, K.~N., \& {Petrov}, P.~P. 2012, Astronomy
  Letters, 38, 783, \dodoi{10.1134/S1063773712110011}

\bibitem[{{Astropy Collaboration} {et~al.}(2013){Astropy Collaboration},
  {Robitaille}, {Tollerud}, {Greenfield}, {Droettboom}, {Bray}, {Aldcroft},
  {Davis}, {Ginsburg}, {Price-Whelan}, {Kerzendorf}, {Conley}, {Crighton},
  {Barbary}, {Muna}, {Ferguson}, {Grollier}, {Parikh}, {Nair}, {Unther},
  {Deil}, {Woillez}, {Conseil}, {Kramer}, {Turner}, {Singer}, {Fox}, {Weaver},
  {Zabalza}, {Edwards}, {Azalee Bostroem}, {Burke}, {Casey}, {Crawford},
  {Dencheva}, {Ely}, {Jenness}, {Labrie}, {Lim}, {Pierfederici}, {Pontzen},
  {Ptak}, {Refsdal}, {Servillat}, \& {Streicher}}]{2013A&A...558A..33A}
{Astropy Collaboration}, {Robitaille}, T.~P., {Tollerud}, E.~J., {et~al.} 2013,
  \aap, 558, A33, \dodoi{10.1051/0004-6361/201322068}

\bibitem[{{Banzatti} {et~al.}(2019){Banzatti}, {Pascucci}, {Edwards}, {Fang},
  {Gorti}, \& {Flock}}]{Banzatti2019}
{Banzatti}, A., {Pascucci}, I., {Edwards}, S., {et~al.} 2019, \apj, 870, 76,
  \dodoi{10.3847/1538-4357/aaf1aa}

\bibitem[{Baraffe {et~al.}(2015)Baraffe, Homeier, Allard, \&
  Chabrier}]{Baraffe_2015}
Baraffe, I., Homeier, D., Allard, F., \& Chabrier, G. 2015, A\&A, 577, A42,
  \dodoi{10.1051/0004-6361/201425481}

\bibitem[{{Bayo} {et~al.}(2012){Bayo}, {Barrado}, {Hu{\'e}lamo},
  {Morales-Calder{\'o}n}, {Melo}, {Stauffer}, \& {Stelzer}}]{Bayo_2012}
{Bayo}, A., {Barrado}, D., {Hu{\'e}lamo}, N., {et~al.} 2012, \aap, 547, A80,
  \dodoi{10.1051/0004-6361/201219374}

\bibitem[{{Bouvier} {et~al.}(1993){Bouvier}, {Cabrit}, {Fernandez}, {Martin},
  \& {Matthews}}]{Bouvier1993}
{Bouvier}, J., {Cabrit}, S., {Fernandez}, M., {Martin}, E.~L., \& {Matthews},
  J.~M. 1993, \aap, 272, 176

\bibitem[{{Brasseur} {et~al.}(2019){Brasseur}, {Phillip}, {Fleming},
  {Mullally}, \& {White}}]{TESSCut}
{Brasseur}, C.~E., {Phillip}, C., {Fleming}, S.~W., {Mullally}, S.~E., \&
  {White}, R.~L. 2019, {Astrocut: Tools for creating cutouts of TESS images}.
\newblock \doeprint{1905.007}

\bibitem[{Brice{\~{n}}o {et~al.}(2019)Brice{\~{n}}o, Calvet, Hern{\'{a}}ndez,
  Vivas, Mateu, Downes, Loerincs, P{\'{e}}rez-Blanco, Berlind, Espaillat,
  Allen, Hartmann, Mateo, \& III}]{Briceno_2019}
Brice{\~{n}}o, C., Calvet, N., Hern{\'{a}}ndez, J., {et~al.} 2019, The
  Astronomical Journal, 157, 85, \dodoi{10.3847/1538-3881/aaf79b}

\bibitem[{{Cabrit} {et~al.}(1990){Cabrit}, {Edwards}, {Strom}, \&
  {Strom}}]{Cabrit1990}
{Cabrit}, S., {Edwards}, S., {Strom}, S.~E., \& {Strom}, K.~M. 1990, \apj, 354,
  687, \dodoi{10.1086/168725}

\bibitem[{{Cao} \& {Pinsonneault}(2022)}]{Cao_2022}
{Cao}, L., \& {Pinsonneault}, M.~H. 2022, \mnras, 517, 2165,
  \dodoi{10.1093/mnras/stac270610.48550/arXiv.2209.10549}

\bibitem[{Carpenter {et~al.}(2006)Carpenter, Mamajek, Hillenbrand, \&
  Meyer}]{Carpenter_2006}
Carpenter, J.~M., Mamajek, E.~E., Hillenbrand, L.~A., \& Meyer, M.~R. 2006, The
  Astrophysical Journal, 651, L49, \dodoi{10.1086/509121}

\bibitem[{{Chandrasekhar} \& {M{\"u}nch}(1950)}]{Chandrasekhar_1950}
{Chandrasekhar}, S., \& {M{\"u}nch}, G. 1950, \apj, 111, 142,
  \dodoi{10.1086/145245}

\bibitem[{Cody \& Hillenbrand(2010)}]{Cody2010}
Cody, A.~M., \& Hillenbrand, L.~A. 2010, Astrophys. J. Suppl. Ser., 191, 389,
  \dodoi{10.1088/0067-0049/191/2/389}

\bibitem[{{Cody} {et~al.}(2014){Cody}, {Stauffer}, {Baglin}, {Micela},
  {Rebull}, {Flaccomio}, {Morales-Calder{\'o}n}, {Aigrain}, {Bouvier},
  {Hillenbrand}, {Gutermuth}, {Song}, {Turner}, {Alencar}, {Zwintz},
  {Plavchan}, {Carpenter}, {Findeisen}, {Carey}, {Terebey}, {Hartmann},
  {Calvet}, {Teixeira}, {Vrba}, {Wolk}, {Covey}, {Poppenhaeger}, {G{\"u}nther},
  {Forbrich}, {Whitney}, {Affer}, {Herbst}, {Hora}, {Barrado}, {Holtzman},
  {Marchis}, {Wood}, {Medeiros Guimar{\~a}es}, {Lillo Box}, {Gillen},
  {McQuillan}, {Espaillat}, {Allen}, {D'Alessio}, \& {Favata}}]{Cody2014}
{Cody}, A.~M., {Stauffer}, J., {Baglin}, A., {et~al.} 2014, \aj, 147, 82,
  \dodoi{10.1088/0004-6256/147/4/82}

\bibitem[{{Collier Cameron} \&
  {Campbell}(1993{\natexlab{a}})}]{CollierCameron_1993}
{Collier Cameron}, A., \& {Campbell}, C.~G. 1993{\natexlab{a}}, \aap, 274, 309

\bibitem[{{Collier Cameron} \&
  {Campbell}(1993{\natexlab{b}})}]{CollierCameron1993}
---. 1993{\natexlab{b}}, \aap, 274, 309

\bibitem[{Cranmer(2008)}]{Cranmer_2008}
Cranmer, S.~R. 2008, The Astrophysical Journal, 689, 316,
  \dodoi{10.1086/592566}

\bibitem[{{Czesla} {et~al.}(2019){Czesla}, {Schr{\"o}ter}, {Schneider},
  {Huber}, {Pfeifer}, {Andreasen}, \& {Zechmeister}}]{pyastronomy}
{Czesla}, S., {Schr{\"o}ter}, S., {Schneider}, C.~P., {et~al.} 2019, {PyA:
  Python astronomy-related packages}.
\newblock \doeprint{1906.010}

\bibitem[{{Donati} \& {Landstreet}(2009)}]{Donati_2019}
{Donati}, J.~F., \& {Landstreet}, J.~D. 2009, \araa, 47, 333,
  \dodoi{10.1146/annurev-astro-082708-101833}

\bibitem[{{Dotter}(2016)}]{Dotter2016}
{Dotter}, A. 2016, \apjs, 222, 8, \dodoi{10.3847/0067-0049/222/1/8}

\bibitem[{{Edwards} {et~al.}(2003){Edwards}, {Fischer}, {Kwan}, {Hillenbrand},
  \& {Dupree}}]{Edwards2003}
{Edwards}, S., {Fischer}, W., {Kwan}, J., {Hillenbrand}, L., \& {Dupree}, A.~K.
  2003, \apjl, 599, L41, \dodoi{10.1086/381077}

\bibitem[{Emeriau-Viard \& Brun(2017)}]{Viard_2017}
Emeriau-Viard, C., \& Brun, A.~S. 2017, The Astrophysical Journal, 846, 8,
  \dodoi{10.3847/1538-4357/aa7b33}

\bibitem[{{Ercolano} {et~al.}(2021){Ercolano}, {Picogna}, {Monsch}, {Drake}, \&
  {Preibisch}}]{Ercolano_2021}
{Ercolano}, B., {Picogna}, G., {Monsch}, K., {Drake}, J.~J., \& {Preibisch}, T.
  2021, \mnras, 508, 1675,
  \dodoi{10.1093/mnras/stab259010.48550/arXiv.2109.04113}

\bibitem[{{Espaillat} {et~al.}(2021){Espaillat}, {Robinson}, {Romanova},
  {Thanathibodee}, {Wendeborn}, {Calvet}, {Reynolds}, \&
  {Muzerolle}}]{Espaillat_2021}
{Espaillat}, C.~C., {Robinson}, C.~E., {Romanova}, M.~M., {et~al.} 2021, \nat,
  597, 41, \dodoi{10.1038/s41586-021-03751-5}

\bibitem[{Finley \& Matt(2017)}]{Finley_2017}
Finley, A.~J., \& Matt, S.~P. 2017, The Astrophysical Journal, 845, 46,
  \dodoi{10.3847/1538-4357/aa7fb9}

\bibitem[{Folsom {et~al.}(2016)Folsom, Petit, Bouvier, Lèbre, Amard, Palacios,
  Morin, Donati, Jeffers, Marsden, \& Vidotto}]{Folsom_2016}
Folsom, C.~P., Petit, P., Bouvier, J., {et~al.} 2016, Monthly Notices of the
  Royal Astronomical Society, 457, 580, \dodoi{10.1093/mnras/stv2924}

\bibitem[{{Gaia Collaboration} {et~al.}(2022){Gaia Collaboration}, {Vallenari},
  {Brown}, {Prusti}, {de Bruijne}, {Arenou}, {Babusiaux}, {Biermann},
  {Creevey}, {Ducourant}, {Evans}, {Eyer}, {Guerra}, {Hutton}, {Jordi},
  {Klioner}, {Lammers}, {Lindegren}, {Luri}, {Mignard}, {Panem}, {Pourbaix},
  {Randich}, {Sartoretti}, {Soubiran}, {Tanga}, {Walton}, {Bailer-Jones},
  {Bastian}, {Drimmel}, {Jansen}, {Katz}, {Lattanzi}, {van Leeuwen}, {Bakker},
  {Cacciari}, {Casta{\~n}eda}, {De Angeli}, {Fabricius}, {Fouesneau},
  {Fr{\'e}mat}, {Galluccio}, {Guerrier}, {Heiter}, {Masana}, {Messineo},
  {Mowlavi}, {Nicolas}, {Nienartowicz}, {Pailler}, {Panuzzo}, {Riclet}, {Roux},
  {Seabroke}, {Sordo{\o}rcit}, {Th{\'e}venin}, {Gracia-Abril}, {Portell},
  {Teyssier}, {Altmann}, {Andrae}, {Audard}, {Bellas-Velidis}, {Benson},
  {Berthier}, {Blomme}, {Burgess}, {Busonero}, {Busso}, {C{\'a}novas}, {Carry},
  {Cellino}, {Cheek}, {Clementini}, {Damerdji}, {Davidson}, {de Teodoro},
  {Nu{\~n}ez Campos}, {Delchambre}, {Dell'Oro}, {Esquej},
  {Fern{\'a}ndez-Hern{\'a}ndez}, {Fraile}, {Garabato}, {Garc{\'\i}a-Lario},
  {Gosset}, {Haigron}, {Halbwachs}, {Hambly}, {Harrison}, {Hern{\'a}ndez},
  {Hestroffer}, {Hodgkin}, {Holl}, {Jan{\ss}en}, {Jevardat de Fombelle},
  {Jordan}, {Krone-Martins}, {Lanzafame}, {L{\"o}ffler}, {Marchal}, {Marrese},
  {Moitinho}, {Muinonen}, {Osborne}, {Pancino}, {Pauwels}, {Recio-Blanco},
  {Reyl{\'e}}, {Riello}, {Rimoldini}, {Roegiers}, {Rybizki}, {Sarro}, {Siopis},
  {Smith}, {Sozzetti}, {Utrilla}, {van Leeuwen}, {Abbas}, {{\'A}brah{\'a}m},
  {Abreu Aramburu}, {Aerts}, {Aguado}, {Ajaj}, {Aldea-Montero}, {Altavilla},
  {{\'A}lvarez}, {Alves}, {Anders}, {Anderson}, {Anglada Varela}, {Antoja},
  {Baines}, {Baker}, {Balaguer-N{\'u}{\~n}ez}, {Balbinot}, {Balog}, {Barache},
  {Barbato}, {Barros}, {Barstow}, {Bartolom{\'e}}, {Bassilana}, {Bauchet},
  {Becciani}, {Bellazzini}, {Berihuete}, {Bernet}, {Bertone}, {Bianchi},
  {Binnenfeld}, {Blanco-Cuaresma}, {Blazere}, {Boch}, {Bombrun}, {Bossini},
  {Bouquillon}, {Bragaglia}, {Bramante}, {Breedt}, {Bressan}, {Brouillet},
  {Brugaletta}, {Bucciarelli}, {Burlacu}, {Butkevich}, {Buzzi}, {Caffau},
  {Cancelliere}, {Cantat-Gaudin}, {Carballo}, {Carlucci}, {Carnerero},
  {Carrasco}, {Casamiquela}, {Castellani}, {Castro-Ginard}, {Chaoul},
  {Charlot}, {Chemin}, {Chiaramida}, {Chiavassa}, {Chornay}, {Comoretto},
  {Contursi}, {Cooper}, {Cornez}, {Cowell}, {Crifo}, {Cropper}, {Crosta},
  {Crowley}, {Dafonte}, {Dapergolas}, {David}, {David}, {de Laverny}, {De
  Luise}, {De March}, {De Ridder}, {de Souza}, {de Torres}, {del Peloso}, {del
  Pozo}, {Delbo}, {Delgado}, {Delisle}, {Demouchy}, {Dharmawardena}, {Di
  Matteo}, {Diakite}, {Diener}, {Distefano}, {Dolding}, {Edvardsson}, {Enke},
  {Fabre}, {Fabrizio}, {Faigler}, {Fedorets}, {Fernique}, {Fienga}, {Figueras},
  {Fournier}, {Fouron}, {Fragkoudi}, {Gai}, {Garcia-Gutierrez},
  {Garcia-Reinaldos}, {Garc{\'\i}a-Torres}, {Garofalo}, {Gavel}, {Gavras},
  {Gerlach}, {Geyer}, {Giacobbe}, {Gilmore}, {Girona}, {Giuffrida}, {Gomel},
  {Gomez}, {Gonz{\'a}lez-N{\'u}{\~n}ez}, {Gonz{\'a}lez-Santamar{\'\i}a},
  {Gonz{\'a}lez-Vidal}, {Granvik}, {Guillout}, {Guiraud},
  {Guti{\'e}rrez-S{\'a}nchez}, {Guy}, {Hatzidimitriou}, {Hauser}, {Haywood},
  {Helmer}, {Helmi}, {Sarmiento}, {Hidalgo}, {Hilger}, {H{\l}adczuk}, {Hobbs},
  {Holland}, {Huckle}, {Jardine}, {Jasniewicz}, {Jean-Antoine Piccolo},
  {Jim{\'e}nez-Arranz}, {Jorissen}, {Juaristi Campillo}, {Julbe}, {Karbevska},
  {Kervella}, {Khanna}, {Kontizas}, {Kordopatis}, {Korn}, {K{\'o}sp{\'a}l},
  {Kostrzewa-Rutkowska}, {Kruszy{\'n}ska}, {Kun}, {Laizeau}, {Lambert},
  {Lanza}, {Lasne}, {Le Campion}, {Lebreton}, {Lebzelter}, {Leccia}, {Leclerc},
  {Lecoeur-Taibi}, {Liao}, {Licata}, {Lindstr{\o}m}, {Lister}, {Livanou},
  {Lobel}, {Lorca}, {Loup}, {Madrero Pardo}, {Magdaleno Romeo}, {Managau},
  {Mann}, {Manteiga}, {Marchant}, {Marconi}, {Marcos}, {Marcos Santos},
  {Mar{\'\i}n Pina}, {Marinoni}, {Marocco}, {Marshall}, {Polo},
  {Mart{\'\i}n-Fleitas}, {Marton}, {Mary}, {Masip}, {Massari},
  {Mastrobuono-Battisti}, {Mazeh}, {McMillan}, {Messina}, {Michalik}, {Millar},
  {Mints}, {Molina}, {Molinaro}, {Moln{\'a}r}, {Monari}, {Mongui{\'o}},
  {Montegriffo}, {Montero}, {Mor}, {Mora}, {Morbidelli}, {Morel}, {Morris},
  {Muraveva}, {Murphy}, {Musella}, {Nagy}, {Noval}, {Oca{\~n}a}, {Ogden},
  {Ordenovic}, {Osinde}, {Pagani}, {Pagano}, {Palaversa}, {Palicio},
  {Pallas-Quintela}, {Panahi}, {Payne-Wardenaar}, {Pe{\~n}alosa Esteller},
  {Penttil{\"a}}, {Pichon}, {Piersimoni}, {Pineau}, {Plachy}, {Plum}, {Poggio},
  {Pr{\v{s}}a}, {Pulone}, {Racero}, {Ragaini}, {Rainer}, {Raiteri}, {Rambaux},
  {Ramos}, {Ramos-Lerate}, {Re Fiorentin}, {Regibo}, {Richards}, {Rios Diaz},
  {Ripepi}, {Riva}, {Rix}, {Rixon}, {Robichon}, {Robin}, {Robin}, {Roelens},
  {Rogues}, {Rohrbasser}, {Romero-G{\'o}mez}, {Rowell}, {Royer}, {Ruz Mieres},
  {Rybicki}, {Sadowski}, {S{\'a}ez N{\'u}{\~n}ez}, {Sagrist{\`a} Sell{\'e}s},
  {Sahlmann}, {Salguero}, {Samaras}, {Sanchez Gimenez}, {Sanna},
  {Santove{\~n}a}, {Sarasso}, {Schultheis}, {Sciacca}, {Segol}, {Segovia},
  {S{\'e}gransan}, {Semeux}, {Shahaf}, {Siddiqui}, {Siebert}, {Siltala},
  {Silvelo}, {Slezak}, {Slezak}, {Smart}, {Snaith}, {Solano}, {Solitro},
  {Souami}, {Souchay}, {Spagna}, {Spina}, {Spoto}, {Steele},
  {Steidelm{\"u}ller}, {Stephenson}, {S{\"u}veges}, {Surdej}, {Szabados},
  {Szegedi-Elek}, {Taris}, {Taylo}, {Teixeira}, {Tolomei}, {Tonello}, {Torra},
  {Torra}, {Torralba Elipe}, {Trabucchi}, {Tsounis}, {Turon}, {Ulla}, {Unger},
  {Vaillant}, {van Dillen}, {van Reeven}, {Vanel}, {Vecchiato}, {Viala},
  {Vicente}, {Voutsinas}, {Weiler}, {Wevers}, {Wyrzykowski}, {Yoldas}, {Yvard},
  {Zhao}, {Zorec}, {Zucker}, \& {Zwitter}}]{GAIADR3}
{Gaia Collaboration}, {Vallenari}, A., {Brown}, A.~G.~A., {et~al.} 2022, arXiv
  e-prints, arXiv:2208.00211.
\newblock \doarXiv{2208.00211}

\bibitem[{Gallet \& Bouvier(2013)}]{Gallet2013}
Gallet, F., \& Bouvier, J. 2013, Astron. Astrophys., 556, A36,
  \dodoi{10.1051/0004-6361/201321302}

\bibitem[{Gallet \& Bouvier(2015)}]{Gallet_2015}
---. 2015, A\&A, 577, A98, \dodoi{10.1051/0004-6361/201525660}

\bibitem[{{Gallet} {et~al.}(2019){Gallet}, {Zanni}, \& {Amard}}]{Gallet2019}
{Gallet}, F., {Zanni}, C., \& {Amard}, L. 2019, \aap, 632, A6,
  \dodoi{10.1051/0004-6361/201935432}

\bibitem[{Garraffo {et~al.}(2015)Garraffo, Drake, \& Cohen}]{Garraffo_2015}
Garraffo, C., Drake, J.~J., \& Cohen, O. 2015, The Astrophysical Journal, 807,
  L6, \dodoi{10.1088/2041-8205/807/1/l6}

\bibitem[{{Gehrig} {et~al.}(2022){Gehrig}, {Steiner}, {Vorobyov}, \&
  {G{\"u}del}}]{Gehrig_2022}
{Gehrig}, L., {Steiner}, D., {Vorobyov}, E., \& {G{\"u}del}, M. 2022, arXiv
  e-prints, arXiv:2208.08852.
\newblock \doarXiv{2208.08852}

\bibitem[{{Ghosh} \& {Lamb}(1978)}]{Ghosh_1978}
{Ghosh}, P., \& {Lamb}, F.~K. 1978, \apjl, 223, L83, \dodoi{10.1086/182734}

\bibitem[{{Ghosh} \& {Lamb}(1979)}]{Ghosh_1979}
---. 1979, \apj, 234, 296, \dodoi{10.1086/157498}

\bibitem[{{Gonneau} {et~al.}(2020){Gonneau}, {Lyubenova}, {Lan{\c{c}}on},
  {Trager}, {Peletier}, {Arentsen}, {Chen}, {Coelho}, {Dries},
  {Falc{\'o}n-Barroso}, {Prugniel}, {S{\'a}nchez-Bl{\'a}zquez}, {Vazdekis}, \&
  {Verro}}]{Gonneau_2020}
{Gonneau}, A., {Lyubenova}, M., {Lan{\c{c}}on}, A., {et~al.} 2020, \aap, 634,
  A133, \dodoi{10.1051/0004-6361/201936825}

\bibitem[{{Gullbring} {et~al.}(1998){Gullbring}, {Hartmann}, {Brice{\~n}o}, \&
  {Calvet}}]{Gullbring1998}
{Gullbring}, E., {Hartmann}, L., {Brice{\~n}o}, C., \& {Calvet}, N. 1998, \apj,
  492, 323, \dodoi{10.1086/305032}

\bibitem[{Harris {et~al.}(2020)Harris, Millman, van~der Walt, Gommers,
  Virtanen, Cournapeau, Wieser, Taylor, Berg, Smith, Kern, Picus, Hoyer, van
  Kerkwijk, Brett, Haldane, del R{'{\i}}o, Wiebe, Peterson,
  G{'{e}}rard-Marchant, Sheppard, Reddy, Weckesser, Abbasi, Gohlke, \&
  Oliphant}]{Harris2020}
Harris, C.~R., Millman, K.~J., van~der Walt, S.~J., {et~al.} 2020, Nature, 585,
  357, \dodoi{10.1038/s41586-020-2649-2}

\bibitem[{{Hartigan} {et~al.}(1995){Hartigan}, {Edwards}, \&
  {Ghandour}}]{Hartigan1995}
{Hartigan}, P., {Edwards}, S., \& {Ghandour}, L. 1995, \apj, 452, 736,
  \dodoi{10.1086/176344}

\bibitem[{{Hartmann} {et~al.}(1982){Hartmann}, {Avrett}, \&
  {Edwards}}]{Hartmann_1982}
{Hartmann}, L., {Avrett}, E., \& {Edwards}, S. 1982, \apj, 261, 279,
  \dodoi{10.1086/160339}

\bibitem[{Hartmann {et~al.}(2016)Hartmann, Herczeg, \& Calvet}]{Hartmann_2016}
Hartmann, L., Herczeg, G., \& Calvet, N. 2016, Annual Review of Astronomy and
  Astrophysics, 54, 135, \dodoi{10.1146/annurev-astro-081915-023347}

\bibitem[{{Hartmann} \& {Stauffer}(1989)}]{Hartmann1989}
{Hartmann}, L., \& {Stauffer}, J.~R. 1989, \aj, 97, 873, \dodoi{10.1086/115033}

\bibitem[{{Herbst} {et~al.}(2002){Herbst}, {Bailer-Jones, C. A. L.}, {Mundt,
  R.}, {Meisenheimer, K.}, \& {Wackermann, R.}}]{Herbst_2002}
{Herbst}, {Bailer-Jones, C. A. L.}, {Mundt, R.}, {Meisenheimer, K.}, \&
  {Wackermann, R.} 2002, A\&A, 396, 513, \dodoi{10.1051/0004-6361:20021362}

\bibitem[{{Herczeg} \& {Hillenbrand}(2008)}]{Herzceg_2008}
{Herczeg}, G.~J., \& {Hillenbrand}, L.~A. 2008, \apj, 681, 594,
  \dodoi{10.1086/586728}

\bibitem[{{Hern{\'a}ndez} {et~al.}(2008){Hern{\'a}ndez}, {Hartmann}, {Calvet},
  {Jeffries}, {Gutermuth}, {Muzerolle}, \& {Stauffer}}]{Hernandez_2008}
{Hern{\'a}ndez}, J., {Hartmann}, L., {Calvet}, N., {et~al.} 2008, \apj, 686,
  1195, \dodoi{10.1086/591224}

\bibitem[{Hern{\'{a}}ndez {et~al.}(2014)Hern{\'{a}}ndez, Calvet, Perez,
  Brice{\~{n}}o, Olguin, Contreras, Hartmann, Allen, Espaillat, \&
  Hernan}]{Hernandez_2014}
Hern{\'{a}}ndez, J., Calvet, N., Perez, A., {et~al.} 2014, The Astrophysical
  Journal, 794, 36, \dodoi{10.1088/0004-637x/794/1/36}

\bibitem[{Hunter(2007)}]{Hunter2007}
Hunter, J.~D. 2007, Computing in Science \& Engineering, 9, 90,
  \dodoi{10.1109/MCSE.2007.55}

\bibitem[{{Husser} {et~al.}(2013){Husser}, {Wende-von Berg}, {Dreizler},
  {Homeier}, {Reiners}, {Barman}, \& {Hauschildt}}]{Husser2013}
{Husser}, T.~O., {Wende-von Berg}, S., {Dreizler}, S., {et~al.} 2013, \aap,
  553, A6, \dodoi{10.1051/0004-6361/201219058}

\bibitem[{{Iben}(1965)}]{Iben_1965}
{Iben}, Icko, J. 1965, \apj, 141, 993, \dodoi{10.1086/148193}

\bibitem[{Ingleby {et~al.}(2013)Ingleby, Calvet, Herczeg, Blaty, Walter,
  Ardila, Alexander, Edwards, Espaillat, Gregory, Hillenbrand, \&
  Brown}]{Ingleby_2013}
Ingleby, L., Calvet, N., Herczeg, G., {et~al.} 2013, The Astrophysical Journal,
  767, 112, \dodoi{10.1088/0004-637x/767/2/112}

\bibitem[{Ireland {et~al.}(2020)Ireland, Zanni, Matt, \&
  Pantolmos}]{Ireland_2020}
Ireland, L.~G., Zanni, C., Matt, S.~P., \& Pantolmos, G. 2020, The
  Astrophysical Journal, 906, 4, \dodoi{10.3847/1538-4357/abc828}

\bibitem[{Jayawardhana {et~al.}(2006)Jayawardhana, Coffey, Scholz, Brandeker,
  \& van Kerkwijk}]{Jayawardhana_2006}
Jayawardhana, R., Coffey, J., Scholz, A., Brandeker, A., \& van Kerkwijk, M.~H.
  2006, The Astrophysical Journal, 648, 1206, \dodoi{10.1086/506171}

\bibitem[{Johns-Krull(2007)}]{Johns_Krull_2007}
Johns-Krull, C.~M. 2007, The Astrophysical Journal, 664, 975,
  \dodoi{10.1086/519017}

\bibitem[{{Kausch} {et~al.}(2015){Kausch}, {Noll}, {Smette}, {Kimeswenger, S.},
  {Barden, M.}, {Szyszka, C.}, {Jones, A. M.}, {Sana, H.}, {Horst, H.}, \&
  {Kerber, F.}}]{Kausch_2015}
{Kausch}, W., {Noll}, S., {Smette}, A., {et~al.} 2015, A\&A, 576, A78,
  \dodoi{10.1051/0004-6361/201423909}

\bibitem[{{Koenigl}(1991)}]{Koenigl1991}
{Koenigl}, A. 1991, \apjl, 370, L39, \dodoi{10.1086/185972}

\bibitem[{Kounkel {et~al.}(2018)Kounkel, Covey, Su{\'{a}}rez,
  Rom{\'{a}}n-Z{\'{u}}{\~{n}}iga, Hernandez, Stassun, Jaehnig, Feigelson,
  Ram{\'{\i}}rez, Roman-Lopes, Rio, Stringfellow, Kim, Borissova,
  Fern{\'{a}}ndez-Trincado, Burgasser, Garc{\'{\i}}a-Hern{\'{a}}ndez, Zamora,
  Pan, \& Nitschelm}]{Kounkel_2018}
Kounkel, M., Covey, K., Su{\'{a}}rez, G., {et~al.} 2018, The Astronomical
  Journal, 156, 84, \dodoi{10.3847/1538-3881/aad1f1}

\bibitem[{{Kounkel} {et~al.}(2019){Kounkel}, {Covey}, {Moe}, {Kratter},
  {Su{\'a}rez}, {Stassun}, {Rom{\'a}n-Z{\'u}{\~n}iga}, {Hernand ez}, {Kim},
  {Pe{\~n}a Ram{\'\i}rez}, {Roman-Lopes}, {Stringfellow}, {Jaehnig},
  {Borissova}, {Tofflemire}, {Krolikowski}, {Rizzuto}, {Kraus}, {Badenes},
  {Longa-Pe{\~n}a}, {G{\'o}mez Maqueo Chew}, {Barba}, {Nidever}, {Brown}, {De
  Lee}, {Pan}, {Bizyaev}, {Oravetz}, \& {Oravetz}}]{Kounkel_2019}
{Kounkel}, M., {Covey}, K., {Moe}, M., {et~al.} 2019, \aj, 157, 196,
  \dodoi{10.3847/1538-3881/ab13b1}

\bibitem[{Kounkel {et~al.}(2021)Kounkel, Covey, Stassun, Price-Whelan,
  Holtzman, Chojnowski, Longa-Pe{\~{n}}a, Rom{\'{a}}n-Z{\'{u}}{\~{n}}iga,
  Hernandez, Serna, Badenes, Lee, Majewski, Stringfellow, Kratter, Moe,
  Frinchaboy, Beaton, Fern{\'{a}}ndez-Trincado, Mahadevan, Minniti, Beers,
  Schneider, Barba, Brownstein, Garc{\'{\i}}a-Hern{\'{a}}ndez, Pan, \&
  Bizyaev}]{Kounkel_2021}
Kounkel, M., Covey, K.~R., Stassun, K.~G., {et~al.} 2021, The Astronomical
  Journal, 162, 184, \dodoi{10.3847/1538-3881/ac1798}

\bibitem[{{Kounkel} {et~al.}(2023){Kounkel}, {Zari}, {Covey}, {Tkachenko},
  {Rom{\'a}n Z{\'u}{\~n}iga}, {Stassun}, {Stutz}, {Stringfellow},
  {Roman-Lopes}, {Hern{\'a}ndez}, {Pe{\~n}a Ram{\'\i}rez}, {Bayo}, {Kim},
  {Cao}, {Wolk}, {Kollmeier}, {L{\'o}pez-Valdivia}, \&
  {Rojas-Ayala}}]{Kounkel_2023}
{Kounkel}, M., {Zari}, E., {Covey}, K., {et~al.} 2023, arXiv e-prints,
  arXiv:2301.07186, \dodoi{10.48550/arXiv.2301.07186}

\bibitem[{{Kuhi}(1964)}]{Kuhi1964}
{Kuhi}, L.~V. 1964, \apj, 140, 1409, \dodoi{10.1086/148047}

\bibitem[{{Kurosawa} \& {Romanova}(2013)}]{Kurosawa2013}
{Kurosawa}, R., \& {Romanova}, M.~M. 2013, \mnras, 431, 2673,
  \dodoi{10.1093/mnras/stt365}

\bibitem[{{Lavail} {et~al.}(2019){Lavail}, {Kochukhov}, \&
  {Hussain}}]{Lavail_2019}
{Lavail}, A., {Kochukhov}, O., \& {Hussain}, G.~A.~J. 2019, \aap, 630, A99,
  \dodoi{10.1051/0004-6361/201935695}

\bibitem[{{Lavail} {et~al.}(2017){Lavail}, {Kochukhov}, {Hussain}, {Alecian},
  {Herczeg}, \& {Johns-Krull}}]{Lavail_2017}
{Lavail}, A., {Kochukhov}, O., {Hussain}, G.~A.~J., {et~al.} 2017, \aap, 608,
  A77, \dodoi{10.1051/0004-6361/201731889}

\bibitem[{{Lindegren} {et~al.}(2020){Lindegren}, {Klioner}, {Hern{\'a}ndez},
  {Bombrun}, {Ramos-Lerate}, {Steidelm{\"u}ller}, {Bastian}, {Biermann}, {de
  Torres}, {Gerlach}, {Geyer}, {Hilger}, {Hobbs}, {Lammers}, {McMillan},
  {Stephenson}, {Casta{\~n}eda}, {Davidson}, {Fabricius}, {Gracia-Abril},
  {Portell}, {Rowell}, {Teyssier}, {Torra}, {Bartolom{\'e}}, {Clotet},
  {Garralda}, {Gonz{\'a}lez-Vidal}, {Torra}, {Abbas}, {Altmann}, {Anglada
  Varela}, {Balaguer-N{\'u}{\~n}ez}, {Balog}, {Barache}, {Becciani}, {Bernet},
  {Bertone}, {Bianchi}, {Bouquillon}, {Brown}, {Bucciarelli}, {Busonero},
  {Butkevich}, {Buzzi}, {Cancelliere}, {Carlucci}, {Charlot}, {Cioni},
  {Crosta}, {Crowley}, {del Peloso}, {del Pozo}, {Drimmel}, {Esquej}, {Fienga},
  {Fraile}, {Gai}, {Garcia-Reinaldos}, {Guerra}, {Hambly}, {Hauser},
  {Jan{\ss}en}, {Jordan}, {Kostrzewa-Rutkowska}, {Lattanzi}, {Liao}, {Licata},
  {Lister}, {L{\"o}ffler}, {Marchant}, {Masip}, {Mignard}, {Mints}, {Molina},
  {Mora}, {Morbidelli}, {Murphy}, {Pagani}, {Panuzzo}, {Pe{\~n}alosa Esteller},
  {Poggio}, {Re Fiorentin}, {Riva}, {Sagrist{\`a} Sell{\'e}s}, {Sanchez
  Gimenez}, {Sarasso}, {Sciacca}, {Siddiqui}, {Smart}, {Souami}, {Spagna},
  {Steele}, {Taris}, {Utrilla}, {van Reeven}, \& {Vecchiato}}]{Lindegreen2020a}
{Lindegren}, L., {Klioner}, S.~A., {Hern{\'a}ndez}, J., {et~al.} 2020, arXiv
  e-prints, arXiv:2012.03380.
\newblock \doarXiv{2012.03380}

\bibitem[{{Lindegren} {et~al.}(2021){Lindegren}, {Bastian}, {Biermann},
  {Bombrun}, {de Torres}, {Gerlach}, {Geyer}, {Hern{\'a}ndez}, {Hilger},
  {Hobbs}, {Klioner}, {Lammers}, {McMillan}, {Ramos-Lerate},
  {Steidelm{\"u}ller}, {Stephenson}, \& {van Leeuwen}}]{Lindegren2021}
{Lindegren}, L., {Bastian}, U., {Biermann}, M., {et~al.} 2021, \aap, 649, A4,
  \dodoi{10.1051/0004-6361/202039653}

\bibitem[{{Liu} {et~al.}(2020){Liu}, {Fu}, {Shi}, {Wu}, {Han}, {Chen}, {Dong},
  {Zhao}, {Chen}, {Zhang}, {Bai}, {Chen}, {Cui}, {Du}, {Hsia}, {Jiang}, {Hou},
  {Hou}, {Li}, {Li}, {Li}, {Liu}, {Liu}, {Luo}, {Ren}, {Tian}, {Tian}, {Wang},
  {Wu}, {Xie}, {Yan}, {Yang}, {Yu}, {Zhang}, {Zhang}, {Zhang}, {Zhang}, {Zhao},
  {Zhong}, {Zong}, \& {Zuo}}]{Liu_2020}
{Liu}, C., {Fu}, J., {Shi}, J., {et~al.} 2020, arXiv e-prints,
  arXiv:2005.07210.
\newblock \doarXiv{2005.07210}

\bibitem[{{Lomb}(1976)}]{Lomb_1976}
{Lomb}, N.~R. 1976, \apss, 39, 447, \dodoi{10.1007/BF00648343}

\bibitem[{Luo {et~al.}(2015)Luo, Zhao, Zhao, Deng, Liu, Jing, Wang, Zhang, Shi,
  Cui, Chu, Li, Bai, Wu, Cai, Cao, Cao, Carlin, Chen, Chen, Chen, Chen, Chen,
  Chen, Chen, Christlieb, Chu, Cui, Dong, Du, Fan, Feng, Fu, Gao, Gong, Gu,
  Guo, Han, He, Hou, Hou, Hou, Hu, Hu, Hu, Huo, Jia, Jiang, Jiang, Jiang, Jin,
  Kong, Kong, Lei, Li, Li, Li, Li, Li, Li, Li, Li, Li, Li, Li, Li, Liang, Lin,
  Liu, Liu, Liu, Liu, Lu, Luo, Mao, Newberg, Ni, Qi, Qi, Shen, Shi, Song, Song,
  Su, Su, Tang, Tao, Tian, Wang, Wang, Wang, Wang, Wang, Wang, Wang, Wang,
  Wang, Wang, Wang, Wang, Wang, Wang, Wang, Wang, Wang, Wang, Wang, Wang, Wei,
  Wei, Wu, Wu, Wu, Wu, Xing, Xu, Xu, Xu, Yan, Yang, Yang, Yang, Yang, Yao, Yu,
  Yuan, Yuan, Yuan, Yuan, Zhai, Zhang, Zhang, Zhang, Zhang, Zhang, Zhang,
  Zhang, Zhang, Zhao, Zhou, Zhou, Zhu, Zhu, Zou, \& Zuo}]{Luo_2015}
Luo, A.-L., Zhao, Y.-H., Zhao, G., {et~al.} 2015, Research in Astronomy and
  Astrophysics, 15, 1095, \dodoi{10.1088/1674-4527/15/8/002}

\bibitem[{{MacGregor} \& {Brenner}(1991)}]{MacGregor1991}
{MacGregor}, K.~B., \& {Brenner}, M. 1991, \apj, 376, 204,
  \dodoi{10.1086/170269}

\bibitem[{{Majewski} {et~al.}(2017){Majewski}, {Schiavon}, {Frinchaboy},
  {Allende Prieto}, {Barkhouser}, {Bizyaev}, {Blank}, {Brunner}, {Burton},
  {Carrera}, {Chojnowski}, {Cunha}, {Epstein}, {Fitzgerald}, {Garc{\'\i}a
  P{\'e}rez}, {Hearty}, {Henderson}, {Holtzman}, {Johnson}, {Lam}, {Lawler},
  {Maseman}, {M{\'e}sz{\'a}ros}, {Nelson}, {Nguyen}, {Nidever}, {Pinsonneault},
  {Shetrone}, {Smee}, {Smith}, {Stolberg}, {Skrutskie}, {Walker}, {Wilson},
  {Zasowski}, {Anders}, {Basu}, {Beland}, {Blanton}, {Bovy}, {Brownstein},
  {Carlberg}, {Chaplin}, {Chiappini}, {Eisenstein}, {Elsworth}, {Feuillet},
  {Fleming}, {Galbraith-Frew}, {Garc{\'\i}a}, {Garc{\'\i}a-Hern{\'a}ndez},
  {Gillespie}, {Girardi}, {Gunn}, {Hasselquist}, {Hayden}, {Hekker}, {Ivans},
  {Kinemuchi}, {Klaene}, {Mahadevan}, {Mathur}, {Mosser}, {Muna}, {Munn},
  {Nichol}, {O'Connell}, {Parejko}, {Robin}, {Rocha-Pinto}, {Schultheis},
  {Serenelli}, {Shane}, {Silva Aguirre}, {Sobeck}, {Thompson}, {Troup},
  {Weinberg}, \& {Zamora}}]{Majewski2017}
{Majewski}, S.~R., {Schiavon}, R.~P., {Frinchaboy}, P.~M., {et~al.} 2017, \aj,
  154, 94, \dodoi{10.3847/1538-3881/aa784d}

\bibitem[{{Manzo-Mart{\'\i}nez} {et~al.}(2020){Manzo-Mart{\'\i}nez}, {Calvet},
  {Hern{\'a}ndez}, {Lizano}, {Hern{\'a}ndez}, {Miller}, {Mauc{\'o}},
  {Brice{\~n}o}, \& {D'Alessio}}]{Manzo_2020}
{Manzo-Mart{\'\i}nez}, E., {Calvet}, N., {Hern{\'a}ndez}, J., {et~al.} 2020,
  \apj, 893, 56, \dodoi{10.3847/1538-4357/ab7ead}

\bibitem[{Marjoram {et~al.}(2015)Marjoram, Hamblin, \& Foley}]{Marjoram2015}
Marjoram, P., Hamblin, S., \& Foley, B. 2015, in Proceedings of the Conference
  on Summer Computer Simulation, SummerSim '15 (San Diego, CA, USA: Society for
  Computer Simulation International), 1–8

\bibitem[{{Matt} \& {Pudritz}(2005{\natexlab{a}})}]{Matt_2005a}
{Matt}, S., \& {Pudritz}, R.~E. 2005{\natexlab{a}}, \mnras, 356, 167,
  \dodoi{10.1111/j.1365-2966.2004.08431.x}

\bibitem[{{Matt} \& {Pudritz}(2005{\natexlab{b}})}]{Matt2005}
---. 2005{\natexlab{b}}, \apjl, 632, L135, \dodoi{10.1086/498066}

\bibitem[{Matt \& Pudritz(2005)}]{Matt_2005b}
Matt, S., \& Pudritz, R.~E. 2005, The Astrophysical Journal, 632, L135,
  \dodoi{10.1086/498066}

\bibitem[{{Matt} \& {Pudritz}(2008{\natexlab{a}})}]{Matt_2008a}
{Matt}, S., \& {Pudritz}, R.~E. 2008{\natexlab{a}}, \apj, 681, 391,
  \dodoi{10.1086/587453}

\bibitem[{{Matt} \& {Pudritz}(2008{\natexlab{b}})}]{Matt_2008b}
---. 2008{\natexlab{b}}, \apj, 678, 1109, \dodoi{10.1086/533428}

\bibitem[{{Matt} {et~al.}(2010){Matt}, {Pinz{\'o}n}, {de la Reza}, \&
  {Greene}}]{Matt_2010}
{Matt}, S.~P., {Pinz{\'o}n}, G., {de la Reza}, R., \& {Greene}, T.~P. 2010,
  \apj, 714, 989, \dodoi{10.1088/0004-637X/714/2/989}

\bibitem[{{Matt} {et~al.}(2012){Matt}, {Pinz{\'o}n}, {Greene}, \&
  {Pudritz}}]{Matt2012}
{Matt}, S.~P., {Pinz{\'o}n}, G., {Greene}, T.~P., \& {Pudritz}, R.~E. 2012,
  \apj, 745, 101, \dodoi{10.1088/0004-637X/745/1/101}

\bibitem[{{Natta} {et~al.}(2014){Natta}, {Testi}, {Alcal{\'a}}, {Rigliaco},
  {Covino}, {Stelzer}, \& {D'Elia}}]{Natta2014}
{Natta}, A., {Testi}, L., {Alcal{\'a}}, J.~M., {et~al.} 2014, \aap, 569, A5,
  \dodoi{10.1051/0004-6361/201424136}

\bibitem[{Nguyen {et~al.}(2009)Nguyen, Jayawardhana, van Kerkwijk, Brandeker,
  Scholz, \& Damjanov}]{Nguyen_2009}
Nguyen, D.~C., Jayawardhana, R., van Kerkwijk, M.~H., {et~al.} 2009, The
  Astrophysical Journal, 695, 1648, \dodoi{10.1088/0004-637x/695/2/1648}

\bibitem[{{Olney} {et~al.}(2020){Olney}, {Kounkel}, {Schillinger}, {Scoggins},
  {Yin}, {Howard}, {Covey}, {Hutchinson}, \& {Stassun}}]{Olney_2020}
{Olney}, R., {Kounkel}, M., {Schillinger}, C., {et~al.} 2020, \aj, 159, 182,
  \dodoi{10.3847/1538-3881/ab7a97}

\bibitem[{{Pantolmos} {et~al.}(2020){Pantolmos}, {Zanni}, \&
  {Bouvier}}]{Pantolmos_2020}
{Pantolmos}, G., {Zanni}, C., \& {Bouvier}, J. 2020, A\&A, 643, A129,
  \dodoi{10.1051/0004-6361/202038569}

\bibitem[{{Pascucci} {et~al.}(2022){Pascucci}, {Cabrit}, {Edwards}, {Gorti},
  {Gressel}, \& {Suzuki}}]{Pascucci2022}
{Pascucci}, I., {Cabrit}, S., {Edwards}, S., {et~al.} 2022, arXiv e-prints,
  arXiv:2203.10068.
\newblock \doarXiv{2203.10068}

\bibitem[{{Pecaut} \& {Mamajek}(2013)}]{Pecaut2013}
{Pecaut}, M.~J., \& {Mamajek}, E.~E. 2013, \apjs, 208, 9,
  \dodoi{10.1088/0067-0049/208/1/9}

\bibitem[{Pinz{\'{o}}n {et~al.}(2021)Pinz{\'{o}}n, Hern{\'{a}}ndez, Serna,
  Garc{\'{\i}}a, Manzo-Mart{\'{\i}}nez, Roman-Lopes,
  Rom{\'{a}}n-Z{\'{u}}{\~{n}}iga, Batista, Ram{\'{\i}}rez-V{\'{e}}lez, Osorio,
  \& Avenda{\~{n}}o}]{Pinzon2021}
Pinz{\'{o}}n, G., Hern{\'{a}}ndez, J., Serna, J., {et~al.} 2021, The
  Astronomical Journal, 162, 90, \dodoi{10.3847/1538-3881/ac04ae}

\bibitem[{R{\'{e}}ville {et~al.}(2015)R{\'{e}}ville, Brun, Matt, Strugarek, \&
  Pinto}]{Reville_2015}
R{\'{e}}ville, V., Brun, A.~S., Matt, S.~P., Strugarek, A., \& Pinto, R.~F.
  2015, The Astrophysical Journal, 798, 116,
  \dodoi{10.1088/0004-637x/798/2/116}

\bibitem[{{Roggero} {et~al.}(2021){Roggero}, {Bouvier}, {Rebull}, \&
  {Cody}}]{Roggero_2021}
{Roggero}, N., {Bouvier}, J., {Rebull}, L.~M., \& {Cody}, A.~M. 2021, \aap,
  651, A44, \dodoi{10.1051/0004-6361/202140646}

\bibitem[{{Scargle}(1982)}]{Scargle_1982}
{Scargle}, J.~D. 1982, \apj, 263, 835, \dodoi{10.1086/160554}

\bibitem[{{Schatzman}(1962)}]{Schatzman_1962}
{Schatzman}, E. 1962, Annales d'Astrophysique, 25, 18

\bibitem[{Scott(1979)}]{Scott_1979}
Scott, D.~W. 1979, Biometrika, 66, 605, \dodoi{10.1093/biomet/66.3.605}

\bibitem[{{Serna} {et~al.}(2021){Serna}, {Hernandez}, {Kounkel},
  {Manzo-Mart{\'\i}nez}, {Roman-Lopes}, {Rom{\'a}n-Z{\'u}{\~n}iga}, {Batista},
  {Pinz{\'o}n}, {Calvet}, {Brice{\~n}o}, {Tapia}, {Su{\'a}rez}, {Ram{\'\i}rez},
  {G. Stassun}, {Covey}, {Vargas-Gonz{\'a}lez}, \&
  {Fern{\'a}ndez-Trincado}}]{Serna_2021}
{Serna}, J., {Hernandez}, J., {Kounkel}, M., {et~al.} 2021, \apj, 923, 177,
  \dodoi{10.3847/1538-4357/ac300a}

\bibitem[{{Shu} {et~al.}(1988){Shu}, {Lizano}, {Ruden}, \& {Najita}}]{Shu_1988}
{Shu}, F.~H., {Lizano}, S., {Ruden}, S.~P., \& {Najita}, J. 1988, \apjl, 328,
  L19, \dodoi{10.1086/185152}

\bibitem[{{Simon} {et~al.}(2016){Simon}, {Pascucci}, {Edwards}, {Feng},
  {Gorti}, {Hollenbach}, {Rigliaco}, \& {Keane}}]{simon2016}
{Simon}, M.~N., {Pascucci}, I., {Edwards}, S., {et~al.} 2016, \apj, 831, 169,
  \dodoi{10.3847/0004-637X/831/2/169}

\bibitem[{{Smette} {et~al.}(2015){Smette}, {Sana}, {Noll}, {Horst}, {Kausch},
  {Kimeswenger}, {Barden}, {Szyszka}, {Jones}, {Gallenne}, {Vinther},
  {Ballester}, \& {Taylor}}]{smette2015}
{Smette}, A., {Sana}, H., {Noll}, S., {et~al.} 2015, \aap, 576, A77,
  \dodoi{10.1051/0004-6361/201423932}

\bibitem[{{Somers} {et~al.}(2020){Somers}, {Cao}, \&
  {Pinsonneault}}]{Somers2020}
{Somers}, G., {Cao}, L., \& {Pinsonneault}, M.~H. 2020, \apj, 891, 29,
  \dodoi{10.3847/1538-4357/ab722e}

\bibitem[{{Stassun} \& {Torres}(2021)}]{Stassun_2021}
{Stassun}, K.~G., \& {Torres}, G. 2021, \apjl, 907, L33,
  \dodoi{10.3847/2041-8213/abdaad}

\bibitem[{{Still} \& {Barclay}(2012)}]{PyKE}
{Still}, M., \& {Barclay}, T. 2012, {PyKE: Reduction and analysis of Kepler
  Simple Aperture Photometry data}.
\newblock \doeprint{1208.004}

\bibitem[{Strugarek {et~al.}(2015)Strugarek, Brun, Matt, \&
  R{\'{e}}ville}]{Strugarek_2015}
Strugarek, A., Brun, A.~S., Matt, S.~P., \& R{\'{e}}ville, V. 2015, The
  Astrophysical Journal, 815, 111, \dodoi{10.1088/0004-637x/815/2/111}

\bibitem[{Swain \& Ballard(1991)}]{Swain_1991}
Swain, M.~J., \& Ballard, D.~H. 1991, International Journal of Computer Vision,
  7, 11.
\newblock \url{https://api.semanticscholar.org/CorpusID:8167136}

\bibitem[{{Thanathibodee} {et~al.}(2022){Thanathibodee}, {Calvet},
  {Hern{\'a}ndez}, {Mauc{\'o}}, \& {Brice{\~n}o}}]{Thanathibodee_2022}
{Thanathibodee}, T., {Calvet}, N., {Hern{\'a}ndez}, J., {Mauc{\'o}}, K., \&
  {Brice{\~n}o}, C. 2022, \aj, 163, 74, \dodoi{10.3847/1538-3881/ac3ee6}

\bibitem[{Tokovinin {et~al.}(2020)Tokovinin, Petr-Gotzens, \&
  Briceño}]{Tokovinin_2020}
Tokovinin, A., Petr-Gotzens, M.~G., \& Briceño, C. 2020, The Astronomical
  Journal, 160, 268, \dodoi{10.3847/1538-3881/abc2d6}

\bibitem[{{Tout} \& {Pringle}(1992)}]{Tout1992}
{Tout}, C.~A., \& {Pringle}, J.~E. 1992, \mnras, 256, 269,
  \dodoi{10.1093/mnras/256.2.269}

\bibitem[{Turner \& {Van Zandt}(2012)}]{Turner2012}
Turner, B.~M., \& {Van Zandt}, T. 2012, Journal of Mathematical Psychology, 56,
  69, \dodoi{https://doi.org/10.1016/j.jmp.2012.02.005}

\bibitem[{{Uzdensky}(2004)}]{Uzdensky2004}
{Uzdensky}, D.~A. 2004, \apss, 292, 573,
  \dodoi{10.1023/B:ASTR.0000045064.93078.87}

\bibitem[{{Vernet} {et~al.}(2011){Vernet}, {Dekker}, {D\'{}Odorico}, {Kaper,
  L.}, {Kjaergaard, P.}, {Hammer, F.}, {Randich, S.}, {Zerbi, F.}, {Groot, P.
  J.}, {Hjorth, J.}, {Guinouard, I.}, {Navarro, R.}, {Adolfse, T.}, {Albers, P.
  W.}, {Amans, J.-P.}, {Andersen, J. J.}, {Andersen, M. I.}, {Binetruy, P.},
  {Bristow, P.}, {Castillo, R.}, {Chemla, F.}, {Christensen, L.}, {Conconi,
  P.}, {Conzelmann, R.}, {Dam, J.}, {De Caprio, V.}, {De Ugarte Postigo, A.},
  {Delabre, B.}, {Di Marcantonio, P.}, {Downing, M.}, {Elswijk, E.}, {Finger,
  G.}, {Fischer, G.}, {Flores, H.}, {Fran\c{c}ois, P.}, {Goldoni, P.},
  {Guglielmi, L.}, {Haigron, R.}, {Hanenburg, H.}, {Hendriks, I.}, {Horrobin,
  M.}, {Horville, D.}, {Jessen, N. C.}, {Kerber, F.}, {Kern, L.}, {Kiekebusch,
  M.}, {Kleszcz, P.}, {Klougart, J.}, {Kragt, J.}, {Larsen, H. H.}, {Lizon,
  J.-L.}, {Lucuix, C.}, {Mainieri, V.}, {Manuputy, R.}, {Martayan, C.}, {Mason,
  E.}, {Mazzoleni, R.}, {Michaelsen, N.}, {Modigliani, A.}, {Moehler, S.},
  {M\o{}ller, P.}, {Norup S\o{}rensen, A.}, {N\o{}rregaard, P.}, {P\'eroux,
  C.}, {Patat, F.}, {Pena, E.}, {Pragt, J.}, {Reinero, C.}, {Rigal, F.}, {Riva,
  M.}, {Roelfsema, R.}, {Royer, F.}, {Sacco, G.}, {Santin, P.}, {Schoenmaker,
  T.}, {Spano, P.}, {Sweers, E.}, {Ter Horst, R.}, {Tintori, M.}, {Tromp, N.},
  {van Dael, P.}, {van der Vliet, H.}, {Venema, L.}, {Vidali, M.}, {Vinther,
  J.}, {Vola, P.}, {Winters, R.}, {Wistisen, D.}, {Wulterkens, G.}, \&
  {Zacchei, A.}}]{Vernet_2011}
{Vernet}, J., {Dekker}, H., {D\'{}Odorico}, S., {et~al.} 2011, A\&A, 536, A105,
  \dodoi{10.1051/0004-6361/201117752}

\bibitem[{Vidotto {et~al.}(2014)Vidotto, Gregory, Jardine, Donati, Petit,
  Morin, Folsom, Bouvier, Cameron, Hussain, Marsden, Waite, Fares, Jeffers, \&
  do~Nascimento}]{vidotto_2014}
Vidotto, A.~A., Gregory, S.~G., Jardine, M., {et~al.} 2014, Monthly Notices of
  the Royal Astronomical Society, 441, 2361, \dodoi{10.1093/mnras/stu728}

\bibitem[{Watson {et~al.}(2016)Watson, Calvet, Fischer, Forrest, Manoj,
  Megeath, Melnick, Najita, Neufeld, Sheehan, Stutz, \& Tobin}]{Watson_2016}
Watson, D.~M., Calvet, N.~P., Fischer, W.~J., {et~al.} 2016, The Astrophysical
  Journal, 828, 52, \dodoi{10.3847/0004-637x/828/1/52}

\bibitem[{{Weber} \& {Davis}(1967)}]{Weber1967}
{Weber}, E.~J., \& {Davis}, Leverett, J. 1967, \apj, 148, 217,
  \dodoi{10.1086/149138}

\bibitem[{{Yi}(1994)}]{Yi1994}
{Yi}, I. 1994, \apj, 428, 760, \dodoi{10.1086/174283}

\bibitem[{{Zanni} \& {Ferreira}(2013)}]{Zanni2013}
{Zanni}, C., \& {Ferreira}, J. 2013, \aap, 550, A99,
  \dodoi{10.1051/0004-6361/201220168}

\end{thebibliography}

\end{document}